\PassOptionsToPackage{table}{xcolor}
\documentclass[]{article} %draft
\usepackage{arxiv}

\usepackage[utf8]{inputenc} % allow utf-8 input
\usepackage[T1]{fontenc}    % use 8-bit T1 fonts
\usepackage[english]{babel}
\usepackage{hyperref}       % hyperlinks
\usepackage{url}            % simple URL typesetting
\usepackage{doi}
\usepackage{microtype}      % microtypography

\usepackage{graphicx}
\usepackage{tikz}
\usepackage{tcolorbox}
\usepackage{pgfplots}
\usetikzlibrary{patterns, arrows, shapes}
\usetikzlibrary{intersections, pgfplots.fillbetween}
\pgfplotsset{compat=1.18} 

\usepackage{caption}
\usepackage{subcaption}
\usepackage{rotating}

\usepackage{dsfont, mathrsfs, bm}
\usepackage{nicefrac}      
\usepackage{amsmath, amssymb, amsthm, mathtools}
\usepackage{threeparttable}
\newcolumntype{L}{>{\raggedright\arraybackslash}X}

\usepackage[ruled,linesnumbered]{algorithm2e}

\usepackage{makecell, multirow, tabularx}
\usepackage{tabularray}
\usepackage{booktabs}     

\newcommand{\rowgroup}[1]{\hspace{-0.4em}#1}

\usepackage[round, authoryear]{natbib}
\title{Goodness-of-fit tests for spatial point processes:\\A review}
\date{}	
\author{Chiara~Fend\\
	Department of Mathematics\\
	RPTU University Kaiserslautern-Landau\\
	Kaiserslautern, Germany \\
	\texttt{chiara.fend@rptu.de} \\
	\And Claudia~Redenbach \\
	Department of Mathematics\\
	RPTU University Kaiserslautern-Landau\\
	Kaiserslautern, Germany \\
	\texttt{claudia.redenbach@rptu.de}
}

\renewcommand{\shorttitle}{Goodness-of-fit tests for spatial point processes}

\hypersetup{pdftitle={GOF tests for spatial point processes: A review},
            pdfsubject={A review of the classical approaches and recent developments in goodness-of-fit testing for spatial point processes},
            pdfauthor={Chiara~Fend, Claudia~Redenbach},
            pdfkeywords={spatial point process, model validation, goodness-of-fit, TDA}
}

\DeclareMathOperator{\R}{\mathbb{R}}
\DeclareMathOperator{\N}{\mathbb{N}}

\DeclareMathOperator{\calP}{\mathcal{P}}
\DeclareMathOperator{\calR}{\mathcal{R}}
\DeclareMathOperator{\calN}{\mathcal{N}}

\newcommand{\1}[1]{\mathds{1}\!\left(#1\right)}

\DeclarePairedDelimiterX{\norm}[1]{\lVert}{\rVert}{#1}

\DeclareMathOperator{\X}{\mathbf{X}}        % point process
\DeclareMathOperator{\Y}{\mathbf{Y}}        % point process
\DeclareMathOperator{\x}{\mathbf{x}}        % point pattern
\DeclareMathOperator{\y}{\mathbf{y}}
\DeclareMathOperator{\z}{\mathbf{z}}

\begin{document}
\maketitle

% ........................... Abstract .............................

\begin{abstract}
In this review, the state-of-the-art for goodness-of-fit testing for spatial point processes is summarized. Test statistics based on classical functional summary statistics and recent contributions from topological data analysis are considered. Different approaches to derive test statistics from functional summary statistics are categorized in a unifying notation. We discuss additional aspects such as the graphical representation in terms of global envelopes and the selection of the parameters in the individual tests.
\end{abstract}

\keywords{model validation \and Monte Carlo tests\and topological data analysis \and global envelope test \and complete spatial randomness}

% ............................ Intro ...............................

\section{Introduction}
Spatial point patterns are observed in a wide range of applications. Planar point patterns are formed by the locations of trees in a forest, of accidents in a city or of disease outbreaks in a county. Point patterns in 3D are obtained when recording particle centers in a material or coordinates of stars in a galaxy. Analysis and modelling of spatial point processes is an active field of research in spatial statistics. In recent years, statistics for spatial point processes has seen many new developments. In particular, new parametric model families have been proposed whose parameters can be estimated by novel approaches in the context of estimating functions \citep[c.f.][]{moller2017}. Consequently, model selection and validation are as important as ever. 

Here, we focus on model validation via goodness-of-fit testing. In this field, the main new developments are the following. The first is the introduction of global envelope tests \citep{myllymaki2017,mrkvicka2017, mrkvicka2022, GET}. These Monte Carlo tests quickly became popular in spatial statistics as they provide graphical insights into the fit of a statistical model. The second development comes from the field of topological data analysis (TDA). TDA methods are used to define new functional summary statistics that extract topological characteristics of point patterns \citep{robins2016, biscio2019, biscio2020}. A third development lies in new (functional) central limit theorems for several empirical functional summary statistics that can be used for asymptotic goodness-of-fit tests \citep{blaszczyszyn2019, biscio2020, biscio2022}. Last but not least, the application of scoring rules such as the continuous ranked probability score to functional summary statistics allows  the set of test statistics to be extended \citep{heinrichmertsching2024}.

All these contributions form a large set of possible goodness-of-fit tests that one can use to infer on the fit of a spatial point process model. The first aim of this review is therefore to bring the different contributions to a unifying notation. Then, we established a generic framework which includes the popular and widely used tests as specific combinations of a functional summary statistic, a test statistic and a test procedure as individual components.

The remainder of this review is organized as follows. In Section~\ref{sec:setting} we introduce the main setting and fundamental concepts related to spatial point processes. Then, we describe both the classical and new functional summary statistics for stationary and isotropic point processes (Section~\ref{sec:summary-stat}). In Section~\ref{sec:test-statistic} we introduce three categories of test statistics, in particular incorporating the continuous ranked probability score and the test statistics used in both the global envelope tests and in the new asymptotic tests. The next aspect that we discuss is how the test decision is made. This consists of a brief introduction to Monte Carlo tests which includes orderings of sets of vectors (Section~\ref{sec:mc}) and the asymptotic results for functional summary statistics (Section~\ref{sec:CLT}). 
Further aspects of goodness-of-fit are discussed in Section~\ref{sec:additional}. This includes the special hypothesis of complete spatial randomness and the graphical representation of test statistics.
Finally, we present an overview of how previous power studies for the hypothesis of complete spatial randomness fit into our general framework (Section~\ref{sec:powerstudy}). This allows us to identify the combinations of the individual components that have -- to the best of our knowledge -- not yet been studied.
We conclude with a discussion (Section~\ref{sec:conclusion}).

A forthcoming comparative power study will accompany this review. The aim is to provide better recommendations for the selection of individual components that provide powerful tests against many alternatives. In particular, it will be used to fill the gaps highlighted in Section~\ref{sec:powerstudy}.

% ........................... Setting ..............................

\section{Preliminaries}
\label{sec:setting}

% - - - - - - - - - - - - Point Processes - - - - - - - - - - - - - -

Throughout the paper, we will use the following notation.
A spatial point process $\X$ on $\R^d$, $d > 1$, is a random locally finite counting measure. We denote the number of points of $\X$ observed in any Borel set $B \in \mathscr{B}^d$ by $\X(B)$. We assume that $\X$ is a simple point process which means that $\X(\{x\}) \leq 1$ almost surely for all $x\in \R^d$. We often identify the simple point process $\X$ with its support $\{x \in \R^d \mid \X(\{x\}) > 0\} = \{X_1, \dots, X_n\} \subset \R^d$ with $n \in \N \cup \{\infty\}$ which we call a (random) point pattern. The set of all such locally finite random subsets is denoted by $\calN$. Note that the identification between simple counting measures and point patterns is a bijection \citep[see e.g.][]{schneider2008}. Consequently, we use both interpretations interchangeably throughout this paper. 
We call a spatial point process on $\R^d$ stationary if its distribution is invariant under translations and isotropic if its distribution is invariant under rotations around the origin.

The distribution of the counts is characterized by the $k$th order factorial moment measures $\alpha^{(k)}$ for $k\in\N$. 
These are measures defined on $\R^{dk}$ and for Borel sets $B_1,\dots, B_k \in \mathscr{B}^d$ we have \begin{equation*}
    \alpha^{(k)}(B_1 \times \dots \times B_k) = \mathbb{E}\left( \quad \sideset{}{^{\neq}} \sum_{x_1, \dots, x_k \in \X} \1{x_1 \in B_1} \cdot \,\dots\, \cdot \1{x_k \in B_k}\right)
\end{equation*}
where $\sum^{\neq}$ denotes the summation only over $k$-tuples of pairwise distinct points and $\1{\cdot \in A}$ denotes the indicator function of the set $A$.

We assume here that $\alpha^{(k)}$ is absolutely continuous and has a density with respect to the Lebesgue measure. This density will be denoted by $\rho^{(k)}$ and is called the $k$th order product density or $k$th order joint intensity function. Intuitively, we can think of $\rho^{(k)}(x_1, \dots, x_k) \mathrm{d}x_1 \cdots \mathrm{d}x_k$ for pairwisely distinct positions $x_1, \dots, x_k$ as the probability of simultaneously observing a point in each of the $k$ infinitesimal sets around $x_1, \dots, x_k$. For $k=1$, we call $\rho^{(1)} = \rho$ the intensity function. For stationary spatial point processes, $\rho$ is a constant function whose value $\lambda > 0$ is called the intensity of the point process.

In practice we do not observe $\X$ in the entire Euclidean space $\R^d$ but restrict the observation to a bounded observation window $W$. We denote the restriction of $\X$ to $W$ by $\X_W$. The set of all (random) point patterns on $W$ is denoted by $\calN_W = \{ \X_W \mid \X \in \calN \}$. By local finiteness of $\X$ and boundedness of $W$, we have $\X(W) = \X_W(W) < \infty$.

In point process statistics, specific properties of a point process are often summarized using functional summary statistics. These statistics are then used for various tasks such as parameter estimation or, as in our setting, hypothesis testing. A general functional summary statistic or simply summary statistic $T$ can hereby be defined as a mapping \[
T\!:\calN \times \calR\to \R
\] where $\calR$ contains all possible evaluation points for the specific summary statistic. Usually, $\calR$ is either a bounded interval in $\R_{\geq 0}$ representing spatial scales or $\calR$ coincides with the observation window $W$. \\ An estimator $\widehat{T}$ of the summary statistic $T$ computed from a single observed point pattern $\x_0 \in \calN_W$ has the form \[
\widehat{T}\!:\calN_W \times \calR \to \R.
\]

% - - - - - - - - - - - - - - Hypotheses - - - - - - - - - - - - - -

In this review we focus on goodness-of-fit testing. For this, we generally have to distinguish between simple and composite null hypotheses.
In a simple hypothesis setting, have a point process $\X$ with unknown distribution $P$ and test the hypothesis \begin{equation*}
	H_0: P = P_0 \quad \text{vs.} \quad H_1: P \neq P_0
\end{equation*}
where $P_0$ is a fully specified model distribution. In this case, we do not need to estimate any parameters of the proposed model. 

For a composite hypothesis we consider a parametric model family $\mathcal{P}_\Theta = \{P_\theta \mid \theta \in \Theta\}$ with some parameter set $\Theta \subset \R^p$, $p \geq 1$ and want to test
\begin{equation*}
	H_0: P \in \mathcal{P}_\Theta \quad \text{vs.} \quad H_1: P \notin \mathcal{P}_\Theta.
\end{equation*}
Note that the general composite hypothesis also includes cases where some parameters of the model are fixed and some need to be estimated. 

We are interested in testing different kinds of possible null models. Consequently, we do not want to restrict the null hypotheses $P_0$ and $\mathcal{P}_\Theta$ to specific models, e.g., complete spatial randomness. For this reason, our review will only consider goodness-of-fit tests that are applicable to general hypotheses or -- as in the case of asymptotic tests -- are applicable for a wide range of spatial point process models. Specialized tests that can only be used to test the null hypothesis of complete spatial randomness are summarized in Section~\ref{sec:csr-hypo}.

Such general goodness-of-fit tests that have been proposed in the literature all have in common that they are based on a functional summary statistic $T$. This summary statistic specifies the characteristic for which the goodness-of-fit of the spatial point process is to be tested. Since this characteristic alone does not fully specify the point process model, the exact hypotheses tested depend on the chosen summary statistic. As a result, only point process models that differ with respect to this specific characteristic can be distinguished.

For the test, we need an empirical version of the functional summary statistic in the form of a non-parametric estimator $\widehat{T}$ as well as a subset $\calR^*$ of the domain $\calR$ of $T$ that captures the spatial scales of interest. This functional estimator is then aggregated into a suitable test statistic $D$.

Eventually, we need to decide whether the observed value of this test statistic is extreme under the null hypothesis. To do this, we need an ordering $\preceq$ on the range of $D$ as well as methods for computing or approximating either the critical value or the $p$-value. In what follows, we will first review proposals for all aspects separately.

% ........................ Summary statistics ........................................

\section{Functional summary statistics}\label{sec:summary-stat}

Functional summary statistics are used for both model fitting and model validation. Each statistic focuses on a different characteristic of the point process in question. We will not discuss the estimation of the individual characteristics in this paper, but will concentrate only on the different summary statistics themselves. In the context of hypothesis testing, the choice of the test statistic and the relevant spatial scales $\calR^*$ used in combination with the summary statistic often have a higher influence on the power of the test than the choice of estimator \citep[c.f.][p.327]{baddeley2000, ho2009}. Nevertheless, the influence of $\calR^*$ on the power can sometimes be reduced by choosing specific estimators \citep[e.g.][]{ho2006}.

In Section~\ref{sec:classical-summary} we provide an overview of the classical functional summary statistics defined for stationary and isotropic point process models. A review of statistics for inhomogeneous or anisotropic models can be found in \citet{moller2017}. Section~\ref{sec:tda-summary} focuses on the recent advances in using topological data analysis for characterizing spatial point processes. Finally, Section~\ref{sec:add-summary} shortly discusses the general approach of using characteristics from spatial structures such as graphs or tessellations associated with the point process.

% - - - - - - - - - - - - Classical statistics - - - - - - - - - - - - - - - - - - - -

\subsection{Classical summary statistics}\label{sec:classical-summary}

All definitions in this section are taken from \citet{moller2003}. This book provides interpretations of the functions and suggestions for nonparametric estimators. We assume here that the point process $\X$ is isotropic and stationary with intensity $\lambda > 0$. The frequently used classical summary statistics for spatial point processes describe either second-order characteristics or are distance-based. 

The second-order structure of a spatial point process is usually analyzed using Ripley's $K$-function \citep{ripley1976, ripley1977}. The quantity $\lambda K(\X,r)$ is the expected number of further points in a ball with radius $r$ around an arbitrary point of the point process. Formally, this can be defined in terms of the expectation $\mathbb{E}^{!,0}$ with respect to the reduced Palm distribution $\mathbb{P}^{!,0}$. For a stationary point process, this distribution can be seen as the conditional distribution of $\X \setminus \{0\}$ given that $\X$ has a point in the origin $0$. The distribution $\mathbb{P}^{!,0}$ is often also interpreted as the distribution of the further points of $\X$ given a \textit{typical} point of $\X$ \citep[Appendix~C of][]{moller2003}.
Using the reduced Palm distribution and the corresponding expectation we can write 
 \begin{equation}
	K(\X, r) = \frac{1}{\lambda}\mathbb{E}^{!,0}\!\left[\X(B_r(0))\right], \quad r\geq 0
\end{equation}
where $B_r(x) = \{ y\in \R^d \mid \norm{y-x} \leq r\}$ denotes the closed ball with radius $r$ centered in $x\in\R^d$.

\citet{besag1977L} proposed a transformation of Ripley's $K$-function which results in what is called the $L$-function
\begin{equation}
	L(\X, r) = (K(\X, r)/\omega_d)^{1/d} , \quad r\geq 0,
\end{equation} where $\omega_d$ is the volume of the $d$-dimensional unit ball. 
This transformation stabilizes the variance of the common nonparametric estimators of the $K$-function w.r.t. the range variable $r$.

The joint distribution of pairs of points is also characterized by the pair correlation function $pcf$ which is a normalization of the second-order product density $\rho^{(2)}$.  If both $\rho$ and $\rho^{(2)}$ exist then for any $x,y\in \R^d$
\begin{equation}
    pcf(\X, (x,y)) = \frac{\rho^{(2)}(\X, (x,y))}{\rho(\X, x)\rho(\X, y)}.
\end{equation}

For stationary and isotropic point processes as considered here, the $pcf$ only depends on the distance $r=\norm{x-y} \geq 0$ between the points $x$ and $y$ and we write \begin{equation}
    pcf(\X, r) = pcf(\X, (x,y)).
\end{equation}

Compared to the $K$- and $L$-function, the pair correlation function provides a non-cumulative analysis of the second-order structure.

Other classical summary statistics are defined by using distances from a \textit{test location} to the closest point of the point process. For a stationary point process, we can choose the origin as test location. The empty space function $F$ is then defined as the distribution function of the random distance from the origin to the closest point in $\X$, i.e., \begin{equation}
	F(\X, r) = \mathbb{P}(\X(B_0(r)) > 0), \quad r\geq 0.
\end{equation}

If we are interested in the distance from a point of $\X$ to the closest other point in $\X$ we can define the nearest neighbor distance distribution function $G$. In terms of the aforementioned reduced Palm distribution this results in \begin{equation}
	G(\X, r) = \mathbb{P}^{!,0}(\X(B_0(r)) > 0), \quad r\geq 0.
\end{equation}
The difference to the $F$-function is that we condition on $\X$ having a point at the origin.
For similar reasons as for the $L$-function, \citet{baddeley2014} additionally consider the variance-stabilized $G$-function
\begin{equation}
	G^\bigstar\!(\X, r) = \arcsin(\sqrt{G(\X, r)}), \quad r\geq 0.
\end{equation}

Finally, the $J$-function was introduced in \citet{vanLieshout1996} as a combination of $G$ and $F$ which is defined as \begin{equation}
	J(\X,r) = \frac{1-G(\X, r)}{1-F(\X, r)}, \quad r \geq 0 \text{ such that } F(\X, r) < 1.
\end{equation}

% - - - - - - - - - - - - TDA - - - - - - - - - - - - - - - - - - - - - - - - - - - -

\subsection{TDA-based summary statistics}\label{sec:tda-summary}

Topological data analysis (TDA) combines aspects of algebraic topology with algorithms from computational geometry to investigate topological and geometric properties of high-dimensional structures. Recently, tools from TDA have also gained popularity in the field of spatial statistics \citep[e.g.][]{robins2016,robinson2017, biscio2019, biscio2020, dogas2024}.  With TDA it is possible to define new functional summary statistics for spatial point processes that focus on the shape of the point patterns.

We will not rigorously explain the algebraic concepts behind TDA here, but rather give an intuitive geometric interpretation. For a detailed (mathematical) introduction see the influential book by \citet{edelsbrunner2010}. For a more statistical view of TDA we refer to the recent review by \citet{chazal2021}. An overview of the topology of random geometric complexes can be found in \citet{bobrowski2018}.

The topological structure being analyzed is often provided by a filtration of simplicial complexes, which are built from the collection of points and represent the space at different scales. A simplicial complex $\mathcal{K}$ is a collection of simplices, which means that $\mathcal{K}$ contains single points, line segments, triangles, tetrahedrons and $k$-simplices for higher dimensions $k \geq 4$. The dimension $k$ of a simplex is hereby the number of vertices minus one, i.e. a single point is a $0$-dimensional simplex. Furthermore, for a simplicial complex $\mathcal{K}$ we require that every face of a simplex in $\mathcal{K}$ is again a simplex in $\mathcal{K}$ and that the intersection of two simplices in $\mathcal{K}$ is either the empty set or a common face of both. 

There are multiple ways how a filtration of complexes can be built from a finite observed point pattern $\x_0~=~\{x_1,\dots, x_n\} \subset W$. The most popular approaches are the \v{C}ech, the Vietoris-Rips and the alpha complexes \citep{edelsbrunner2010}. All three complexes have an intuitive geometric construction which we explain in the following. Let $V(\sigma) = \{x_{j_0}, \dots, x_{j_k}\} \subset \x_0$ be the set of vertices of the $k$-simplex $\sigma$. 

\begin{figure}
	\subfloat[point pattern]{\includegraphics[width=0.3\textwidth]{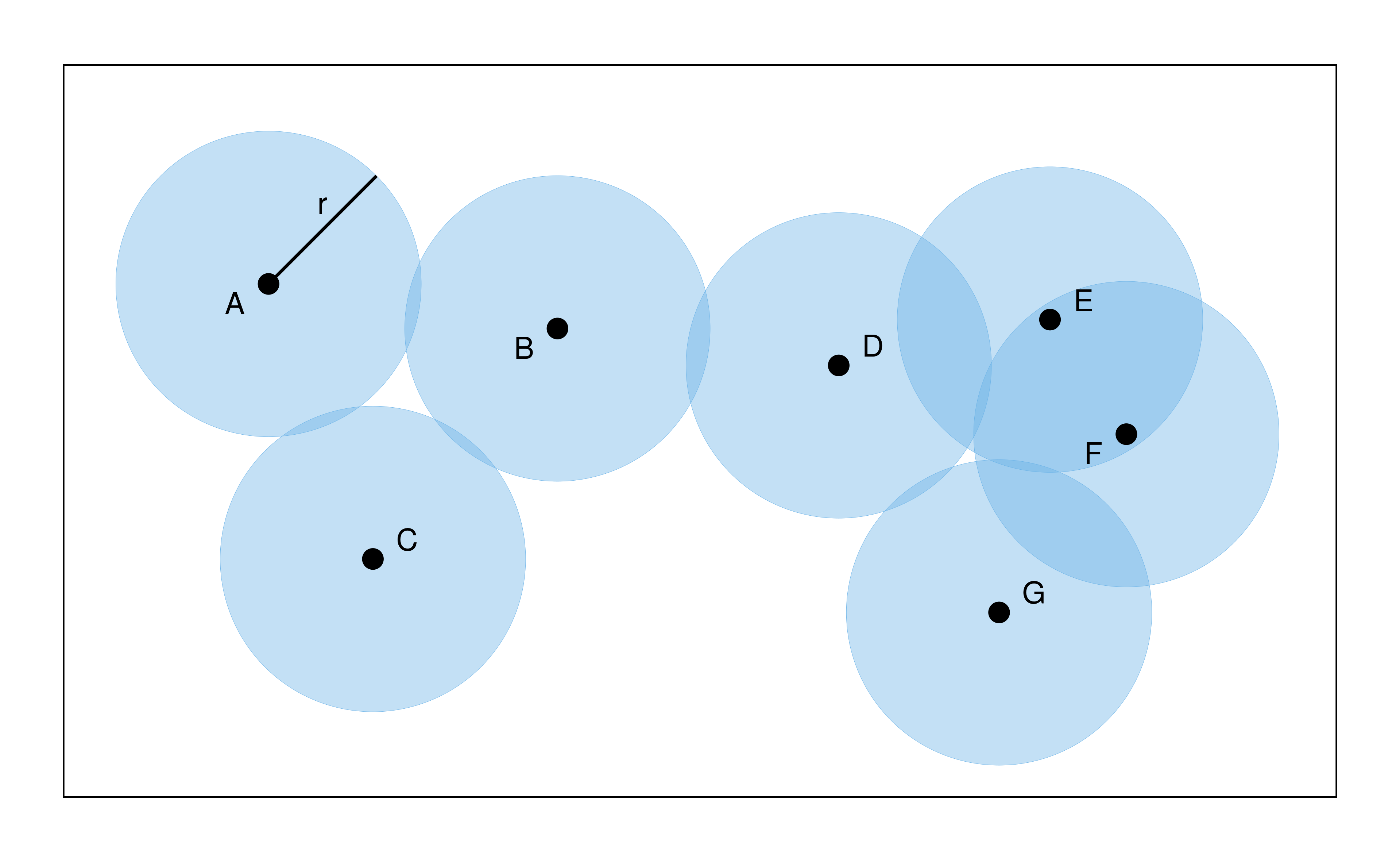}} \hfill
	\subfloat[\v{C}ech complex $\operatorname{Cech}(r)$]{\includegraphics[width=0.34\textwidth]{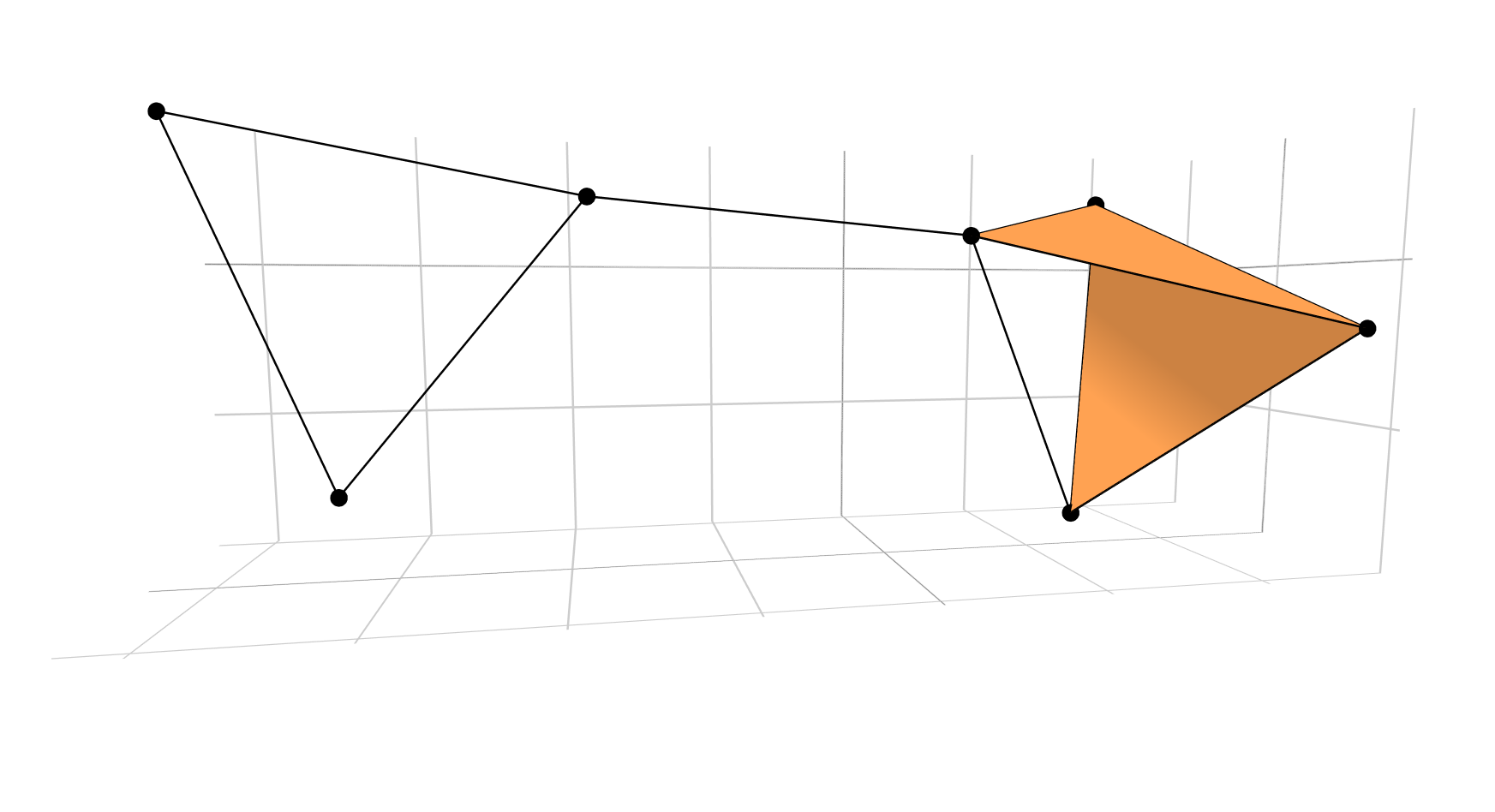}}\hfill 
	\subfloat[Vietoris-Rips complex $\operatorname{VR}(r)$]{\includegraphics[width=0.34\textwidth]{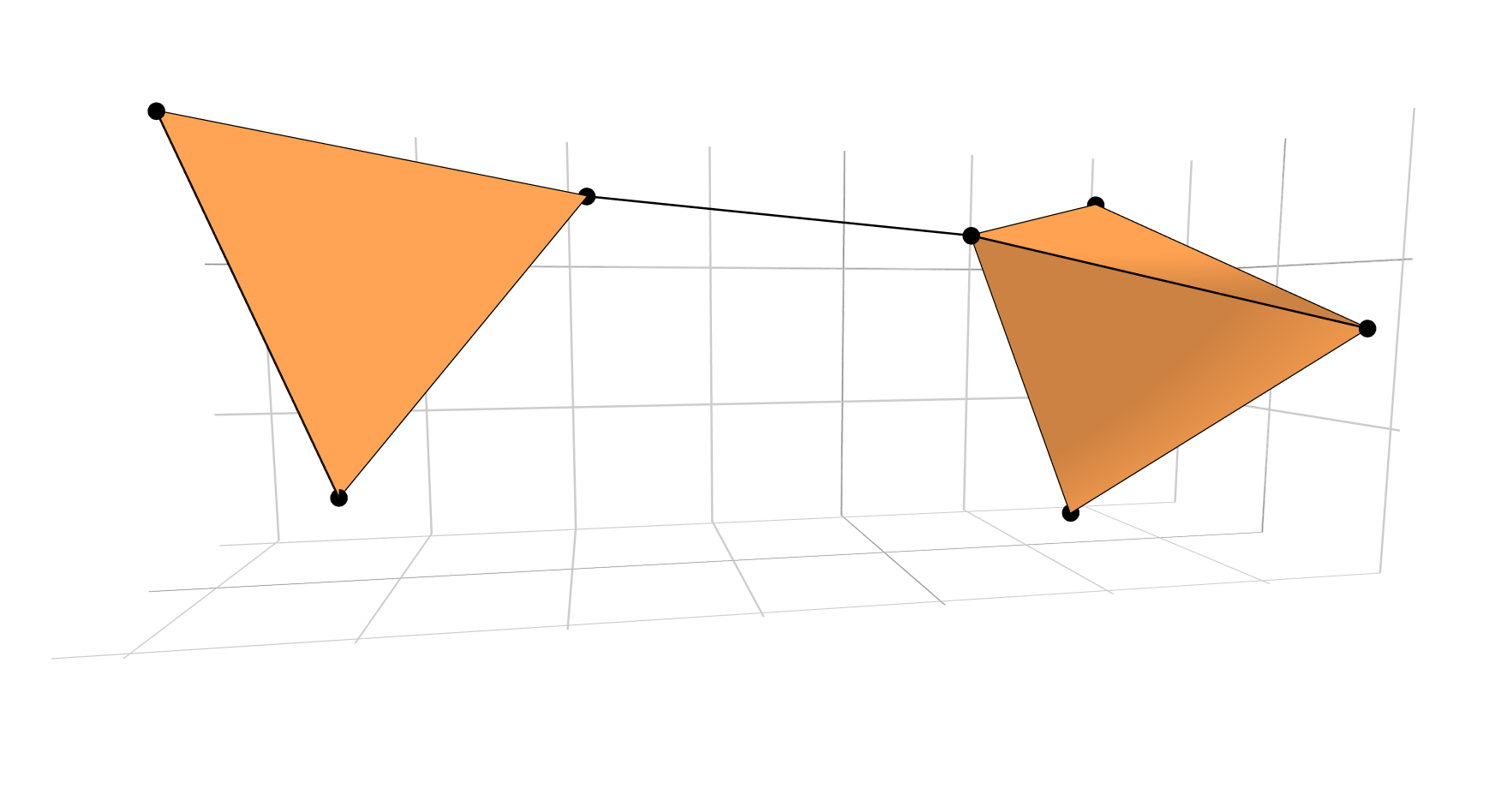}}
	\caption{Comparison of the \v{C}ech complex and the Vietoris-Rips complex for the same scale $r$: (a) the point pattern overlaid with the closed balls with radius $r$, visualizations of (b) the corresponding \v{C}ech complex and (c) the Vietoris-Rips complex in $\R^3$. The $2$-simplices (triangles) spanned by the vertices $D, E$ and $G$ and $D$, $F$ and $G$, are only elements of the Vietoris-Rips complex. Additionally also the $3$-simplex (tetrahedron) spanned by $D,E,F$ and $G$ is part of the Vietoris-Rips complex but not of the \v{C}ech complex.}
	\label{fig:cech-vietoris-diff}
\end{figure}

The \v{C}ech complex at scale $r$ is defined as \begin{equation}
	\operatorname{Cech}(r) = \{ \sigma \text{ simplex with vertices in } \x_0 \mid \bigcap_{x_i \in V(\sigma)} B_r(x_i) \; \neq \emptyset \}
\end{equation} while the Vietoris-Rips complex at the same scale is 
\begin{equation}
	\operatorname{VR}(r) = \{ \sigma \text{ simplex with vertices in } \x_0 \mid \max_{x_i, x_j \in V(\sigma)} \norm{x_i-x_j} < 2r \}.
\end{equation}
The two constructions differ in the way that we require for a simplex $\sigma \in C(r)$ that all closed balls with radius $r$ centered in the vertices of $\sigma$ intersect simultaneously, whereas this condition only needs to be fulfilled pairwise in the Vietoris-Rips complex. Figure~\ref{fig:cech-vietoris-diff} shows an example of a planar point pattern, where the two complexes differ. In particular, the \v{C}ech complex contains only simplices up to dimension $2$ while the Vietoris-Rips complex contains a $3$-simplex, i.e., a tetrahedron. Consequently, we visualize the complexes in three dimensions.

The dimension of the simplices in each of the two complexes is bounded only by the number of points in the point pattern. This can yield large complexes with high-dimensional simplices. A remedy to this is using the alpha complex. The idea is that we further restrict which points can form simplices by intersecting each closed ball $B_r(x_i)$ with the corresponding cell of the Voronoi diagram of the whole point pattern. The Voronoi cell of $x_i \in \x_0$ with respect to the point pattern $\x_0$ is defined as \begin{equation*}
	\mathrm{Voronoi}(x_i, \x_0) = \{ p \in \mathbb{R}^d \mid \norm{p-x_i} \leq \norm{p- x_k} \, \text{for all } x_k \in \x_0\}.
\end{equation*}
Then, the alpha complex at scale $r$ is defined as \begin{equation}
	\operatorname{Alpha}(r) = \{ \sigma \text{ simplex with vertices in } \x_0 \mid \bigcap_{x_i \in V(\sigma)} (B_r(x_i) \cap \mathrm{Voronoi}(x_i, \x_0)) \; \neq \emptyset \}.
\end{equation} 
The alpha complex at each scale $r$ is a subcomplex of the Delaunay triangulation of the point pattern. For point patterns in general position this means that the maximal degree of a simplex in the filtration is bounded. In most point process models, the points are almost surely in general position. General position of a planar point pattern implies that the alpha complex at any scale consists only of simplices up to dimension $2$. The restriction to the Voronoi cells does not change the topology captured by the \v{C}ech-complex. This result is known as the Nerve theorem \citep[e.g.][]{edelsbrunner2010} which states that the \v{C}ech-complex and the alpha complex have the same homotopy type as the union of the closed balls centered at the points of the point pattern. 

\begin{figure}
	\subfloat[]{\includegraphics[width=0.245\textwidth]{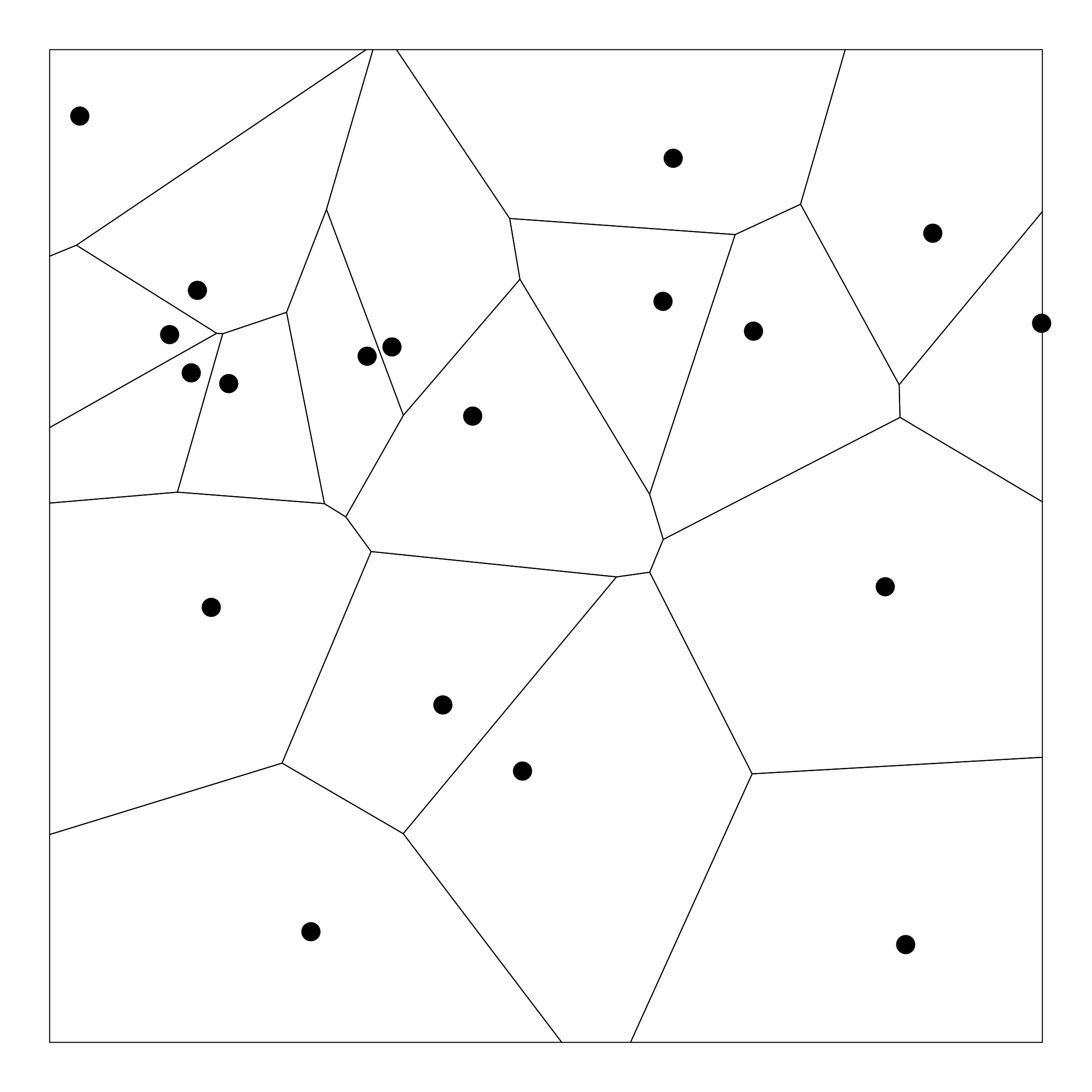} 
	\label{fig:pointpattern}} \hfill
	\subfloat[]{\includegraphics[width=0.245\textwidth]{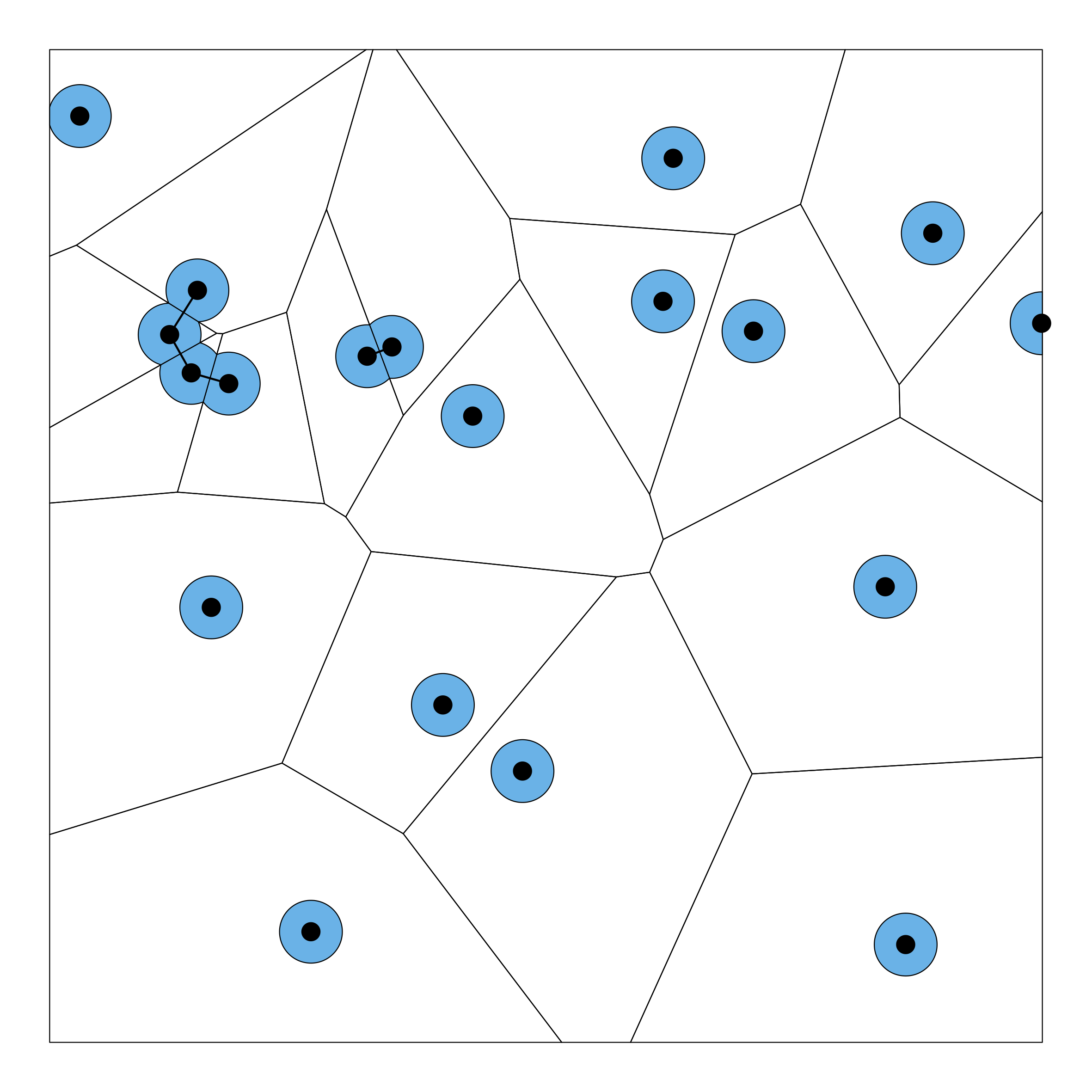}\label{fig:alpha-2}}\hfill 
	\subfloat[]{\includegraphics[width=0.245\textwidth]{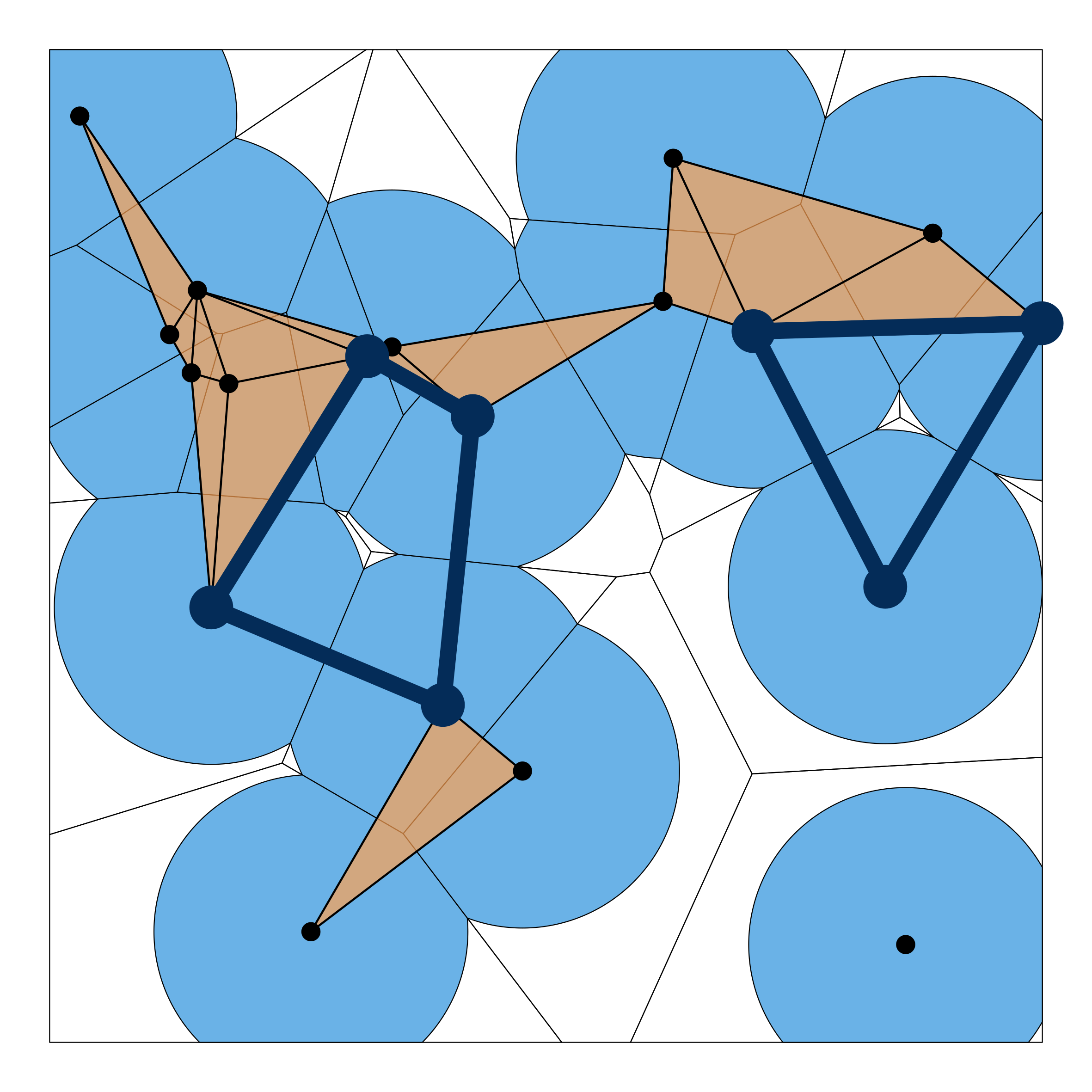}\label{fig:alpha-3}}\hfill
	\subfloat[]{\includegraphics[width=0.245\textwidth]{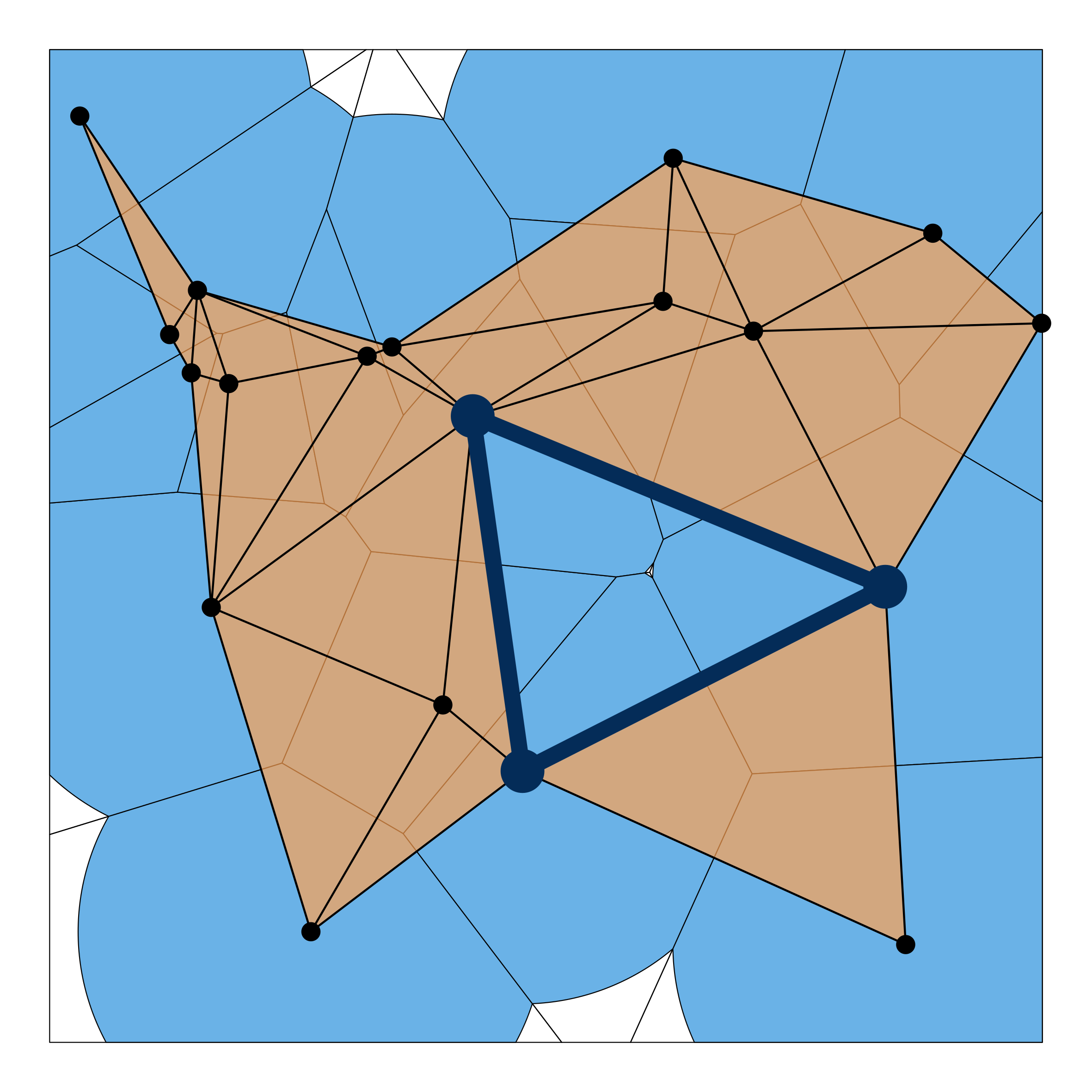}\label{fig:alpha-4}}
	\caption{Visualization of the alpha complex filtration: (a) point pattern and corresponding Voronoi diagram, (b) - (d): alpha complexes for increasing values of the scale $r$. 
    The dark blue solid lines indicate the boundaries of the $1$-dimensional topological features as introduced below.}
	\label{fig:simplicial-complex}
\end{figure}

Figure~\ref{fig:simplicial-complex} shows how the alpha complex filtration is built from a point pattern. As this point pattern is in general position, we can visualize the complex in $\R^2$. A general filtration of simplicial complexes obtained from any point pattern $\x$ is denoted by $\mathbb{F_{\x}} = (\mathcal{K}_r)_{r\in \R^+}$. In computations, we often use the alpha complex because of the reduced dimensions of the simplices, but in theoretical derivations, either the \v{C}ech or the Vietoris-Rips complex is usually used. 

To derive functional summary statistics that capture the topological structure of a point pattern, the idea is to analyze how the simplicial complex evolves over time when the scale increases. The corresponding mathematical framework is called persistent homology. It consists of comparing the homology groups of each simplicial complex in the filtration. Intuitively speaking, we track how topological features, i.e. the $p$-dimensional holes, of a simplicial complex evolve in the filtration. If the point pattern is in $\R^2$, then the topological features of interest are the individual connected components (the $0$-dimensional features) and ``circular'' loops (the $1$-dimensional features). In three dimensions, additionally voids/cavities which are the $2$-dimensional features are of interest. In Figure~\ref{fig:simplicial-complex} the boundaries of all $1$-dimensional features are indicated by thick solid dark blue lines. In Figure~\ref{fig:alpha-3} two $1$-dimensional features are present. For the one in the upper right corner, the balls centered in its three vertices intersect only pairwise, but not yet simultaneously. Thus, the $2$-simplex spanned by the three vertices is not part of the alpha complex on this scale. In Figure~\ref{fig:alpha-4} there is only one $0$-dimensional feature as all vertices are now connected and one remaining $1$-dimensional feature.

We track the scales in the filtration when a feature first appeared (its birth time) and when it disappeared (its death time). In our case, each point of the point pattern creates a connected component at scale $r=0$, that is, the birth time of all $0$-dimensional features is $0$.

\begin{figure}
	\centering
	\subfloat[Barcode]{\includegraphics[width=0.495\textwidth]{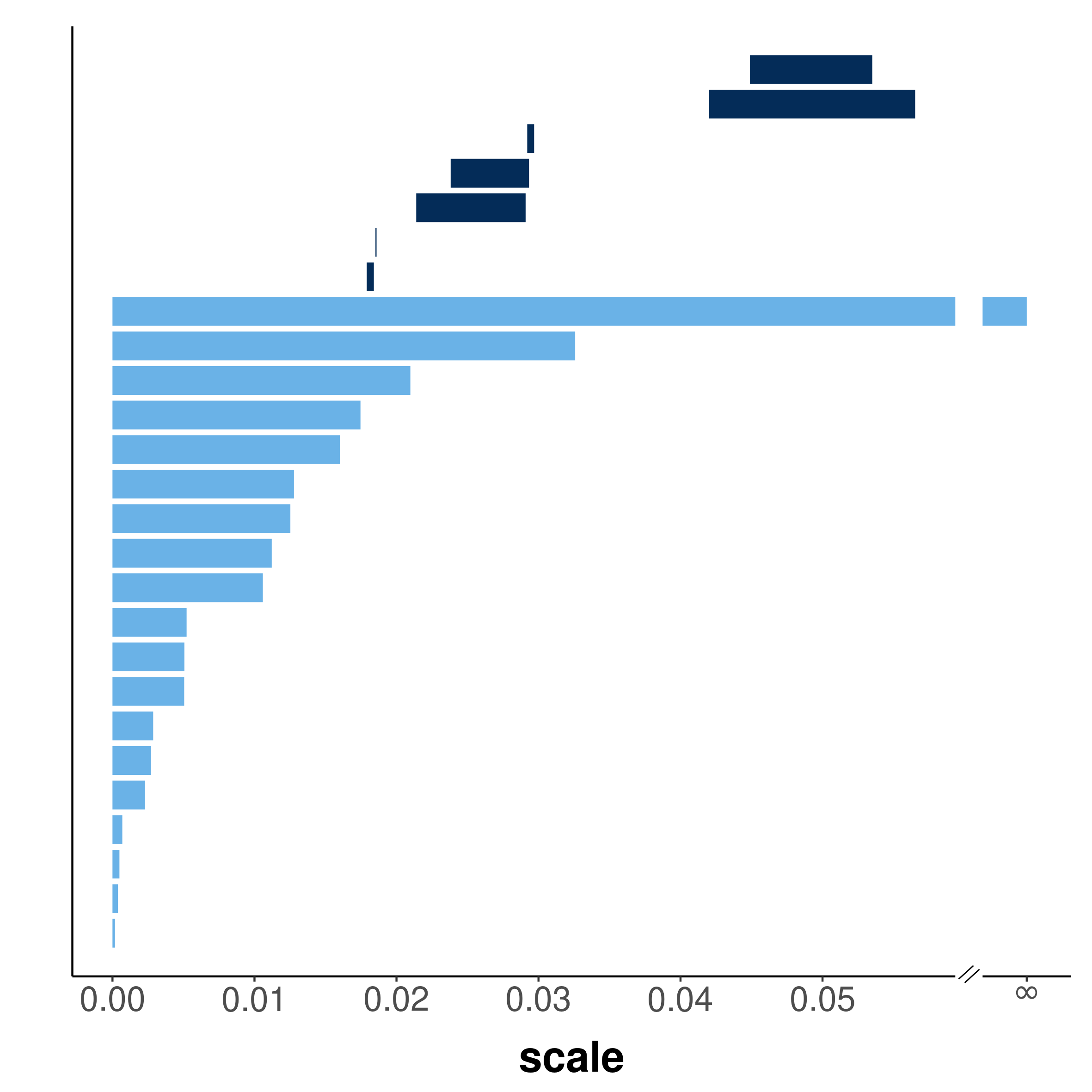}}
	\subfloat[Persistence diagram]{\includegraphics[width=0.495\textwidth]{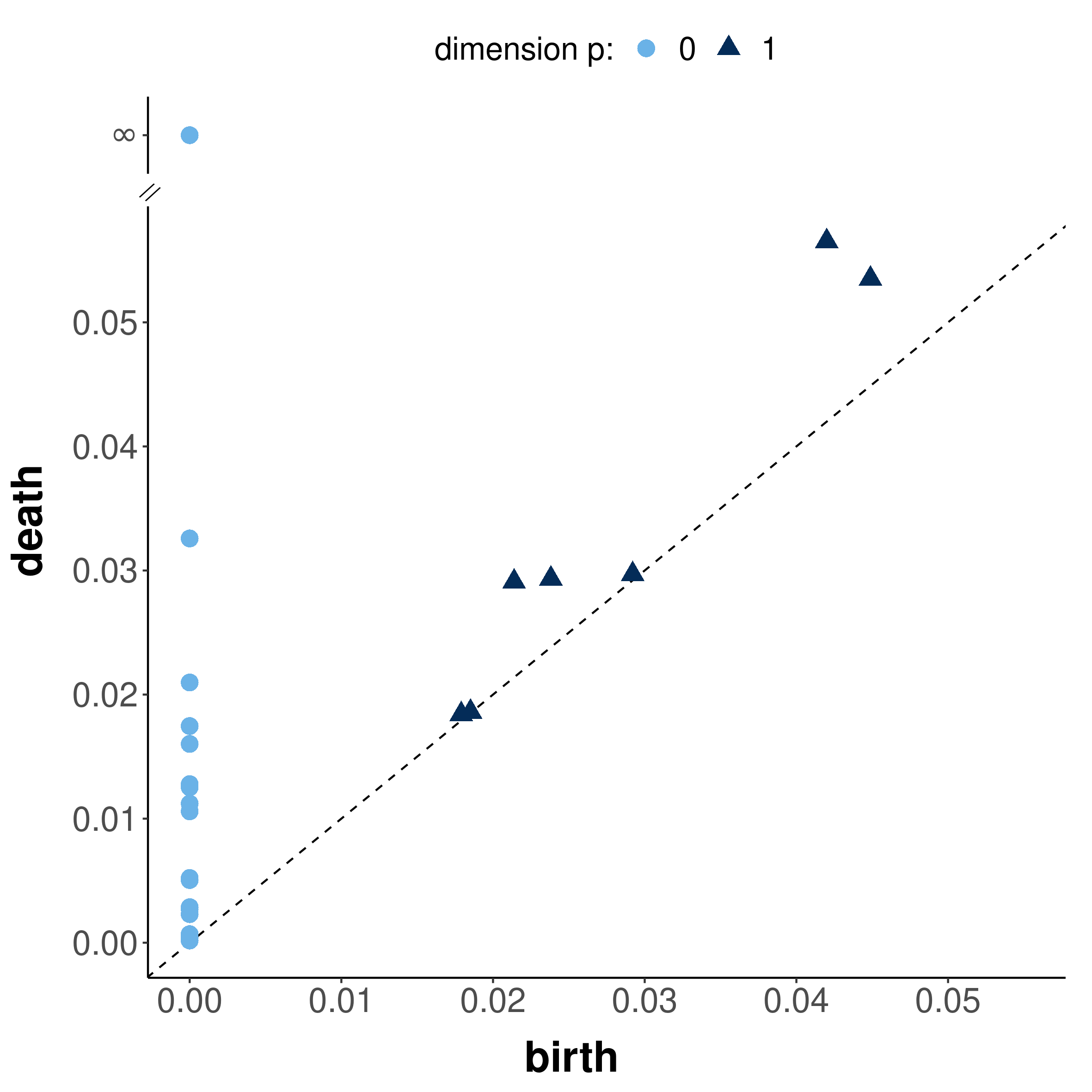}}
	\caption{Barcode and persistence diagram corresponding to the alpha complex filtration of the point pattern shown in Figure~\ref{fig:pointpattern}. In light blue the persistence of the connected components ($0$-dimensional features) is shown and in dark blue the persistence of the loops ($1$-dimensional features).}
	\label{fig:persistence-diagram}
\end{figure}

Standard graphical methods for the visualization of persistent homology are the barcode and the persistence diagram. The formal definition of the persistence diagram is given in \citet{edelsbrunner2010} as a multiset in the extended real plane $\R \times (\R \cup \{+\infty\})$. It is defined separately for each dimension $p$ and contains the points $(b_j,d_j)$ with multiplicity $m_j$ if there are $m_j$ distinct $p$-dimensional features with birth time $b_j$ and death time $d_j$ in the structure. Note that some of the points may have infinite death times. In the barcode, each feature is represented as an interval of its lifetime $[b_j,d_j]$. Figure~\ref{fig:persistence-diagram} shows both graphical representations for the filtration of the point pattern shown in Figure~\ref{fig:simplicial-complex}.

If the persistence diagram is considered random, i.e., constructed from a point process $\X$, it can be seen as a spatial point process with multiplicities \citep{hiraoka2018, biscio2019}. 
With this interpretation we can define the persistence diagram $\operatorname{PD}^p$ of the $p$-dimensional topological features of the point process $\X$ as the following point process on the extended positive real plane $\R_{\geq 0} \times (\R_{\geq 0} \cup \{+\infty\})$ \begin{equation}
	\operatorname{PD}^p(\mathbb{F_{\X}}) = \sum_{j \in \mathcal{J}^{p}(\mathbb{F_{\X}})} \delta_{(b_j, d_j)}
\end{equation} 
where $\mathcal{J}^{p}(\mathbb{F_{\X}})$ is an index set running over all $p$-dimensional topological features of the filtration $\mathbb{F_{\X}}$ and $\delta_x$ denotes
the Dirac measure that assigns unit mass to the point $x$. It is possible that the birth and the death times coincide for $i \neq j$, i.e. $b_i = b_j$ and $d_i = d_j$.

In the literature, there have been many proposals for functional summaries derived from the persistence diagram, see e.g. \citet{berry2020} for a recent review. In a goodness-of-fit test setting for point processes the following functional summary statistics have been introduced and used. As the simplicial complexes are constructed from a finite point pattern, we define the functional summary statistics for the point process $\X_W$ that is restricted to the bounded observation window $W$. For ease of notation we drop the subscript in the corresponding filtration.

\citet{robins2016} define the $p$-dimensional persistent homology rank function \begin{equation}
	\beta_p(\X_W, (b,d)) =  \operatorname{PD}^p(\mathbb{F_{\X}})((-\infty,b]\times(d,\infty]) = \sum_{j \in \mathcal{J}^{p}(\mathbb{F_{\X}})} \1{b_j \leq b, d_j > d} \quad \text{for } b \leq d
\end{equation} as a cumulative summary function of the persistence diagram. The individual quantities $\beta_p^{b,d} = \beta_p(\X_W, (b,d))$ are the persistent Betti numbers \citep[e.g.][]{edelsbrunner2010}. Betti numbers are standard characteristics in topology used for distinguishing different topological spaces (see e.g. the analysis of Betti numbers of random alpha complexes in \citet{robins2002}). Since the rank functions have two-dimensional domains, often also the $p$-dimensional Betti curve
\begin{equation*}
    \beta_p(\X_W, r) = \beta_p(\X_W, (r,r))
\end{equation*}
is considered.

\citet{biscio2019} aggregate the individual lifetimes $d_j - b_j$ of the features in the accumulated persistence function $\operatorname{APF}_p(\X_W, r)$ where the scale $r$ represents the so-called mean age $(b_j+d_j)/2$ of the feature $j$. This function was adapted in \citet{biscio2020} such that $r$ represents either the birth times (for $p=1$) or the death times (for $p=0$). By construction of the filtrations, the birth time of all $0$-dimensional features is $0$ which implies \begin{align}
\operatorname{APF}_0(\X_W, r) &= \int_{-\infty}^r (d-b)\cdot \operatorname{PD}^0(\mathbb{F_{\X}})(\R\times \mathrm{d}d) = \sum_{j \in \mathcal{J}^{0}(\mathbb{F_{\X}})} (d_j-b_j) \1{d_j \leq r} = \sum_{j \in \mathcal{J}^{0}(\mathbb{F_{\X}})} d_j \1{d_j \leq r}, \\
\operatorname{APF}_1(\X_W, r) &= \int_{[-\infty,r]\times\R} (d-b)\cdot \operatorname{PD}^1(\mathbb{F_{\X}})(\mathrm{d}b \times \mathrm{d}d) = \sum_{j \in \mathcal{J}^{1}(\mathbb{F_{\X}})} (d_j-b_j) \1{b_j \leq r}.
\end{align} The accumulated persistence is also used in the recent studies by  \citet{botnan2022} and \citet{krebs2022}. In the former in terms of the overall total persistence over all dimensions of features. %in multiparameter filtrations.

In \citet{biscio2020}, the function \begin{equation}
	\operatorname{ND}_0(\X_W, r) = \operatorname{PD}^0(\mathbb{F_{\X}})(\R\times(-\infty,r]) = \operatorname{PD}^0(\mathbb{F_{\X}})(\{0\}\times(-\infty,r]) = \sum_{j \in \mathcal{J}^{0}(\mathbb{F_{\X}})} \1{d_j \leq r}
\end{equation}
that counts the number of deaths of connected components is additionally introduced. 
In their tests, \citet{biscio2020} use a rescaled version of $\operatorname{ND}_0$ which is independent of the intensity and the size of the observation window. 

The summary function $\operatorname{ND}_0$ can be seen as the counterpart to the $0$-dimensional Betti curve as for each $r > 0$  \begin{align*}
    \operatorname{ND}_0(\X_W, r) + \beta_0(\X_W, (r,r)) &=  \operatorname{PD}^0(\mathbb{F_{\X}})(\R\times(-\infty,r]) + \operatorname{PD}^0(\mathbb{F_{\X}})((-\infty,r]\times(r,\infty])\\
    &= \operatorname{PD}^0(\mathbb{F_{\X}})(\{0\}\times(-\infty,r]) + \operatorname{PD}^0(\mathbb{F_{\X}})(\{0\}\times(r,\infty]) \\
    &= \operatorname{PD}^0(\mathbb{F_{\X}})(\{0\}\times(-\infty,\infty]) = \X(W).
\end{align*}

Another standard topological invariant of a simplicial complex $\mathcal{K}$ is the Euler characteristic $\chi(\mathcal{K})$. It is defined as the alternating sum of the number of $k$-simplices in the complex, i.e. 
 \begin{equation}
	\chi(\mathcal{K}) = \sum_{\sigma \in \mathcal{K}} (-1)^{\text{dim}(\sigma)}.
\end{equation}

Recently, \citet{dlotko2023} used the Euler characteristic curve $\chi(\X_W, r) = \chi(\mathcal{K}_r)$ as topological summary statistic for goodness-of-fit tests in the setting of binomial point processes with different intensity functions. Their processes are also not necessarily stationary. 

% - - - - - - - - - - - - - - Secondary structures - - - - - - - - - - - - - - - - -

\subsection{Higher order and secondary structure based summary statistics} \label{sec:add-summary}

Further approaches for defining functional summary statistics can be found in the literature. We will only give a brief overview while more details can be found in \citet{illian2008}. 

Considering triplets rather than pairs of points, third-order characteristics can be defined in analogy to second-order characteristics \citep{schladitz2000}.   

Other statistics are derived from characteristics of spatial structures such as random closed sets, random fields, tessellations and networks/graphs that are constructed from the point process \citep{illian2008}. The TDA-based statistics can be seen as an example of this approach. Here, we associate the filtration of (simplicial) complexes with the point process. 

In \citet{chiu2003} nine different distribution functions of characteristics extracted from the Voronoi diagram or the Delaunay tessellation of the point pattern are used. Examples include the distribution function of the area or the minimum angle of the Delaunay triangles.

The minimal spanning tree constructed from the point pattern and the Delaunay tessellation are used in the analysis of \citet{hoffman1983}. Their summary statistics include the edge length distribution in the minimal spanning tree and the interior angle distribution in the tessellation.

The construction of the Vietoris-Rips complex also reminds of neighboring graphs. These graphs are used in \citet{rajala2010} to define summary statistics such as an extension of the mean vertex degree, different connectivity functions or the so-called clustering function. \citet{illian2008} state that the connectivity function relates to the Euler characteristic curve and thus could also be seen as a topological characteristic.

Random closed sets are used as secondary structures in \citet{mecke2005} for defining morphological characteristics. The authors construct the associated random closed set as union of closed $r$-balls centered in the points of the point process. As stated above, the \v{C}ech complex and the alpha complex capture the same topological information as this construction.
The random closed set is then described using Minkowski functionals such as the Euler characteristic that was also discussed from the TDA perspective in \citet{dlotko2023}. The other Minkowski functionals, the area and the boundary length of the random closed set, are investigated in \citet{mecke2005}. The authors state that the area statistic is related to the empty space function $F$, thus a similar behavior is expected. Furthermore, they advise to use either the Euler characteristic or the boundary length as a supplement to second-order summary statistics such as the pair correlation function. 

The Minkowski functionals are also investigated in \citet{ebner2018}. Their approach consists of thresholding a count map to create a binarized image of the point pattern. This can be seen as a superlevel filtration \citep[c.f.][]{edelsbrunner2010} on the random field that contains the counts.

% ............................. Test statistics ...................................

\section{Test statistics}\label{sec:test-statistic}

Having introduced a variety of functional summary statistics in the previous sections, we now need test statistics that compare these functions to define goodness-of-fit tests. 
In this section, we give an overview of the different types of test statistics that have been proposed. They can be grouped into three categories in terms of how the test statistic $D$ is computed from the functional summary statistic $T$. 

For the first type A, we compute pointwise deviations of the empirical summary statistic for the observed point pattern from the corresponding reference value under the null hypothesis. The reference value is either the known theoretical summary statistic or an estimate obtained from simulations under the null model. Pointwise deviations are then summarized into a scalar discrepancy measure $D$. Tests known in the literature as deviation tests \citep{myllymaki2015} belong to this type A.

The second type B directly gives a scalar-valued summary of the empirical summary statistic for the observed point pattern. These test statistics appear most often in asymptotic tests as e.g. in \citet{biscio2020}.

The third type C uses the summary statistic in its functional form. For test statistics in this category, the measure of discrepancy is defined using an ordering $\preceq$ for functions that is often induced by a statistical depth measure. The rank envelope tests of \citet{myllymaki2017} belong to this category. 

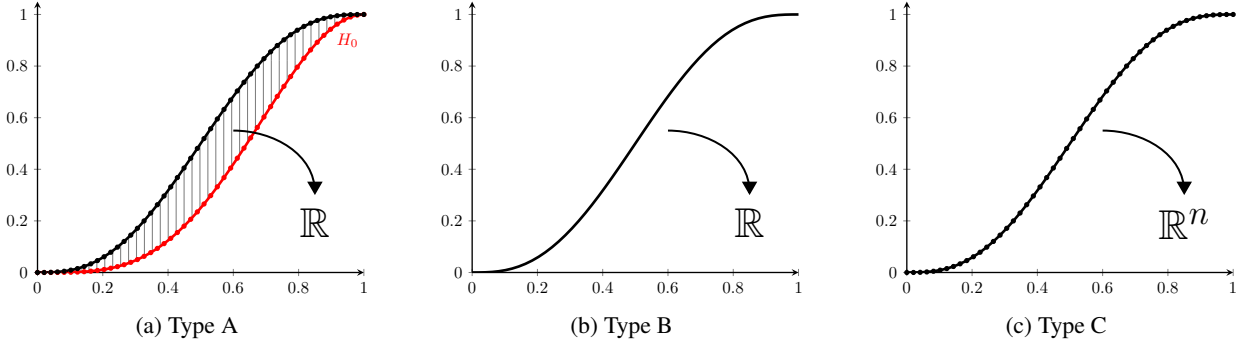
\begin{figure}
	\subfloat[Type A]{\resizebox{0.3\linewidth}{!}{\begin{tikzpicture}
		\begin{axis}[axis lines = left, domain=0:1, xmax=1, ymax=1.05, samples=50,
			axis line style = thick]
			\addplot[red, ultra thick, name path=A, mark=*, mark size=0.75pt] (x,4*x*x*x*x*x*x-12*x*x*x*x*x+9*x*x*x*x);
			\addplot[black,  ultra thick, name path=B, mark=*, mark size=0.75pt] (x, 10*x*x*x-15*x*x*x*x+6*x*x*x*x*x);
			\addplot[pattern color=gray, pattern=vertical lines] fill between[of=A and B];
			\draw[->,-triangle 60, very thick, black] (axis cs:0.6, 0.55) to [out=0,in=90] (axis cs:0.85, 0.3);
			\node[anchor=north] at (axis cs:0.85,0.27){{\Huge $\mathbb{R}$}};
			\node[anchor=north, red] at (axis cs:0.95,0.95){$H_0$};
		\end{axis}
	\end{tikzpicture}}}\hfill
\subfloat[Type B]{\resizebox{0.3\linewidth}{!}{\begin{tikzpicture}
	\begin{axis}[axis lines = left, domain=0:1, xmax=1,ymax=1.05, samples=50,
		axis line style = thick]
		\addplot[black,  ultra thick] (x, 10*x*x*x-15*x*x*x*x+6*x*x*x*x*x);
		\draw[->,-triangle 60, very thick, black] (axis cs:0.6, 0.55) to [out=0,in=90] (axis cs:0.85, 0.3);
		\node[anchor=north] at (axis cs:0.85,0.27){{\Huge $\mathbb{R}$}};
	\end{axis}
\end{tikzpicture}}}\hfill
\subfloat[Type C]{\resizebox{0.3\linewidth}{!}{\begin{tikzpicture}
	\begin{axis}[axis lines = left, domain=0:1, xmax=1,ymax=1.05, samples=50,
		axis line style = thick]
		\addplot[black,  ultra thick, mark=*, mark size=0.75pt] (x, 10*x*x*x-15*x*x*x*x+6*x*x*x*x*x);
		\draw[->,-triangle 60, very thick, black] (axis cs:0.6, 0.55) to [out=0,in=90] (axis cs:0.85, 0.3);
		\node[anchor=north] at (axis cs:0.85,0.27){{\Huge $\mathbb{R}^n$}};
	\end{axis}
\end{tikzpicture}}}
\caption{Overview of types of test statistic constructions: Scalar-valued and based on pointwise deviations (Type A), scalar-valued without taking reference values into account (Type B), vector-valued (Type C).}
\label{fig:test-statistics}
\end{figure}

The three different constructions for test statistics are visualized in Figure~\ref{fig:test-statistics}.

In the following, let $\x_0 \in \calN_W$ be the observed point pattern and denote by $\Y$ and $\Y'$ independent point processes with distribution $P_0$. Furthermore, denote by $T$ an arbitrary functional summary statistic with domain $\calR$ and by $\widehat{T}$ a nonparametric estimator of $T$. Let $\calR^* \subseteq \calR$ be the subset of the domain that should be taken into account.

\subsection{Type A: Scalar-valued and deviation-based}\label{sec:test-deviation}

The natural approach for testing the goodness-of-fit of a point process model is to use Kolmogorov--Smirnov or Cramér--von Mises-like test statistics with some distribution function. Test statistics for general functional summary statistics can be defined in the following way.

\begin{itemize}
	\item Maximum absolute deviation (MAD) \citep{diggle1979} \begin{equation}\label{eq:mad}
		D_{\operatorname{MAD}} = \sup_{r \, \in \mathcal{R}^*} |\widehat{T}(\bm{x}_0, r) - T(\Y, r)|
	\end{equation}
	\item Integrated squared deviation or Diggle-Cressie-Loosmore-Ford test (DCLF) \citep{diggle1979, diggle1986, cressie1993, loosmore2006, baddeley2014} \begin{equation}\label{eq:dclf}
		D_{\operatorname{DCLF}} = \int_{\mathcal{R}^*} (\widehat{T}(\bm{x}_0, r) - T(\Y, r))^2 \; \mathrm{d}r
	\end{equation}
\end{itemize}

In his original work, \citet{diggle1979} also considers absolute deviations $|\widehat{T}(\bm{x}_0, r) - T(\Y, r)|$ rather than the squared deviations in the integral statistic \eqref{eq:dclf}. Both \citet{diggle2013} and \citet{cressie1993} define the statistic \eqref{eq:dclf} with an additional power parameter $c$, i.e. with $\widehat{T}(\bm{x}_0, r)^c$ and $T(\Y, r)^c$ instead of $\widehat{T}(\bm{x}_0, r)$ and $T(\Y, r)$, respectively. This parameter $c$ is meant to stabilize the possibly unequal pointwise variances for different $r$. Additionally, \citet{diggle2013} introduces a pointwise weighting of the deviations. Recommendations for the choice of $c$ and a weight function for the $K$-function can be found in \citet[p. 666f]{cressie1993} and \citet[Section~7.2.1 and Eq. (7.5)]{diggle2013}. Many other multiplicative weight functions that can be used in both test statistics are also discussed and demonstrated for the $L$-function in \citet{ho2009}. Both \citet{loosmore2006} and \citet{mrkvicka2009} use a discretized version of the test statistic \eqref{eq:dclf} in their studies, where the integral is replaced by a Riemann sum.

The problem of unequal variances of the raw pointwise deviations $\widehat{T}(\bm{x}_0, r) - T(\Y, r)$ is also discussed by \citet{baddeley2000}, \citet{moller2012} and \citet{myllymaki2015}. The authors propose to take the distribution of the residuals under the null hypothesis into account by scaling the raw deviations before summarizing them into a deviation measure. \citet{baddeley2000} use a studentized scaling, while \citet{moller2012} use the difference between the upper and lower $2.5\%$-quantiles as scaling. \citet{myllymaki2015} refine the quantile scaling by taking possible asymmetries in the distribution into account. The authors additionally incorporate the scalings into the two  general deviation-based test statistics $D_{\operatorname{MAD}}$ and $D_{\operatorname{DCLF}}$. For the maximum absolute deviation measure $D_{\operatorname{MAD}}$ this results in the following test statistics. Here, $\widehat{T}(\bm{Y}, r)_{\beta}$ denotes the $\beta$-quantile of the distribution of $\widehat{T}(\bm{Y}, r)$ under the null model.

\begin{itemize}
	\item Maximum studentized deviation \citep{myllymaki2015} \begin{equation}\label{eq:st}
		D_{\operatorname{ST}} = \sup_{r \, \in \mathcal{R}^*} \left|\frac{\widehat{T}(\bm{x}_0, r) - T(\Y, r)}{\operatorname{sd}(\widehat{T}(\bm{Y}, r))} \right|
	\end{equation}
	\item Maximum directional quantile scaled deviation \citep{myllymaki2015} \begin{align} \label{eq:qdir} \begin{split}
			D_{\operatorname{QDIR}} &= \sup_{r \, \in \mathcal{R}^*} \Bigg\{ \mathds{1}\!\left(\widehat{T}(\bm{x}_0, r) \geq T(\Y, r)\right)\cdot \frac{\widehat{T}(\bm{x}_0, r) - T(\Y, r)}{|\widehat{T}(\bm{Y}, r)_{1-\frac{0.05}{2}} - T(\Y, r)|} \\ & \qquad\quad + \mathds{1}\!\left(\widehat{T}(\bm{x}_0, r) < T(\Y, r)\right)\cdot \frac{T(\Y, r)-\widehat{T}(\bm{x}_0, r)}{|\widehat{T}(\bm{Y}, r)_{\frac{0.05}{2}}- T(\Y, r)|}\; \Bigg\} \end{split}
	\end{align} 
\end{itemize}

The scalings applied to the Cramér--von Mises-like test statistic $D_{\operatorname{DCLF}}$ yield the next test statistics.

\begin{itemize}
	\item Integrated squared studentized deviation \citep{baddeley2000, myllymaki2015} \begin{equation}\label{eq:st-dclf}
		D_{\operatorname{ST}, \operatorname{DCLF}} =\int_{\mathcal{R}^*} \left(\frac{\widehat{T}(\bm{x}_0, r) - T(\Y, r)}{\operatorname{sd}(\widehat{T}(\bm{Y}, r))} \right)^2 \; \mathrm{d}r
	\end{equation}
	\item Integrated squared directional scaled deviation \citep{myllymaki2015} \begin{align} \label{eq:qdir-dclf} \begin{split}
			D_{\operatorname{QDIR}, \operatorname{DCLF}} &= \int_{\mathcal{R}^*} \Bigg( \Bigg\{ \mathds{1}\!\left(\widehat{T}(\bm{x}_0, r) \geq T(\Y, r)\right)\cdot \frac{\widehat{T}(\bm{x}_0, r) - T(\Y, r)}{|\widehat{T}(\bm{Y}, r)_{1-\frac{0.05}{2}} - T(\Y, r)|} \\ & \qquad\quad + \mathds{1}\!\left(\widehat{T}(\bm{x}_0, r) < T(\Y, r)\right)\cdot \frac{T(\Y, r)-\widehat{T}(\bm{x}_0, r)}{|\widehat{T}(\bm{Y}, r)_{\frac{0.05}{2}}- T(\Y, r)|}\; \Bigg\} \Bigg)^2 \; \mathrm{d}r \end{split}
	\end{align}
\end{itemize}

A different approach for a deviation-based test statistic comes from the theory of scoring rules. \citet{heinrichmertsching2024} construct what they call the summary statistic score for point processes which is a proper scoring rule. This scoring rule is derived from the continuous ranked probability score (CRPS) \citep{hersbach2000, gneiting2007} which is a well-known scoring rule for distributions of real-valued random variables. Scoring rules are also used in \citet{brehmer2024} where the focus is on validating nonstationary point process forecasts.
\citet{heinrichmertsching2024} use their score to compare a set of competing point process models. In particular, they test the significance of differences in mean scores using permutation tests. We suggest using their summary statistic score as the following test statistic in a goodness-of-fit test.
\begin{itemize}
	\item Integrated continuous ranked probability score \begin{equation}\label{eq:crps}
	D_{\operatorname{CRPS}} = \mathbb{E}\!\left[\int_{\mathcal{R}^*} |\widehat{T}(\bm{x}_0, r) - \widehat{T}(\bm{Y}, r)| \; \mathrm{d}r \right] - \frac{1}{2} \cdot\mathbb{E}\!\left[ \int_{\mathcal{R}^*} |\widehat{T}(\bm{Y}, r) - \widehat{T}(\bm{Y}', r)| \; \mathrm{d}r \right]
\end{equation}
\end{itemize}

\citet{heinrichmertsching2024} include an optional multiplicative weight function in the theoretical derivation that can be used to scale the raw absolute deviations before computing the integral. This option can be used in the same way as the weight function from \citet{diggle2013} mentioned above.

Most of the test statistics in this section make use of the true theoretical functional summary statistic for the null model. This quantity is only rarely known in a closed form expression. In practice, it is replaced by an estimate of $T(\Y, r)$. Additionally, the distribution of the estimator under the null model is relevant for the scaling approaches and the continuous ranked probability score. We need to estimate the mean, the variance and certain quantiles. 
We will discuss estimation techniques later in Section~\ref{sec:numerics}. 

\subsection{Type B: Scalar-valued without deviations}\label{sec:test-no-deviation}

\citet{ripley1977} and \citet{diggle1979} consider pointwise envelopes of the functional summary statistic as a graphical supplement to the tests formed by using either the MAD or the DCLF test statistics. These envelopes are formed separately at each evaluation point $r \in \calR^*$  by taking the pointwise minimal and maximal value of the empirical summary statistic of simulations under the null hypothesis as lower and upper envelope, respectively. To fit this approach into our framework, we consider the following test statistic, which is also discussed in \citet{baddeley2014}.

\begin{itemize}
	\item Point Evaluation \begin{equation}\label{eq:point}
		\operatorname{D}_{\operatorname{POINT}} = \widehat{T}(\bm{x}_0, r^*)
	\end{equation}
\end{itemize}

\citet{ripley1977} points out that the derived pointwise Monte Carlo test (see Section~\ref{sec:mc}) is only valid, i.e. has the correct size, if the evaluation point $r^*$ is chosen before observing the data. If one uses the pointwise envelopes simultaneously for multiple evaluation points without any corrections for multiple testing, the derived goodness-of-fit tests are invalid \citep[see e.g.][]{loosmore2006, grabarnik2011, baddeley2014}.

Recently, various versions of large volume limiting theorems have been proven for certain functional summary statistics under specific point process models (see Section~\ref{sec:CLT}). The quantity used in these theorems is often either the maximum absolute deviation $D_{\operatorname{MAD}}$ or the integrated squared deviation $D_{\operatorname{DCLF}}$ test statistic but additionally the integral of the estimated functional summary statistic is analyzed. This yields another type of possible test statistic that can also be used in a Monte Carlo test.

\begin{itemize}
	\item Integral Measure \begin{equation}\label{eq:int}
		\operatorname{D}_{\operatorname{INT}} =  \int_{\mathcal{R}^*} \widehat{T}(\bm{x}_0, r) \; \mathrm{d}r
	\end{equation}
\end{itemize}

\subsection{Type C: Function-/Vector-valued}\label{sec:test-fun}

In this section, we summarize approaches that use either vector-valued or functional test statistics. 

\citet{myllymaki2017} and \citet{wiegand2016} construct global envelope tests instead of the pointwise envelopes already discussed. These tests use the entire functional summary statistic at all evaluation points simultaneously in a single test. To fit this approach into our framework we define the test statistic of the global envelope test to be the entire functional summary statistic restricted to the domain $\calR^*$.

\begin{itemize}
	\item Functional Summary Statistic \citep{myllymaki2017} \begin{equation}
		\operatorname{D}_{\operatorname{FUN}} =  \widehat{T}(\bm{x}_0, \cdot)\big|_{\calR^*}
	\end{equation}
\end{itemize}

An alternative approach to vector-valued test statistics is derived from the continuous ranked probability score that we introduced earlier. The score was initially computed pointwise but then integrated over $\calR^*$. Omitting the integral yields the following test statistic.

\begin{itemize}
    	\item Pointwise Continuous Ranked Probability Score \begin{equation}
		\operatorname{D}_{\operatorname{SCORE}} = \left(\mathbb{E}\!\left[ |\widehat{T}(\bm{x}_0, \cdot) - \widehat{T}(\bm{Y}, \cdot)| \right] - \frac{1}{2} \cdot\mathbb{E}\!\left[  |\widehat{T}(\bm{Y}, \cdot) - \widehat{T}(\bm{Y}', \cdot)|  \right] \right)\Bigg|_{\calR^*}
	\end{equation}
\end{itemize}

To detect extreme values of the test statistic with respect to the null hypothesis, we need to be able to order the values that the test statistic can attain. For vector-valued or functional test statistics, this is not straightforward. Consequently, the emphasis in these cases is on constructing an ordering on the range of the test statistic. 
The different types of orderings that e.g. \citet{myllymaki2017} introduce and that can be used for the statistics $\operatorname{D}_{\operatorname{FUN}}$ and $\operatorname{D}_{\operatorname{SCORE}}$ are described in Section~\ref{sec:mc}.

\section{Test procedures}
\label{sec:procedure}
After having discussed the different types of functional summary statistics and test statistics, the last remaining part is the test decision. For that, we need the distribution of the test statistic under the null hypothesis.
If the distribution is not known explicitly and if there are no approximations available, then Monte Carlo tests can be used. We will give a brief overview of this testing procedure in the next Section~\ref{sec:mc}.
For specific null models and functional summary statistics, asymptotic results for increasing observation windows have recently become available. With these results asymptotic tests can be defined which are discussed in Section~\ref{sec:CLT}.

\subsection{Monte Carlo tests}
\label{sec:mc}

Monte Carlo tests were introduced by \citet{barnard1963} for simple hypotheses and quickly gained popularity in spatial statistics \citep[c.f.][]{besag1977, ripley1977} as the only main requirement is that one can generate samples from the null model. In the following, we describe the general Monte Carlo testing procedure for simple hypotheses as well as corrections in the case of composite hypotheses. 

% ------------------------------------ simple ------------------------------------------------

Assume that we are given a simple null hypothesis, a test statistic $D$ and a total order $\preceq$. Denote by $D_0$ the value of the test statistic for the observed point pattern $\x_0$. Additionally, we sample point patterns $\x_1, \dots, \x_m$, $m < \infty$, from the point process model under the null hypothesis and compute the respective values $D_1, \dots, D_m$ of the test statistic. We follow the convention that $D_k \preceq D_l$ means that $D_k$ is at least as extreme as $D_l$. In other words, \textit{small} values of the test statistic according to the ordering are extreme.

The general rationale of a Monte Carlo test is as follows. Under the null hypothesis, all $m+1$ values of the test statistic are interchangeable and identically distributed samples of $D$. If there are no ties in the test statistics, the probability that any of the values is the $l$th smallest value is exactly $\frac{1}{m+1}$ for $1 \leq l \leq m+1$. Consequently, the probability of being one of the $l$ smallest is $\frac{l}{m+1}$. The exact Monte Carlo test rejects the null hypothesis at level $\alpha = (k+1)/(m+1)$ for $k=0,\dots,m$ if exactly $k$ of the simulated values are strictly smaller than $D_0$. 

For example, if we let $m=19$ and set $\alpha=0.05$, thus $k=0$, then we reject the null hypothesis if and only if $D_0$ is the unique most extreme value of the test statistic in terms of $\preceq$. 

In practice, it is possible that there exists at least one $1 \leq i \leq m$ such that $D_0 = D_i$. A general convention is to count all simulations that tie with $D_0$ as more extreme. %under the null hypothesis. 
The corresponding Monte Carlo $p$-value as estimate of the true $p$-value is defined in \citet{davison1997} as
\begin{equation}\label{eq:mc-pval}
	p_{\operatorname{MC}} = \frac{1}{m+1} \left(1 + \displaystyle\sum_{i=1}^m \1{D_i \preceq D_0} \right).
\end{equation}
Under a simple null hypothesis and without ties, $p_{\operatorname{MC}}$ is uniformly distributed on $\{\frac{1}{m+1}, \dots, \frac{m}{m+1}, 1\}$.

\subsubsection*{Ordering}

The total ordering that is needed for the Monte Carlo $p$-value estimation determines which values of the test statistic $D$ are considered extreme. For the scalar statistics of type A and B the construction is based on the natural order on $\R$. In this case, we have to order $m\!+\!1$ numbers. The smallest of the $s$ values gets assigned the raw rank $1$ and the largest one gets assigned the raw rank $m\!+\!1$. In case of ties, the corresponding groups with the same value are assigned the average of the raw ranks. Formally, this can be defined as \begin{align}
    r_i &= 1 + \sum_{j=0}^{m} \1{D_j < D_i} + \frac{1}{2} \sum_{j=0, j\neq i}^{m} \1{D_j = D_i} \quad \text{for } i=0, \dots, m.%\\
    %&= S_i + \frac{1}{2}E_i + 1
\end{align}
Then, we say that $D_k \preceq D_l :\Leftrightarrow R_k \leq R_l$ where we have for each $i=0,\dots,m$
\begin{equation}
\label{eq:pointranks}
    R_i = \begin{cases}
    m\!+\!1 - r_i + 1 & \text{for test statistics where only large values are extreme,} \\
    \min(r_i, m\!+\!1 - r_i + 1) & \text{for test statistics where large and small values are extreme.}
    \end{cases}
\end{equation}
The test statistics of type A are all defined in terms of absolute or squared pointwise deviations. Consequently, only large values are extreme and result in small ranks $R_i$. For type B test statistics, both small and large values are extreme.

The construction of the ordering for the non-scalar test statistics of type C is more complicated. In practice, we can only compute the empirical functional summary statistics on a finite set of points. Hence, the corresponding test statistics of type C are usually given in a discretized form as vectors. We require that all $m+1$ test statistics be computed at the same finite set of $n$ evaluation points in $\calR^*$ with $n\geq 2$. Denote by $D_i = [D_i^1, \dots, D_i^n] \in \R^n$ the corresponding test statistics for pattern $\x_i$ with $i=0, \dots, m$. The idea of \citet{myllymaki2017} to order these vectors is based on the concept of statistical depth measures. The depth of a vector refers to the centrality of this vector within a set of reference vectors.

As a first step to the ordering \citet{myllymaki2017} compute at each evaluation point a pointwise ranking of the $m\!+\!1$ values. Then, the individual ranks per vector are combined into an overall depth measure. 

The first approach to the pointwise ranks uses the same construction as in Equation~\eqref{eq:pointranks} where the two-sided version was introduced in \citet{mrkvicka2017}. We use the one-sided alternative that counts only large pointwise values as extreme for the test statistic $D_{\operatorname{SCORE}}$ because it already measures absolute deviations. For $D_{\operatorname{FUN}}$, we use the two-sided alternative. We denote by $R_i^j$ the pointwise rank of the $j$th element $D_i^j$ of $D_i$. If all values are pairwise distinct then $R_i^j \in \{1, \dots, m\!+\!1\}$ otherwise $R_i^j \in \{1,1.5,2, \dots,m\!+\!1\}$.  

A second approach to the pointwise ranking are the continuous pointwise ranks $C_i^j$ proposed by \citet{mrkvicka2022}. For this pointwise ranking, not only the order but also the relative position to the next smaller and next larger value are taken into account. This idea was first discussed in \citet{hahn2015}. In both works, the idea is to extend the set of possible pointwise ranks such that ties in the derived depth measure become less likely. 
For the formal definition, denote by $D_{(0)}^j \leq \dots \leq D_{(m)}^j$ the increasingly ordered $m\!+\!1$ values at the $j$th evaluation point. 

If there are no ties involving the $i$th ordered value, then its raw continuous pointwise rank is defined to be
\begin{equation}
    c_{(i)}^j = \begin{cases}
    \exp\!\left(-\frac{D_{(1)}^j - D_{(0)}^j}{D_{(m)}^j - D_{(1)}^j}\right)\cdot\1{D_{(1)}^j < D_{(m)}^j} & \text{if } i=0,\\
    i + \frac{D_{(i)}^j - D_{(i-1)}^j}{D_{(i+1)}^j - D_{(i-1)}^j} & \text{if } 0 < i < m,\\
    m\!+\!1 - \exp\!\left(-\frac{D_{(m)}^j - D_{(m-1)}^j}{D_{(m-1)}^j - D_{(0)}^j}\right)\cdot\1{D_{(0)}^j < D_{(m-1)}^j} & \text{if } i=m.
    \end{cases}
\end{equation}

If there are ties for the $i$th ordered value, i.e. there are some $k\neq l$ with $0 \leq k \leq i \leq l \leq m$ such that $k = \min\{0 \leq k' \leq i \mid D_{(k')}^j = D_{(i)}^j\}$ and $l = \max\{i \leq l' \leq m \mid D_{(l')}^j = D_{(i)}^j\}$, then the raw continuous rank is defined as \begin{equation}
    c_{(i)}^j = \frac{k+l+1}{2}.
\end{equation}
This includes the case where all $m\!+\!1$ values are the same. In this case, the corresponding raw continuous rank is $c_{(i)}^j = \frac{m+1}{2}$ for all $0 \leq i \leq m$. 

Finally, the pointwise continuous ranks are given as
\begin{equation}
\label{eq:contranks}
    C_i^j = \begin{cases}
    m\!+\!1 - c_i^j & \text{for test statistics where only large values are extreme,} \\
    \min(c_i^j, m\!+\!1 - c_i^j) & \text{for test statistics where large and small values are extreme.}
    \end{cases}
\end{equation}
For all $i=0,\dots,m$ and all evaluation points $j=1,\dots,n$ we have $R_i^j - 1 \leq C_i^j \leq R_i^j$. 

Now we have for each test statistic $D_i$, $i=0,\dots, m$, a vector of the corresponding pointwise ranks $[R_i^1,\dots, R_i^n]$ and one containing the pointwise continuous ranks $[C_i^1,\dots, C_i^n]$. In addition, let $[R_i^{(1)},\dots, R_i^{(n)}]$ be the vector of the pointwise ranks sorted in ascending order. Now \citet{myllymaki2017} and \citet{mrkvicka2017} define the extreme rank measure \begin{equation}
    \operatorname{rank}(D_i) = \frac{1}{m+1}\cdot\min \{R_i^j \mid j=1,\dots,n\} = \frac{1}{m+1}R_i^{(1)}
\end{equation}
which takes only the most extreme pointwise rank of $D_i$ into account. Consequently, there is a high probability of ties. Several approaches for breaking these ties have been introduced.

The continuous rank measure \begin{equation}
    \operatorname{cont}(D_i) = \frac{1}{m+1}\cdot \min \{C_i^j \mid j=1,\dots,n\}
\end{equation}
allows by construction more possible values of $\operatorname{cont}(D_i)$ which yields a lower probability of ties.

The extreme rank length measure is defined as \begin{equation}
    \operatorname{erl}(D_i) = \frac{1}{m+1}\cdot \sum_{k=0}^m \1{[R_k^{(1)}, \dots, R_k^{(n)}] \preceq_{\operatorname{lex}} [R_i^{(1)}, \dots, R_i^{(n)}]}
\end{equation}
and resolves the high probability of ties in the extreme rank measure by using the lexicographical order of the sorted pointwise rank vectors \citep[c.f.][]{mrkvicka2017}, i.e. \begin{align}
\begin{split}
    &[R_k^{(1)}, \dots, R_k^{(n)}] \prec_{\operatorname{lex}} [R_i^{(1)}, \dots, R_i^{(n)}] \\ :\Leftrightarrow \quad &\exists\, n_0 \in \{1,\dots,n\}: R_k^{(j)} = R_i^{(j)} \, \text{ for all } 1\leq j < n_0 \text{ and } R_k^{(n_0)} <  R_i^{(n_0)}
\end{split}
\end{align}
and $[R_k^{(1)}, \dots, R_k^{(n)}] =_{\operatorname{lex}} [R_i^{(1)}, \dots, R_i^{(n)}]$ iff $R_k^{(j)} = R_i^{(j)}$ for all $1\leq j \leq n$.

The area rank measure proposed in \citet{mrkvicka2022}
refines the extreme rank measure again in a different way. The idea behind this refinement is to consider for each test statistic $D_i$ the extremeness of its values at all evaluation points $j$ where the pointwise rank $R_i^j$ coincides with the extreme rank $R_i^{(1)}$. By construction of the pointwise continuous ranks we have in these cases \begin{equation*}
   R_i^{(1)} -1 \leq C_i^j \leq R_i^j = R_i^{(1)}.
\end{equation*}
The area measure is computed by averaging the gap between the pointwise continuous and the extreme rank. This results in the following definition
\begin{equation}
    \operatorname{area}(D_i) = \frac{1}{m+1}\left(R_i^{(1)} - \frac{1}{n}\sum_{j=1}^n (R_i^{(1)} - C_i^j) \cdot \1{C_i^j < R_i^{(1)}}\right).
\end{equation}
The factor $1/(m+1)$ appearing in all four measures is used to scale the measures to the interval $[0,1]$ such that values close to $0$ indicate extreme elements of the set of vectors and values close to $1$ belong to the most central elements.

Each depth measure induces the corresponding total ordering between the computed test statistics via \begin{equation}
    D_k \preceq D_l :\Leftrightarrow \operatorname{rank}(D_k) \leq \operatorname{rank}(D_l)
\end{equation} and analogously for the other three measures.

\citet{myllymaki2017} compare their approaches with other depth measures for functional data such as the (modified) band depth \citep{lopez2009} and the (modified) half-region depth \citep{lopez2011}. Both alternative depths induce orderings and can be used in the Monte Carlo test setting. In the simulation studies in \citet{myllymaki2017} for the null hypothesis of complete spatial randomness, the tests using either the extreme rank measure or the extreme rank length measure were in most cases more powerful than the tests using the modified band depth or the modified half-region depth. In this comparative study, neither the continuous rank measure nor the area rank measure were included.

% ------------------------------------ composite ------------------------------------------------
\subsubsection*{Two-stage Monte Carlo tests}

Performing a Monte Carlo test for composite hypotheses requires estimating the unknown parameter $\widehat{\theta} = \widehat{\theta}(\x_0)$ of the parametric model family $\mathcal{P}_\Theta$. For $i=1,\dots,m$, let $D_i^*$ denote the test statistic computed for the $i$th simulation $\x_i$ from the fitted null model $P_{\widehat{\theta}}$.

\citet{davison1997} call the $p$-value approximation 
\begin{equation}\label{eq:boot-pval}
	p_{\operatorname{BOOT}} = \frac{1}{m+1} \left(1 + \sum_{i=1}^m \1{ D_i^* \preceq D_0} \right)
\end{equation}
the bootstrap approximation. The problem with this approximation is that the interpretation of error rates based on the $p$-value is no longer possible. This is due to the $p$-value estimate not being uniformly distributed under the null hypothesis \citep[c.f.][]{barnard1963, davison1997}. 

The remedy discussed by \citet{davison1997} is to construct adjusted $p$-values by considering the bootstrap $p$-value as a random variable in itself. In other words, we use the same bootstrapping idea to estimate the distribution of the $p$-value estimate $p_{\operatorname{BOOT}}$. This yields what they call a double bootstrap test that uses nested Monte Carlo simulations of the (re)fitted null model. The adjusted $p$-value is defined as 
\begin{equation}\label{eq:adj-pval}
	p_{\text{adj}} = \frac{1}{m+1} \left( 1 + \sum_{i=1}^m \1{p_{\operatorname{BOOT}}^i \leq p_{\operatorname{BOOT}}^0}\right)
\end{equation}
where $p_{\operatorname{BOOT}}^0 = p_{\operatorname{BOOT}}$ is the so-called first stage $p$-value and 
\begin{equation}
    p_{\operatorname{BOOT}}^i = \frac{1}{s+1} \left(1 + \sum_{j=1}^s \1{D_{ij}^* \preceq D_i^*} \right), \quad i=1, \dots, m,
\end{equation} are the so-called second stage $p$-values. Again, $D_i^*$ is the test statistic computed for the $i$th simulation $\x_i$ from the null model fitted to the observed point pattern $\x_0$, while $D_{ij}^*$ denotes the test statistic computed for the $j$th realization of the null model fitted to the simulated pattern $\x_i$.

This general two-stage idea has been used in a point process setting by \citet{dao2014} and \citet{baddeley2017}. 
\citet{dao2014} use the specific choice $s=m-1$. Their first and second stage $p$-values are then multiples of $\frac{1}{m+1}$ and $\frac{1}{m}$, respectively, such that there are no ties in \eqref{eq:adj-pval}. 
However, the first and second stage $p$-values are dependent due to the fact that the same realizations of the fitted null model are used in both stages. Therefore, \citet{baddeley2017} propose to use one set of $m$ simulations from the fitted null model to compute $p_{\operatorname{BOOT}}$ and another independent set of $s$ simulations that play the role of $\x_i$ in the computation of the second stage $p$-values.
As they choose $m=s$, both the first and second stage $p$-values are multiples of $\frac{1}{m+1}$ which could result in possible ties when computing the adjusted $p$-value. Instead of counting all ties as done in \eqref{eq:adj-pval}, the number of ties that are counted is randomized by uniformly sampling a number between $1$ and the total number of ties.

Algorithm~\ref{alg:composite} shows a pseudo-code of \citet{baddeley2017}'s balanced independent two-stage (BITS) test.
For better readability, we did not include the tie breaking rule in the pseudo-code.

\citet{baddeley2017} show that the BITS test is exact for simple hypotheses and their simulation study illustrates that it performs better for composite hypotheses than the general two-stage Monte Carlo test in \citet{davison1997} and the special case of \citet{dao2014}. Exact hereby means that $p_{\text{adj}}$ is uniformly distributed on the set $\{\frac{1}{s+1}, \frac{2}{s+1}, \dots, 1\}$. We consider the BITS test as the standard $p$-value estimation for all composite null hypotheses.

\begin{algorithm}[!ht]
	\DontPrintSemicolon
	\caption{BITS for Composite Null Hypothesis \citep{baddeley2017}}\label{alg:composite}
	\KwIn{observed point pattern $\x_{0}$, null model family $\calP_\Theta$, order $\preceq$, test statistic $D$, \linebreak number of samples $m$, $s$}
	\KwOut{adjusted $p$-value $p_{\text{adj}}$}
	\tcc{Stage 1 - Classical Monte Carlo Test}
	$\theta_0 \leftarrow \widehat{\theta}(\x_{0})$, $D_0 \leftarrow D(\x_0)$\;
	\For{$i=1,\dots, m$}{
		generate $\x_i$ from $P_{\theta_0}$\;
		$D_i \leftarrow D(\x_i)$\;
	}
	$p_0 \leftarrow  \left(1 + \sum_{i=1}^m \1{D_i \preceq  D_0} \right)/(m+1)$\;
	\BlankLine
	\tcc{Stage 2 - Distribution of $p$-value}
	\For{$j=1,\dots, s$}{
		generate $\y_j$ from $P_{\theta_0}$\;
		$\theta_j \leftarrow \widehat{\theta}(\y_{j})$, $D_{j0} \leftarrow D(\y_j)$\;
		\For{$i=1,\dots, m$}{
			generate $\z_i$ from $P_{\theta_j}$\;
			$D_{ji} \leftarrow D(\z_i)$\;
		}
		$p_j \leftarrow \left(1 + \sum_{i=1}^m \1{D_{ji} \preceq D_{j0}} \right)/(m+1)$\;
	}
	\tcc{Compute adjusted $p$-value}
	$p_{\text{adj}} \leftarrow \frac{1}{s+1} \left( 1 + \sum_{j=1}^s \1{p_j \leq p_0}\right)$\; 
	\Return $p_{\text{adj}}$
\end{algorithm}

The BITS algorithm needs in total $m + s$ simulations from the model fitted to the observed point pattern $\x_0$ and additionally $s\cdot m$ simulations from a refitted model, resulting in $m + s\cdot(m+1)$ overall simulations and $1+s$ parameter estimations. This comes with a high computational effort for reasonable choices of $s$ and $m$ (see Section~\ref{sec:nsim}). Fortunately, there are cases of composite hypotheses in which the classical Monte Carlo test for simple hypotheses is valid. \citet{davison1997} state two scenarios in which this is the case:
\begin{itemize}
	\item The test statistic $D$ has the same distribution for all possible values of the unknown parameters.
	\item There exists a sufficient statistic which we can condition on, see also \citet{barnard1963}.
\end{itemize}
In spatial statistics, an example of such a composite hypothesis is the standard hypothesis of complete spatial randomness, see Section~\ref{sec:csr-hypo}.

\subsection{Asymptotic tests}
\label{sec:CLT}

As an alternative to simulation based tests, we now consider goodness-of-fit tests that are based on asymptotic results for the distribution of the test statistic under the null hypothesis. Such tests are particularly useful when dealing with point patterns that contain a large number of points. In these cases Monte Carlo tests may be infeasible due to the computational cost of resampling the null model. In particular,  new simulations are required for each observed point pattern even when considering the same (composite) null hypothesis. These shortcomings of Monte Carlo tests are one reason why there are recently new efforts in defining asymptotic tests.

Most of the asymptotic theory for functional summary statistics is formulated for the null hypothesis of complete spatial randomness, either in terms of the homogeneous Poisson point process or the binomial point process with a constant intensity function, see Section~\ref{sec:csr-hypo} for details. In the following, we will first discuss contributions that provide asymptotic tests for a wider range of possible null models. 

Asymptotic theory for stationary point processes can be defined in different regimes. We generally assume large volume asymptotics which means that we observe the point process in a sequence of nested growing observation windows. Formally, the simple stationary point process $\X$ is observed in $W_n = [-n^{1/d}/2, n^{1/d}/2]^d \subset \R^d$ with $|W_n| = n$ for $n \in \N$ and we let $\X_n = \X \cap W_n$.
We are now interested in the properties of the test statistics as $n\to \infty$. In this asymptotic setting, a growing number of observed points does not change the spatial scales such that the parameters of the null model stay the same for all windows. 

Most recent limit theory is based on the fact that functionals of geometric structures can often be written as a sum of scores $\xi(x, \X_n)$ that represent the interaction of a point $x\in\X_n$ with the entire process $\X_n$, formally \begin{equation}\label{eq:score}
    \widehat{T}(\X_n, r) = \sum_{x\in\X_n} \xi(x, \X_n).
\end{equation}
On the right side, the dependence on the spatial scale $r$ is implicitly given in the definition of the score function $\xi$.

To establish the asymptotic normality of $\widehat{T}(\X_n, r)$, several assumptions both on the point process model and the score functions are needed.
These assumptions include certain mixing properties of the point process and the existence of higher order moments and a lower bound on the variance of the score function for the chosen model.

Point process models that fulfill these assumptions are Poisson and binomial point processes, Gibbs processes with finite interaction range such as pairwise interaction point processes or hard-core processes \citep{schreiber_limit_2013},
certain permanental point processes and determinantal point processes \citep[see][Section~2.2]{blaszczyszyn2019} as well as log-Gaussian Cox processes with compactly supported covariance function and Matérn cluster point processes \citep{biscio2020}.

Among the classical summary statistics introduced in Section~\ref{sec:classical-summary}, the $K$-, $G$- and the pair correlation function can be written in the form \eqref{eq:score}, see \citet{biscio2022, svane2024}.
\citet{biscio2020} express the persistent Betti numbers via score functions by assigning to each topological feature the point $x$ of the point process that either led to the birth or the death of the feature. They also extend the pointwise asymptotic normality by proving a functional central limit theorem for the persistent Betti numbers under the assumptions mentioned above. 
This makes asymptotic tests for several topological characteristics derived from the persistent Betti numbers possible. Examples are the accumulated persistence functions that were introduced in Section~\ref{sec:tda-summary}. 
\citet{blaszczyszyn2019} and \citet{schreiber_limit_2013} additionally consider geometric and topological summaries of stochastic structures associated with the point process. In particular, they discuss the total edge-length of the $k$-nearest neighbor graph, the $k$-covered region of a germ-grain model and the count of simplices in a \v{C}ech complex.

\section{Additional aspects of the goodness-of-fit tests}
\label{sec:additional}

The null hypothesis of complete spatial randomness has been studied extensively by proposing specialized goodness-of-fit tests that do not fit directly into our framework based on functional summary statistics. We will examine this hypothesis therefore in the next section.
Additionally, we discuss the estimation of unknown quantities that are used in the test statistics (see Section~\ref{sec:numerics}), the graphical interpretation of the test statistic (see Section~\ref{sec:envelope}), the combination of summary statistics (see Section~\ref{sec:combi}) and the necessary number of simulations from the null model (see Section~\ref{sec:nsim}).

\subsection{Testing complete spatial randomness}
\label{sec:csr-hypo}

Testing the null hypothesis of complete spatial randomness (CSR) is the most common test in the spatial statistics literature and the natural first step in the analysis of spatial point patterns \citep[cf.][]{velazquez2016}. If this hypothesis cannot be rejected then there is no reason to consider more complicated models for the interaction between the observed points or a non-constant intensity function.
Therefore, the goal is to have powerful goodness-of-fit tests that detect different types of deviations from CSR. 

The model for complete spatial randomness is given by the homogeneous Poisson point process. Hence, testing the CSR hypothesis translates to testing
\begin{equation*}
    H_0: P \in \text{Poi}(\lambda) \quad \text{vs.} \quad H_1: P \notin\text{Poi}(\lambda)
\end{equation*}
where $\text{Poi}(\lambda)$ represents the family of stationary Poisson point processes on $\R^d$ with intensity parameter $\lambda > 0$. Consequently, we are dealing with a composite hypothesis as the true intensity $\lambda$ is unknown.

In the context of the general goodness-of-fit tests constructed from functional summary statistics, a composite hypothesis implies that we must use the computationally expensive two-stage Monte Carlo test procedure. In the case of the homogeneous Poisson point process as the null model, we can make the null hypothesis simple by conditioning on a sufficient statistic. The sufficient statistic for the unknown intensity $\lambda$ is the number of points $n$ in a point pattern \citep[e.g.][]{diggle2013, ripley1977}. Thus, conditioning on the number of points in the observed pattern $\x_0$ in $W$ leaves no unknown parameter left. Keeping the number of points fixed transforms the null model into the binomial point process with uniform intensity function on the observation window $W$. In this setting the null model will always be the corresponding binomial point process.

Many goodness-of-fit tests have been proposed precisely for testing the CSR hypothesis, see e.g. \citet[Section~8.2]{cressie1993} and \citet{illian2008} for overviews and \citet{diggle2013} for a comparison of some of the tests on three standard data sets. Many of the specialized tests that do not use functional summary statistics rely on specific, usually scalar-valued, indices of the homogeneous Poisson process. 
However, \citet[Section~2.7]{diggle2013} recommends using functional summary statistics instead of scalar summaries as they additionally allow to interfere on the type of deviation from CSR (see Section~\ref{sec:envelope}). This information can then be used to choose the right type of model family for the observed point pattern when the null hypothesis of CSR is rejected.

Many of the scalar indices are based on the nearest-neighbor distances such as the Clark and Evans test statistic \citep{clark1954} which is a studentization of the average nearest-neighbor distance. We refer the reader to Table~8.6 in \citet{cressie1993} for a survey on nearest-neighbor distance based indices that were proposed before the year 1980. In this overview also the asymptotic distributions of the test statistics are listed.

In the following, we will briefly summarize more recent contributions for specialized tests for CSR. 

Quadrat count tests make use of the fact that the counts of points in disjoint regions of the observation window are - under the null hypothesis of CSR - independent Poisson-distributed random variables. Consequently one can construct goodness-of-fit tests by using dispersion indices such as Pearson's chi-squared statistic. The indices are calculated from the counts in equally sized disjoint subdivisions of the observation window. The dispersion indices are approximately $\chi^2_{k-1}$-distributed where $k$ is the number of subdivisions \citep{cressie1984, illian2008}. Instead of using counts per area, the $Q^2$-test statistic of \citet{grabarnik2002} compares the counts of points having exactly $k$ other points closer than scale $r$ for several integers $k$. The authors propose both a single fixed scale version and a multi-scale test statistic. The box-counting approach of \citet{caballero2022} is based on a subdivision of the window into quadrats. The authors investigate the log-log relationship between the number of quadrats containing at least one point and the side length of the quadrat for several different side lengths. This allows to obtain an estimate of the fractal dimension of a point pattern which can be used as a scalar-valued statistical index. Moreover, the functional relationship is used as a new functional summary statistic. Its interpretation is analogous to the $F$-function.

\citet{liebetrau1977} considers the variance of the counts in a rectangular test set with fixed dimensions. Both \citet{ripley1979} and \citet{zimmerman1993} restrict the test set to a square and call this summary the variance function. In spite of its name, the variance function is a scalar value as they choose a single fixed side length before performing the asymptotic test.

The tests of \citet{zimmerman1993} and \citet{ho2007} use test statistics that measure the discrepancy between the empirical distribution function of the point locations and the distribution function of the uniform distribution. In particular, \citet{zimmerman1993} proposes to use a combined Cramér--von Mises test statistic $\bar{\omega}^2$ which averages the discrepancies obtained when each corner of the rectangular observation window is taken as origin.

Another class of tests for complete spatial randomness is based on spectral analysis of the point process. \citet{mugglestone1990} and \citet{mugglestone2001} propose several tests using test statistics derived as scalar-valued summaries of the periodogram.

For Poisson point processes also residual methods are available \citep{baddeley2005}. \citet{coeurjolly2013} use the residuals to set up asymptotic tests such as a generalization of the quadrat count tests mentioned above. \citet{yang2019} propose a goodness-of-fit test based on a Stein discrepancy which can be seen as a kernelization of the so-called $h$-weighted residuals. 

For the special case of a homogeneous Poisson point process, asymptotic results for empirical functional summary beyond those summarized in Section~\ref{sec:CLT} are available. Asymptotic normality can be proven for the empirical $K$-function used either with the deviation-based test statistics $D_{\operatorname{MAD}}$ or $D_{\operatorname{DCLF}}$ \citep[c.f.][]{heinrich1991, heinrich2015, heinrich2018}. The limit theorem for the $K$-function is used in \citep{marcon2013} to show the asymptotic $\chi^2$-distribution of their test statistic.

The asymptotic normality under CSR has also been shown for the Minkowski functionals of the binary images associated with the point process  \citep{ebner2018}, either individually or in a multivariate setting.

The asymptotic topology of Poisson point processes or binomial point processes is characterized in particular by proving asymptotic normality of the (persistent) Betti numbers \citep[c.f.][]{yogeshwaran2015, yogeshwaran2017, hiraoka2018, bobrowski2018, krebs2022, krebs2023} or the Euler characteristic \citep[c.f.][]{thomas2021, krebs2021, dlotko2023}. 

All these results allow setting up asymptotic goodness-of-fit tests by comparing the computed test statistics with the corresponding quantiles of the limiting distribution. In many cases both the mean and the variance of the limiting Gaussian need to be estimated since closed form expressions are not available or contain unknown quantities. In this case, additional simulations under the null model are needed.

\subsection{Estimation of unknown quantities}
\label{sec:numerics}

Many of the discussed test statistics integrate or compute the supremum over the subset $\calR^* \subset \calR$ of possible evaluation points. In practice, we can estimate the functional summary statistic only at a finite number of evaluation points and thus we have to approximate the test statistics. The integrals are approximated via simple Riemann sums over an equidistant grid of $n$ evaluation points in $\calR^*$. For all the functional summary statistics introduced in this review, $\calR^* = [0,r_{\max}] \subset \R$ is a closed interval with $r_{\max} < \infty$. For certain statistics such as the pair correlation function, it can make sense to restrict the interval to $[\epsilon, r_{\max}]$ for some $\epsilon > 0$ as the nonparametric estimation of the functional summary statistics close to zero is biased and often not even tractable \citep{stoyan1994, moller2003}.

The deviation-based test statistics of type A require that the theoretical summary statistic $T(\Y, r)$ for the null model $\Y\sim P_0$ is known as a closed-form expression at any evaluation point $r$. This is rarely the case for arbitrary null models and arbitrary functional summary statistics. In order to use these test statistics we consequently need to estimate $T(\Y, r) =: \mu$. 

One possible estimator is the sample mean of a set of independent simulations from the null model. Simulations are also needed for the test decision when performing a Monte Carlo test or when estimating the variance of the limiting distribution in an asymptotic test. Estimation of $\mu$ thus implies that a second set of independent realizations of the test statistic has to be simulated. 

In the Monte Carlo setting, the computational cost can be reduced by using a leave-one-out approach \citep{diggle1979, loosmore2006} that needs only a single set of realizations. Let $\x_0$ be the observed point pattern and denote by $\x_1,\dots,\x_m$ the simulations. When we compute the test statistics for the $i$th pattern, the leave-one-out approach estimates the reference value as \begin{equation}
   \widehat{\mu}_{-i}(r)= \frac{1}{m} \sum_{j=0, j\neq i}^m \widehat{T}(\x_j, r).
\end{equation}
Here, the observed point pattern $\x_0$ and the $i$th pattern swap roles and the sample mean is computed over the patterns that are considered to be simulations at that point. This procedure preserves the necessary symmetry for the basic Monte Carlo rationale. However, the estimate of the reference value is different for each computation of the test statistic which again yields a higher computational cost compared to having a single estimate that is used in each test statistic computation.

For the deviation-based test statistics, the raw deviations $\widehat{T}(\x_i, r) - \widehat{\mu}_{-i}(r)$ are summarized by e.g. taking the maximum of all $r$ to obtain $D_{\operatorname{MAD}}$.
\citet{baddeley2014} make use of the representation 
\begin{equation}
    \widehat{T}(\x_i, r) - \widehat{\mu}_{-i}(r) = \frac{m+1}{m} \left(\widehat{T}(\x_i, r) - \frac{1}{m+1} \sum_{j=0}^m \widehat{T}(\x_j, r)\right),
\end{equation}
which allows to compute the raw deviations solely from the observed values and the overall sample mean.
This approach is implemented in the R packages \texttt{spatstat} \citep{spatstat, spatstatR} and \texttt{GET} \citep{GET}.

Other unknown quantities that need to be estimated in some test statistics are related to the distribution of $\widehat{T}(\Y, r)$ (in the case of studentized or directional quantile scaling) or the expectations of the absolute differences in the continuous ranked probability score.
For the scalings, the pointwise standard deviation and selected quantiles are estimated in \texttt{GET} using the respective standard estimators applied to the entire set of $m+1$ values.

For estimating the continuous ranked probability score, we follow the discussion and empirical studies in \citet{zamo2018}. They consider the estimation of the pointwise instantaneous CRPS at a fixed evaluation point $r$ for the observation $\widehat{T}(\x_0, r)$ under the limited information of an ensemble $\widehat{T}(\x_1, r), \dots, \widehat{T}(\x_m, r)$ and call the unbiased estimator
\begin{equation*}
            \widehat{\operatorname{CRPS}}_{\text{fair}} = \frac{1}{m} \sum_{\substack{j=1}}^m |\widehat{T}(\x_0, r) - \widehat{T}(\x_j, r)|  - \frac{1}{2m(m-1)} \sum_{\substack{j=1}}^m \sum_{\substack{k=1}}^m  |\widehat{T}(\x_j, r) - \widehat{T}(\x_k, r)|
\end{equation*}
the fair CRPS estimator. Their simulation studies for several types of distributions show that the empirical relative estimation error goes to zero faster as $m\to\infty$ than for some other estimators. They observe only minor performance improvements when choosing more than $m=100$ elements in the ensemble. 

For Monte Carlo tests, we propose to combine the fair CRPS estimator with the idea of swapping the observed with a simulated point pattern. Consequently, the scalar-valued test statistic $D_{\operatorname{CRPS}}$ for the $i$th pattern, $i=0,\dots,m$, at the evaluation points $r_1 < r_2 < \dots < r_n$ is computed as \begin{equation*}
    D_i = \frac{1}{m} \sum_{j=0}^m \sum_{t=1}^{n-1} |\widehat{T}(\x_i, r_t) - \widehat{T}(\x_j, r_t)|(r_{t+1} - r_t)  - \frac{1}{2m(m-1)} \sum_{\substack{j=0 \\ j \neq i}}^m \sum_{\substack{k=0 \\ k \neq i }}^m \sum_{t=1}^{n-1} |\widehat{T}(\x_j, r_t) - \widehat{T}(\x_k, r_t)| (r_{t+1} - r_{t})
\end{equation*}
which matches the approximation of \citet{heinrichmertsching2024}. 

\subsection{Graphical interpretation}
\label{sec:envelope}
Some of the test statistics in Section~\ref{fig:test-statistics} allow for an intrinsic graphical representation as a global envelope of the chosen functional summary statistic. 
The null hypothesis of the goodness-of-fit test for significance level $\alpha$ is rejected if the observed empirical functional summary statistic is not completely within the global envelope formed by an upper envelope $T_{upp}^\alpha(r)$ and a lower envelope $T_{low}^\alpha(r)$ for $r\in\calR^*$, i.e. we reject $H_0$ if \begin{equation}
    \exists\, r \in \calR^* \, \text{s.t.} \, \widehat{T}(\x_0, r) \notin [T_{low}^\alpha(r), T_{upp}^\alpha(r)].
\end{equation}
The envelopes are discretized at the same evaluation points as the functional summary statistic.

This approach yields tests with a
global level $\alpha$ (that is, a controlled type I error simultaneously on all scales $\calR^*$). This is not fulfilled when using multiple pointwise envelopes simultaneously, which was discussed when introducing the test statistic $\operatorname{D}_{\operatorname{POINT}}$ in Section~\ref{sec:test-no-deviation}.

The graphical representation allows for a visual comparison of the empirical summary statistic with the expectation under the null model. In particular, it is also possible to determine which spatial scales $r\in\calR^*$ lead to the rejection of the null model. 

\citet{wiegand2016} propose an analytical global envelope which requires that under the null model certain assumptions on the chosen estimator of the summary statistic are met. In particular, these include asymptotic normality with known variance and approximate independence between evaluation points that are sufficiently spaced apart. To meet these assumptions, the authors recommend using non-cumulative summary statistics such as the pair correlation function. The main idea is to employ a multiple testing correction to the significance level $\beta$ of the local pointwise envelopes to arrive at the overall significance level $\alpha = 1 - (1-\beta)^n$ where $n$ denotes the total number of evaluation points.
The upper and lower pointwise envelopes are formed as \begin{align}
    T_{low}^\alpha(r_j) = T(\Y,r_j) - q_\beta \cdot \operatorname{sd}(\widehat{T}(\bm{Y}, r_j)) \quad \text{and} \quad
    T_{high}^\alpha(r_j) = T(\Y,r_j) + q_\beta \cdot  \operatorname{sd}(\widehat{T}(\bm{Y}, r_j))
\end{align}
where $q_\beta$ is the critical value of the local test, i.e. the $1-\beta/2$ quantile of a standard normal distribution for a two sided-test, and $\operatorname{sd}(\widehat{T}(\bm{Y}, r_j))$ is the standard deviation of the estimator at evaluation point $r_j$.
To circumvent the asymptotic normality assumption, simulation-based global envelopes are introduced. These envelopes require only the approximate independence assumption, which again limits their applicability to non-cumulative summary statistics. In this case the upper and lower envelopes are given as the $k$th lowest and highest value computed from $m$ simulations of the null model. The number $k=\beta(m+1)/2$ is chosen according to the necessary multiple testing correction. This equality imposes a constraint on the necessary number of simulations when the global level $\alpha$ and the number of evaluation points $n$ are given since $k$ must be an integer. In particular, for $\alpha=0.05$ and $n=50$ we need at least $m=1999$ simulations to obtain $k \geq 1$.

When using cumulative summary statistics or if not all the requirements for the approaches of \citet{wiegand2016} are met, one can use the global envelope approaches of \citet{myllymaki2017, mrkvicka2022}. These global envelopes are the intrinsic graphical representation of the pure Monte Carlo test based on the statistic $D_{\operatorname{FUN}}$ and either the extreme rank, the rank length, the continuous rank or the area rank ordering. Here, the multiple testing problem is resolved by making use of the orderings to directly compute the upper and lower envelopes.

Recall, that all vector orderings were induced by a scalar-valued statistical depth measure which we will denote by $\nu$. By construction, the test statistic computed for the $i$th pattern is simply a vector $D_i = [D_{i1}, \dots, D_{in}] \in \R^n$ containing the values of the empirical functional summary statistic at the $n$ evaluation points. For the global envelope, \citet{GET} first compute a threshold $\nu_\alpha \in \R$ which is given by the largest value of $\nu(D_0), \dots, \nu(D_m)$ such that \begin{equation} \label{eq:threshold-get}
    \sum_{i=0}^m \1{\nu(D_i) < \nu_\alpha} \leq \alpha(m+1).
\end{equation}

Let $I_\alpha = \{i=0,\dots, m\!+\! 1 \mid \nu(D_i) \geq \nu_\alpha\}$ be the indices of the patterns that do not belong to the $\alpha(m+1)$ most extremes with respect to the chosen ordering. The bounds for the global envelope representation of the Monte Carlo test at the $j$th evaluation point are given by \begin{equation}
    T_{low}^\alpha(r_j) = \min_{i \in I_\alpha} D_{ij} \quad \text{and} \quad
    T_{high}^\alpha(r_j) = \max_{i \in I_\alpha} D_{ij}.
\end{equation}
The decision whether to consider one-sided or two-sided tests, and consequently envelopes, is already incorporated in the measure $\nu$.

For the pointwise continuous rank probability score $D_{\operatorname{SCORE}}$ we proposed to use the same orderings as for $D_{\operatorname{FUN}}$ such that we can construct global envelopes in the same way. Here, the pointwise estimated scores take the role of the functional summary statistic.

Finally, also the scalar-valued test statistics based on the maximum absolute deviation, i.e. $D_{\operatorname{MAD}}, D_{\operatorname{ST}}$ and $D_{\operatorname{QDIR}}$, have an intrinsic graphical representation as global envelope \citep{baddeley2014, myllymaki2017, GET}. The global envelopes are constructed by first computing the threshold value $d_\alpha$ of the chosen test statistic. The construction of the threshold works analogously to \eqref{eq:threshold-get}. For these three test statistics only large values are extreme. Therefore, $d_\alpha$ is defined as the smallest value of the test statistics such that the number of test statistics that are strictly larger than $d_\alpha$, i.e. are more extreme, is less than $\alpha(m+1)$. Hence $d_\alpha$ is simply the $\alpha(m+1)$ largest of the $m+1$ test statistics.
Then, the lower and upper envelopes are formed as \begin{equation}
        T_{low}^\alpha(r_j) = T(\Y,r) - d_\alpha\cdot s_{low}(r_j) \quad \text{and} \quad
    T_{high}^\alpha(r_j) =T(\Y,r) + d_\alpha\cdot s_{high}(r_j)
\end{equation}
where the pointwise scaling is either $s_{low} \equiv s_{high} \equiv 1$ in case of the unscaled test statistic $D_{\operatorname{MAD}}$, $s_{low}(r_j)=s_{high}(r_j)=\operatorname{sd}(\widehat{T}(\bm{Y}, r_j))$ in case of $D_{\operatorname{ST}}$ and $s_{low}(r_j) = |\widehat{T}(\bm{Y}, r)_{\frac{0.05}{2}} - T(\Y, r)|$ and $s_{high}(r_j) = |\widehat{T}(\bm{Y}, r)_{1-\frac{0.05}{2}} - T(\Y, r)|$ in case of $D_{\operatorname{QDIR}}$. 
By construction, the global envelope corresponding to the (unscaled) maximum absolute deviation has the same constant width at all evaluation points. In general, the test decision can differ for the different envelope constructions as they visualize different test statistics. 

\begin{figure}
	\subfloat[$D_{\operatorname{FUN}}$ with $L$-function]{\includegraphics[width=0.31\textwidth]{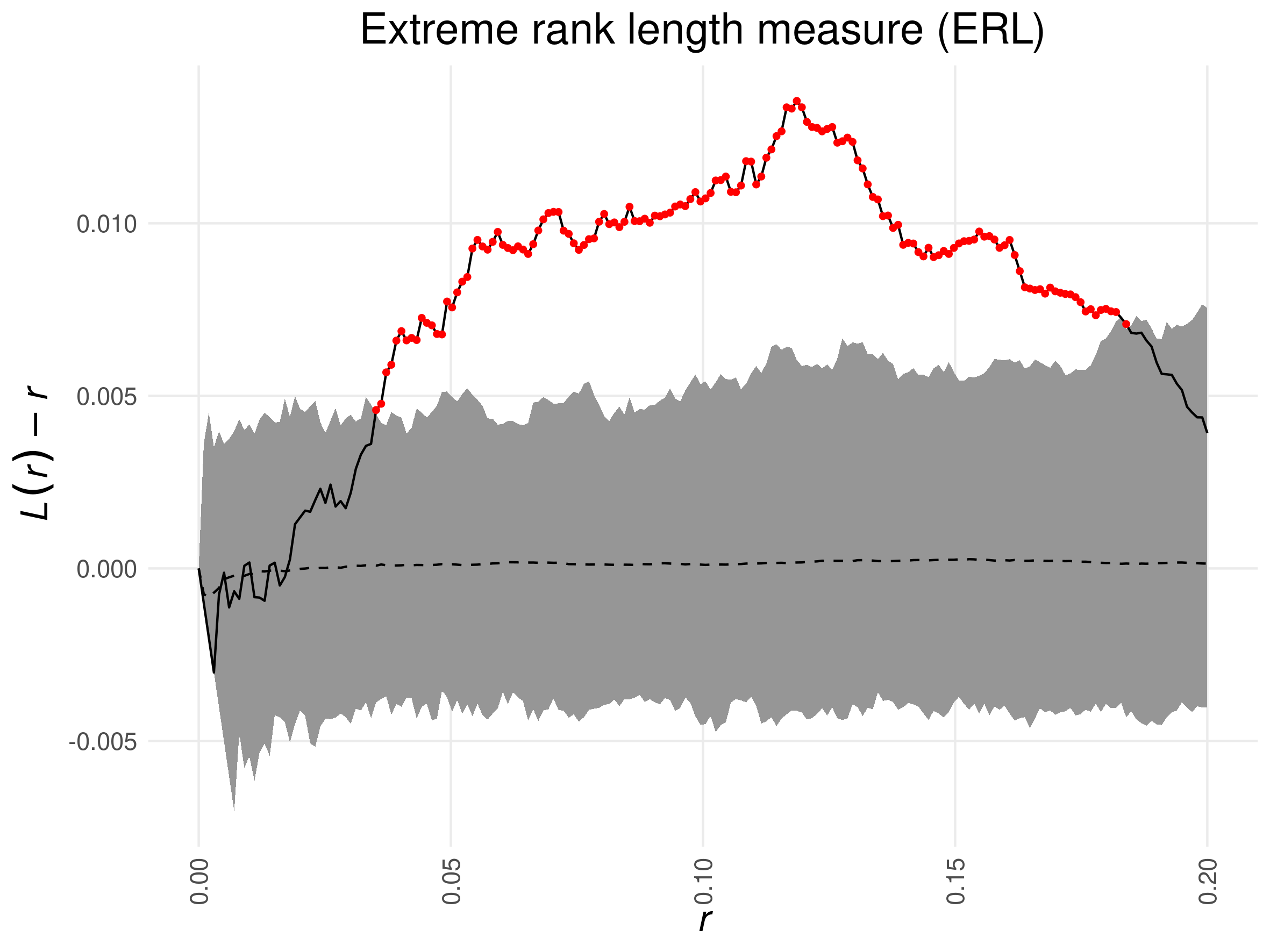}} \hfill
    \subfloat[$D_{\operatorname{FUN}}$ with $J$-function]{\includegraphics[width=0.31\textwidth]{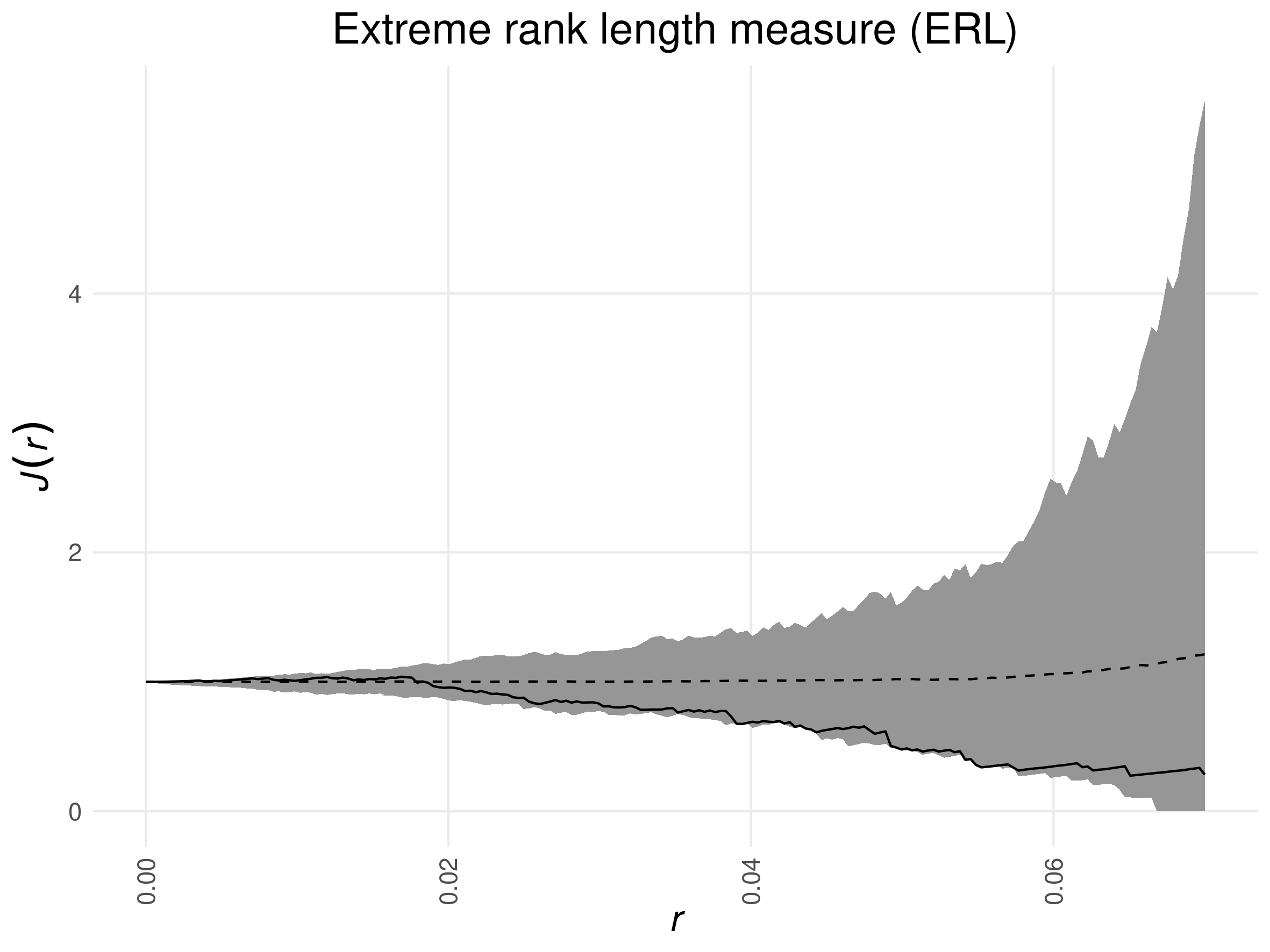}} \hfill
    \subfloat[$D_{\operatorname{FUN}}$ with $\beta_1$]
    {\includegraphics[width=0.31\textwidth]{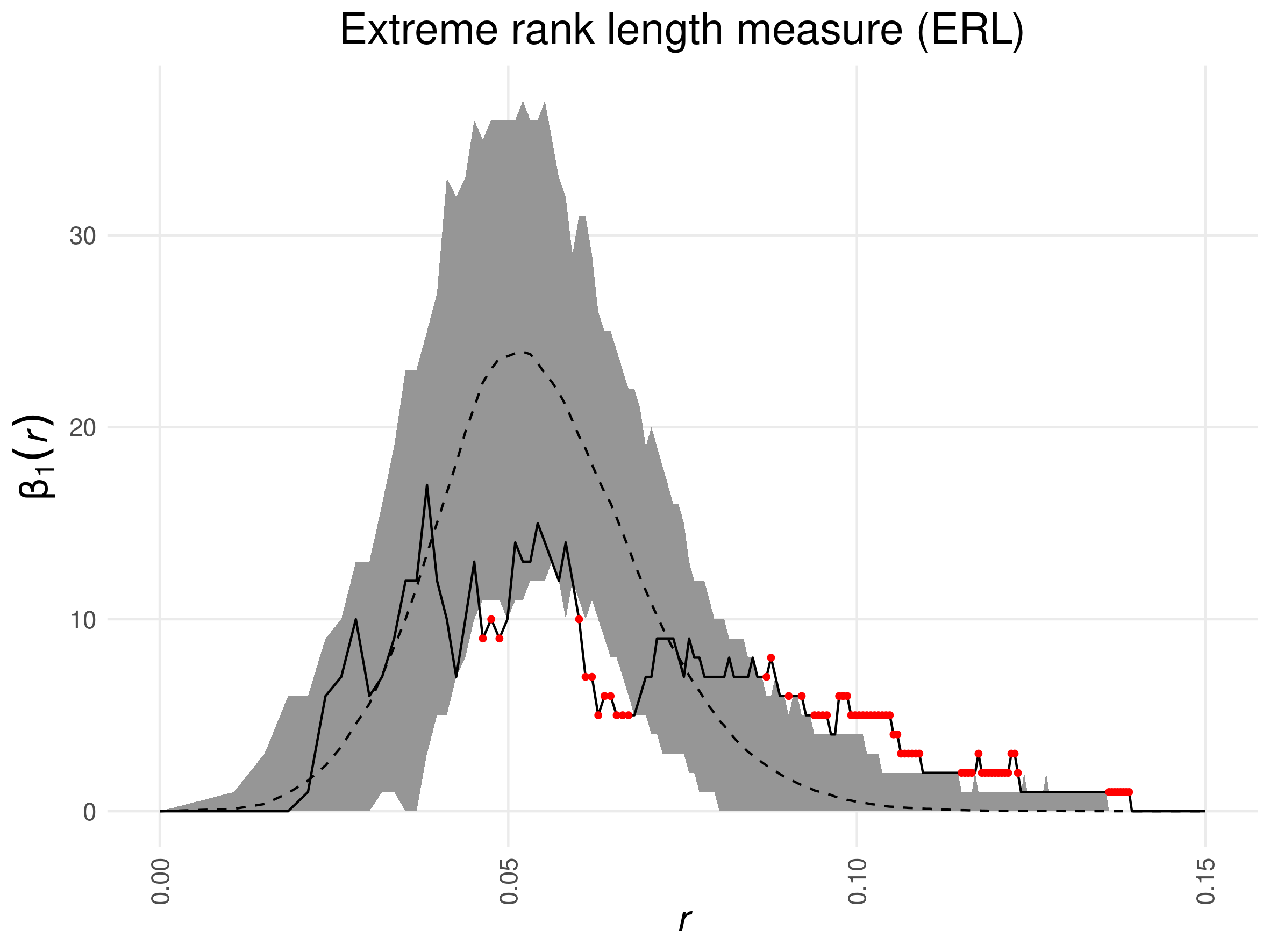}}
    
    \vspace*{3ex}
 	\subfloat[$D_{\operatorname{MAD}}$ with $L$-function]{\includegraphics[width=0.31\textwidth]{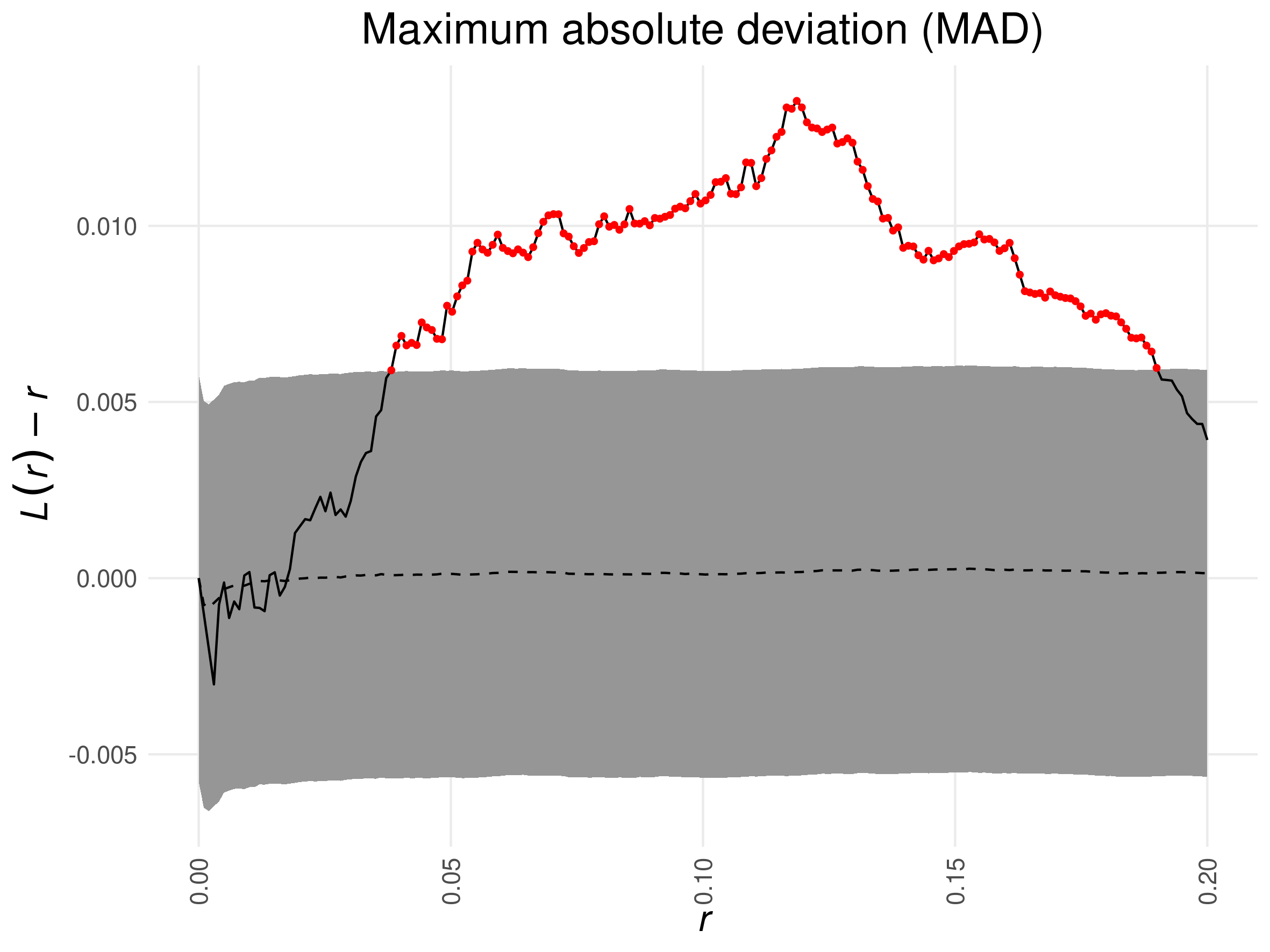}}\hfill
  	\subfloat[$D_{\operatorname{MAD}}$ with $J$-function]{\includegraphics[width=0.31\textwidth]{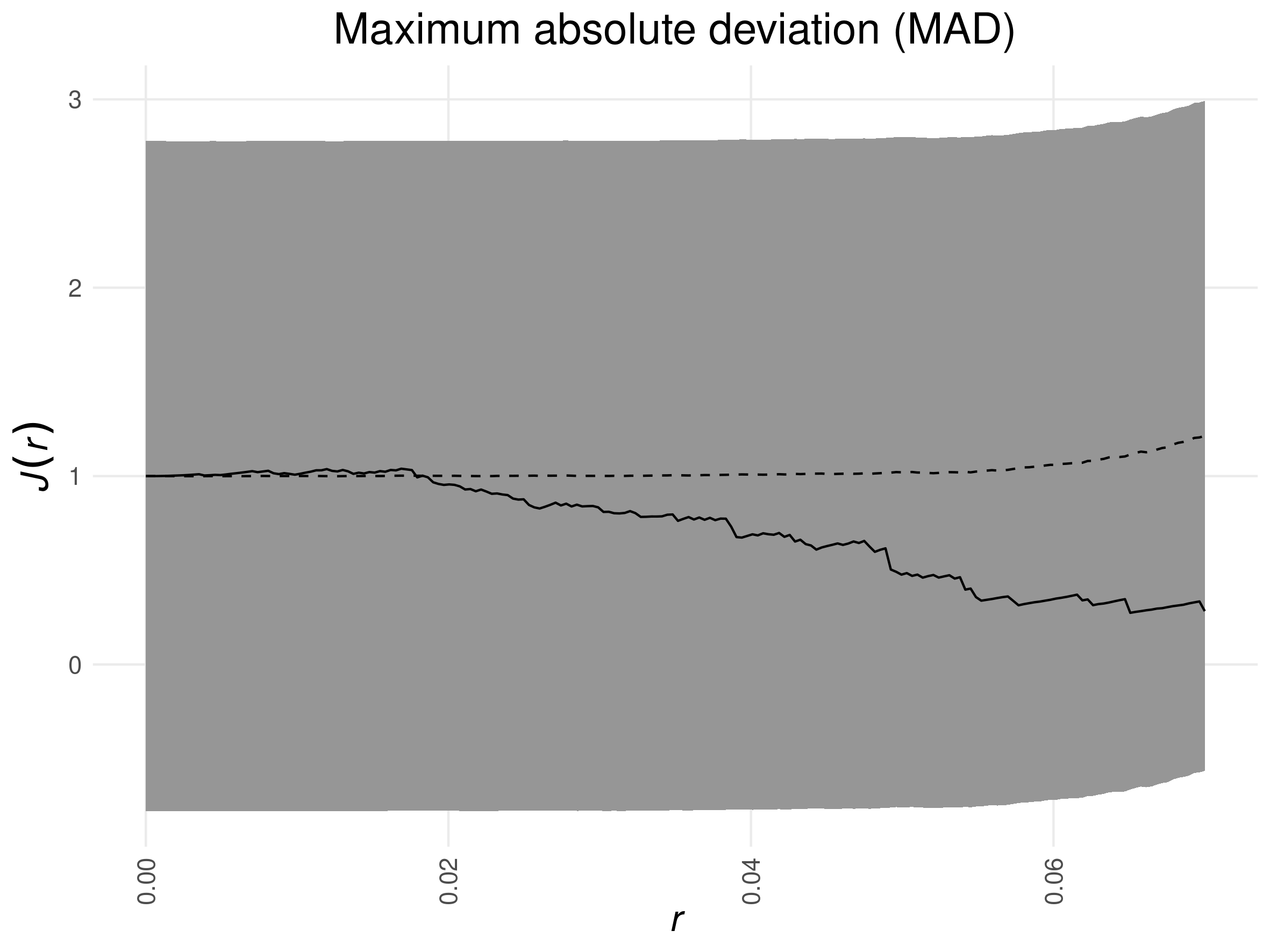}}\hfill
   \subfloat[$D_{\operatorname{MAD}}$ with $\beta_1$]{\includegraphics[width=0.31\textwidth]{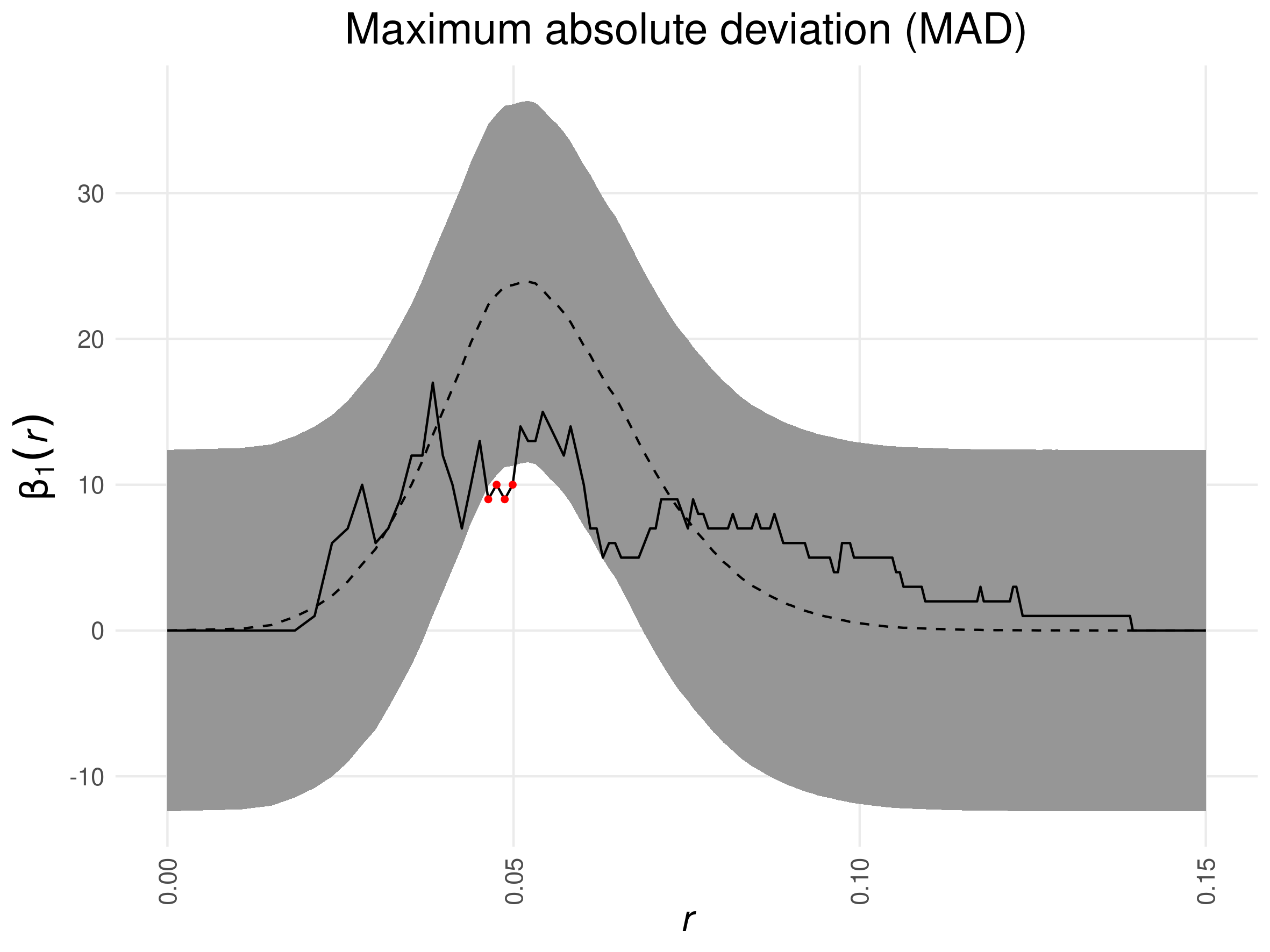}}

   \vspace*{3ex}
	\subfloat[$D_{\operatorname{QDIR}}$ with $L$-function]{\includegraphics[width=0.31\textwidth]{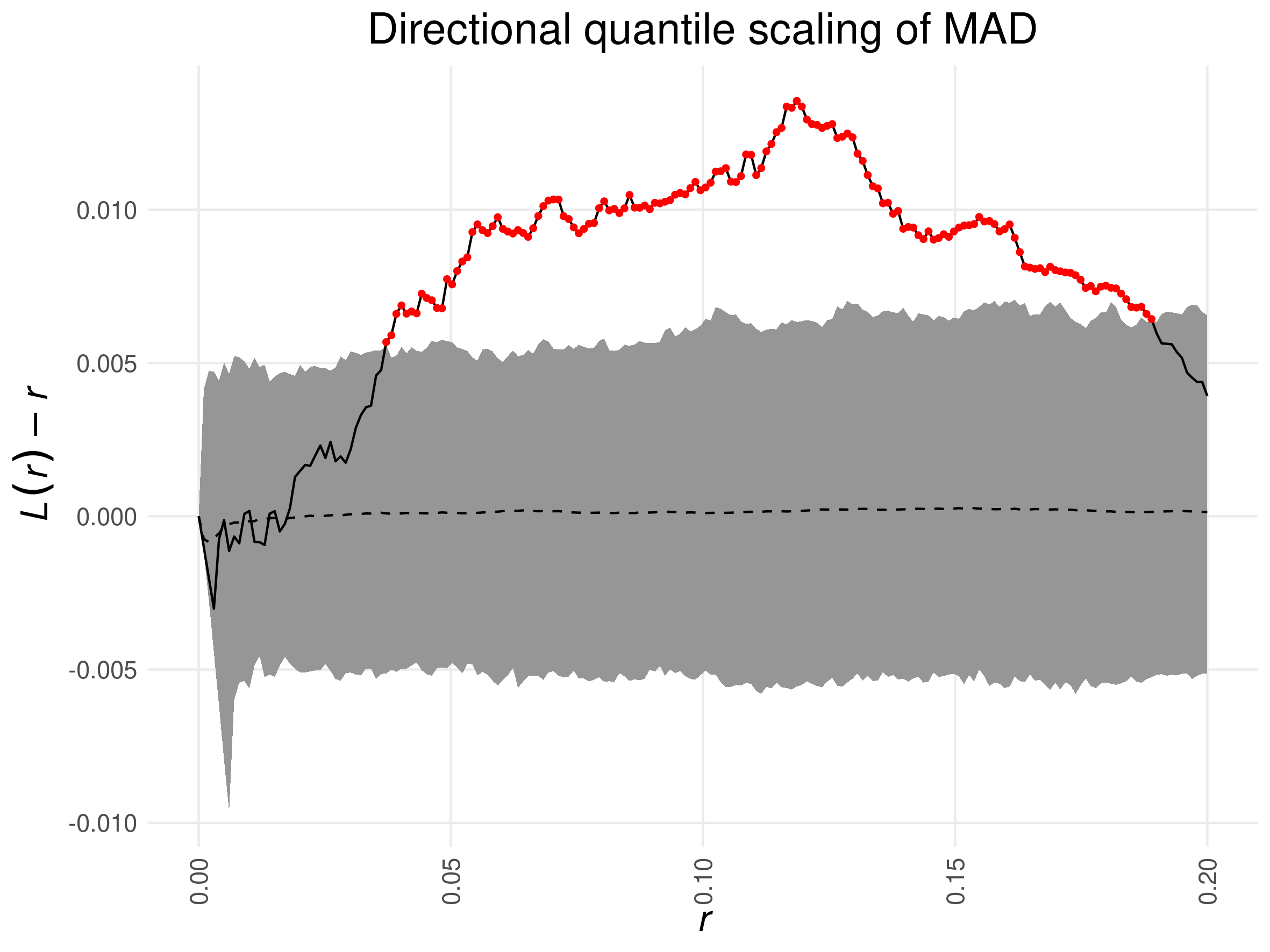}}\hfill
    \subfloat[$D_{\operatorname{QDIR}}$ with $J$-function]{\includegraphics[width=0.31\textwidth]{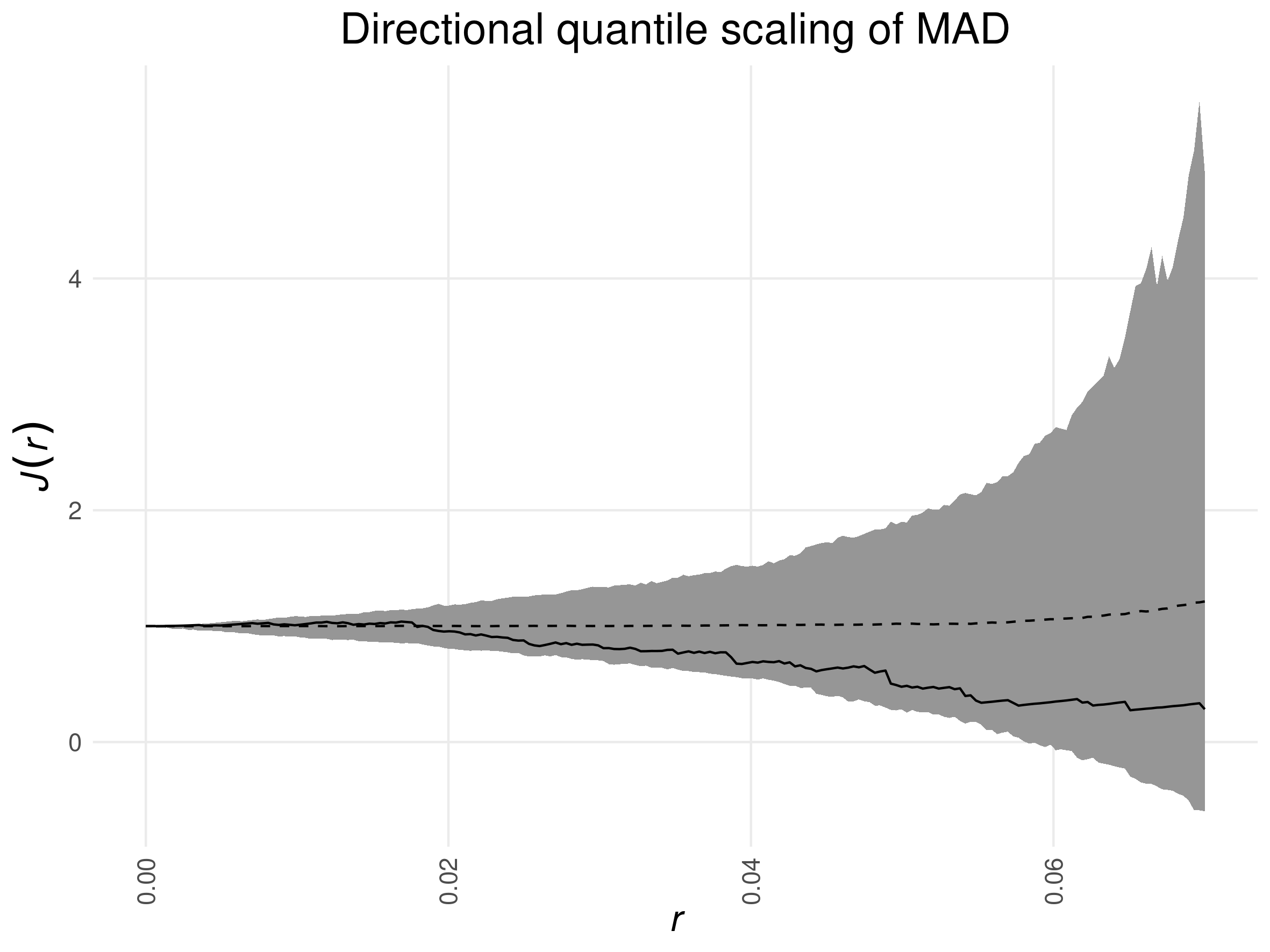}}\hfill
	\subfloat[$D_{\operatorname{QDIR}}$ with $\beta_1$]{\includegraphics[width=0.31\textwidth]{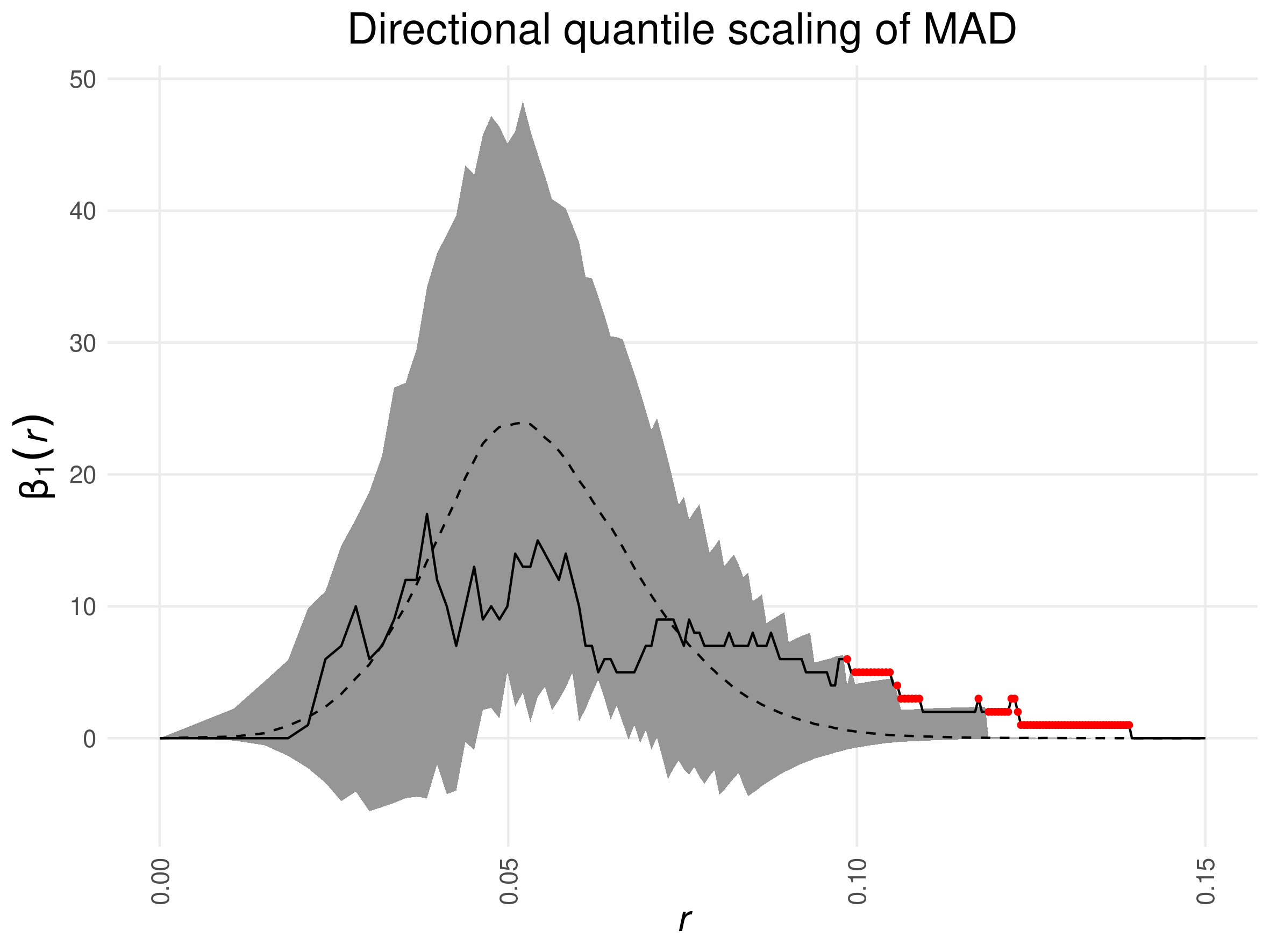}}

    \vspace*{3ex}
  	\subfloat[$D_{\operatorname{SCORE}}$ with $L$-function]{\includegraphics[width=0.31\textwidth]{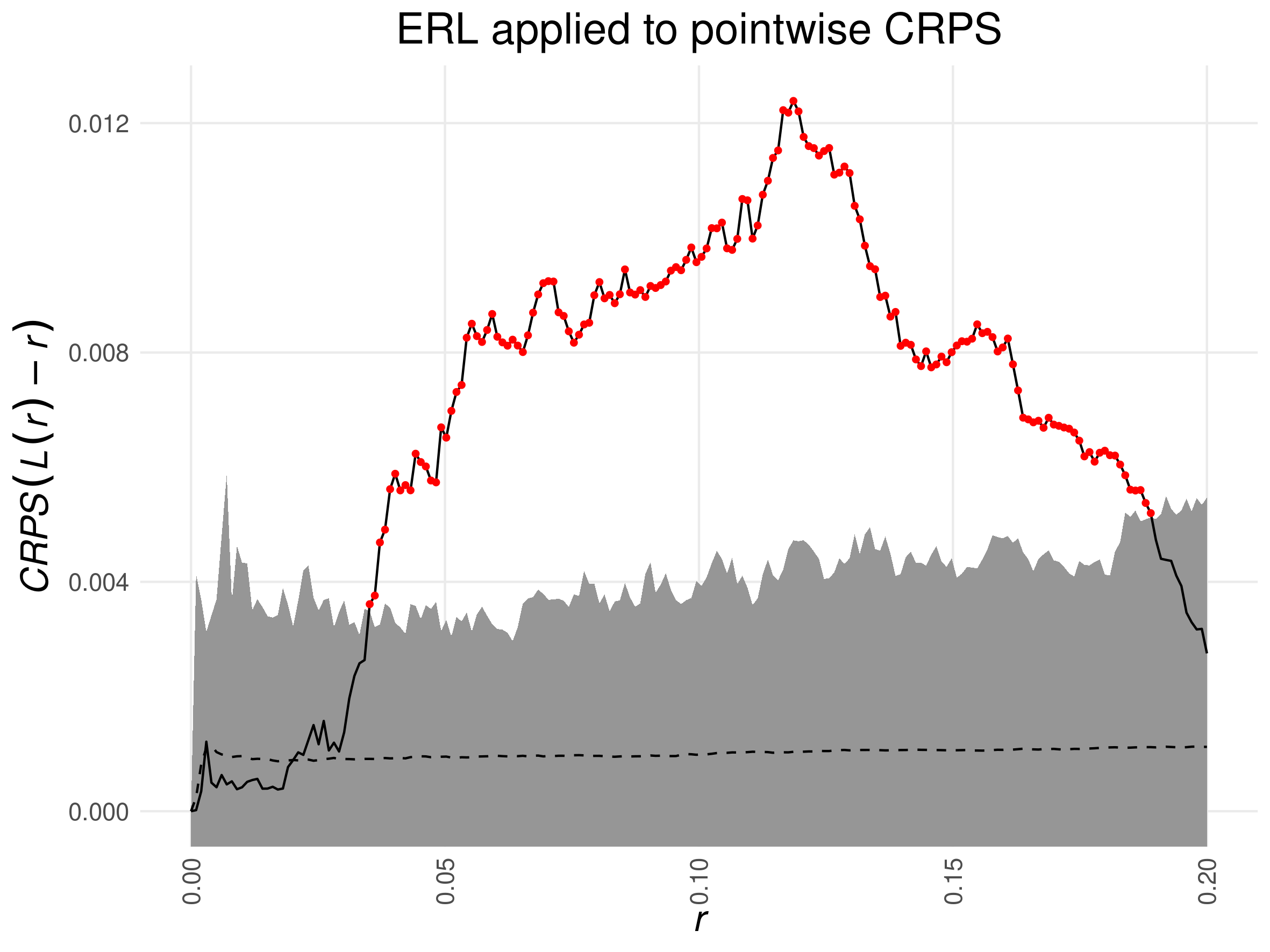}}  \hfill
	\subfloat[$D_{\operatorname{SCORE}}$ with $J$-function]{\includegraphics[width=0.31\textwidth]{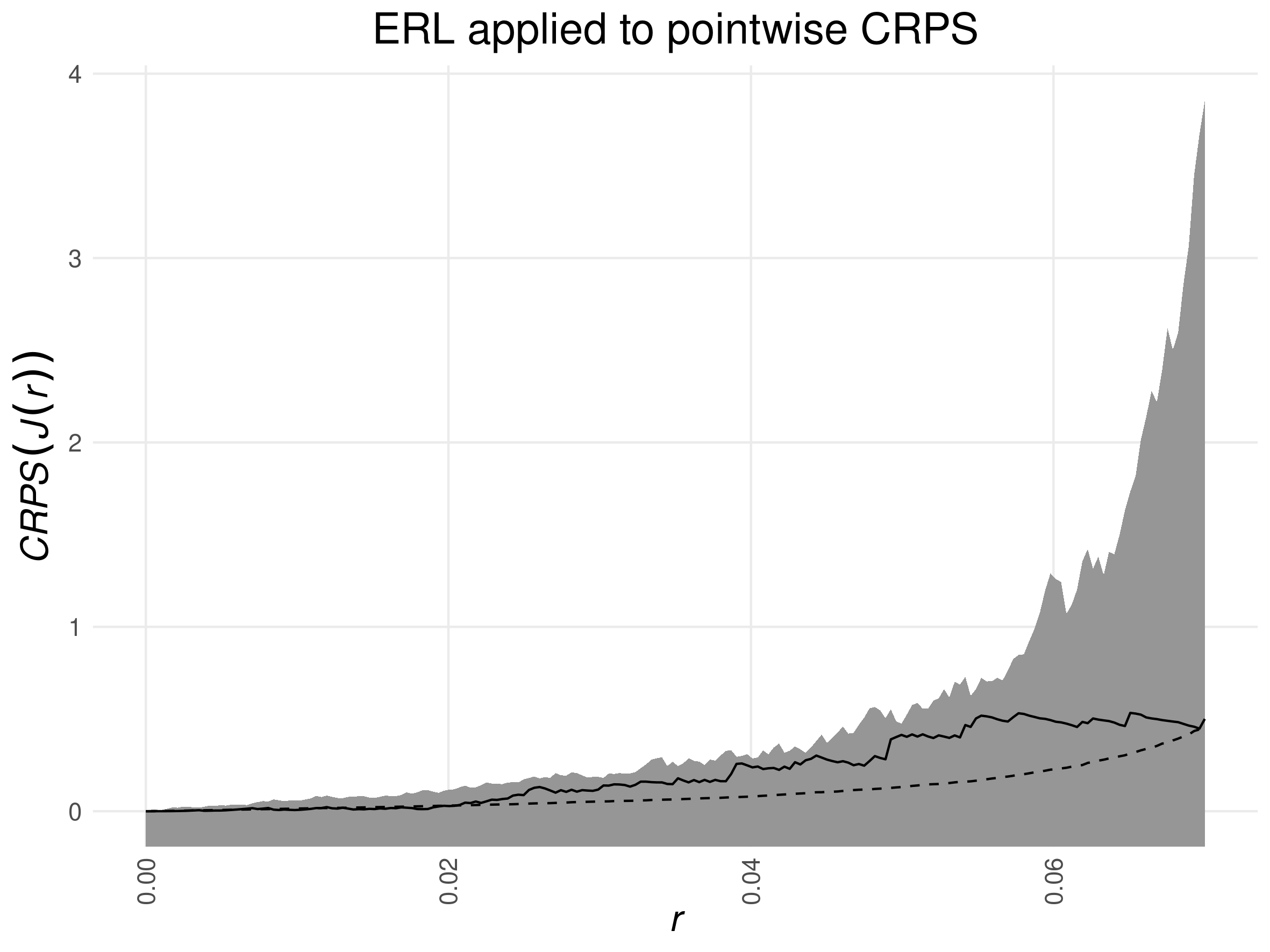}}  \hfill
	\subfloat[$D_{\operatorname{SCORE}}$ with $\beta_1$]{\includegraphics[width=0.31\textwidth]{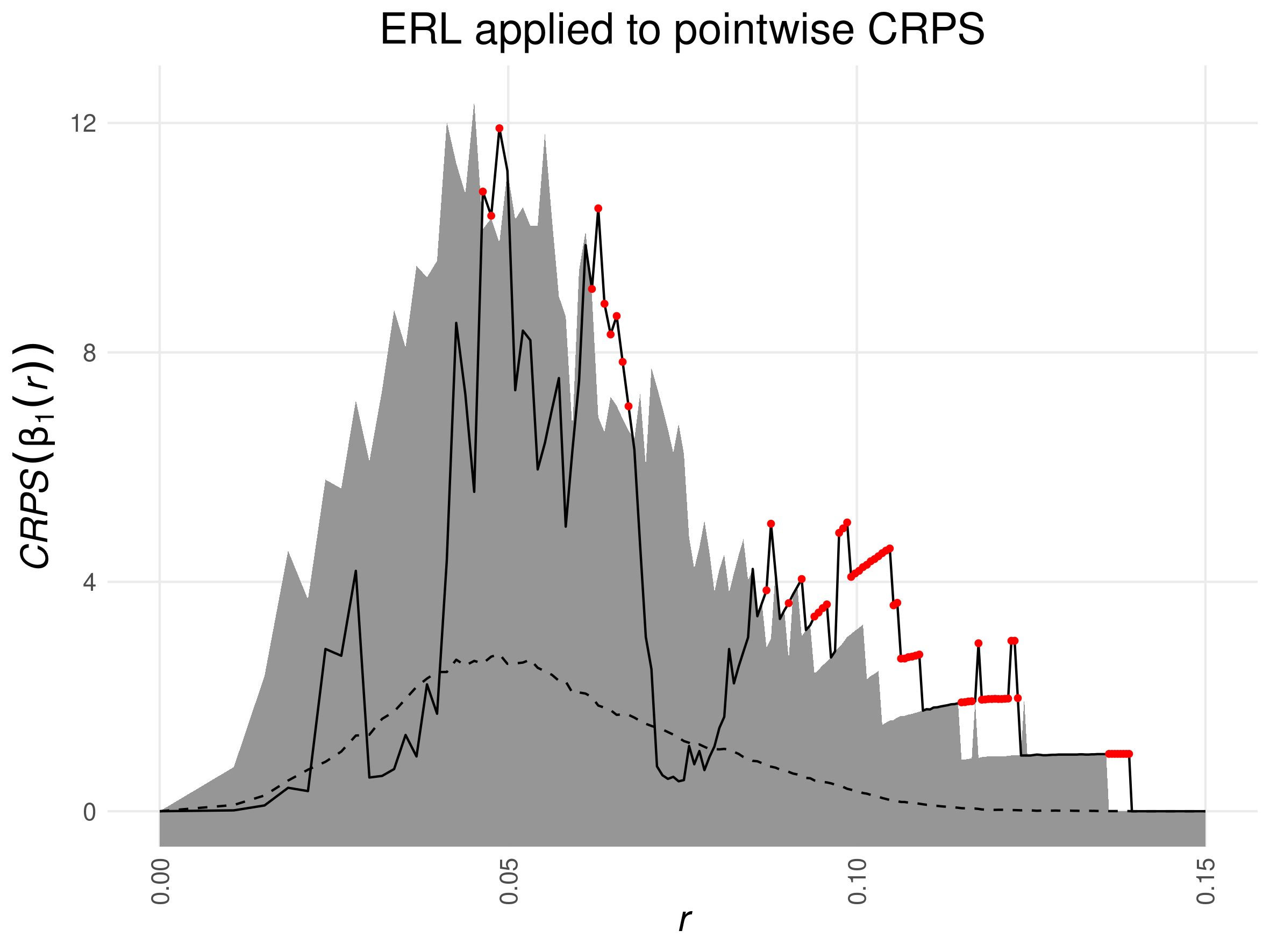}}
	\caption{Graphical representation of the goodness-of-fit tests testing complete spatial randomness as global envelopes. The solid black/red line corresponds to the empirical summary statistic for the observed point pattern shown in Figure~\ref{fig:get-obs-pattern}. The dashed line represents the pointwise sample mean of all the individual summary statistics. We used the same $m=499$ simulations of the null hypothesis for all tests. In the first column, the $L$-function was used as functional summary statistic, in the second column the $J$-function and in the third column the $1$-dimensional Betti curve.}
	\label{fig:get-example}
\end{figure}

Figure~\ref{fig:get-example} shows an example of the different global envelopes that are implemented in the R-package \texttt{GET}. The observed pattern shown in Figure~\ref{fig:get-obs-pattern} is a realization of a Matérn cluster point process with cluster center intensity $\kappa=50$, mean number of points per cluster $\mu=5$ and cluster radius $R=0.1$. The observation window is $W = [0,1]^2$. The null hypothesis is complete spatial randomness. We use as functional summary statistic either the $L$-function estimate using the isotropic edge correction implemented in the \texttt{spatstat} function \texttt{Lest}, the $J$-function with the Kaplan-Meier estimator implemented in the \texttt{spatstat} function \texttt{Jest} or the one-dimensional persistent Betti curve $\beta_1$ evaluated at the diagonal $\{(r,r) \mid r \in \R_{+}\}$. For the TDA-based statistic, we used the persistence diagram of the alpha complex filtration computed in the R-package \texttt{TDA} \citep{TDA}. No additional edge-correction was used.

The graphical representations of $D_{\operatorname{QDIR}}$, $D_{\operatorname{ST}}$ and $D_{\operatorname{FUN}}$ with the extreme rank ordering turned out to be very similar in the study of \citet{myllymaki2017}. Additionally, also the extreme rank ordering yields approximately the same graphical representation and hence eventually the same test decision. For a good approximation of the common global envelope, fewer simulations were needed in \citet{myllymaki2017}, when the scaling approaches $D_{\operatorname{QDIR}}$ and $D_{\operatorname{ST}}$ are used. However, this approximation is not always guaranteed. The scaling approaches use comparably little information on the pointwise distribution. In particular, in comparison with the extreme rank length ordering, they do not take pointwise ties into account. Ties occur often if the chosen functional summary statistics is integer-valued. Consequently, the envelopes obtained from the measures can differ a lot, see e.g. Figure~\ref{fig:get-example} (c), (f) where the summary statistic is the $1$-dimensional Betti curve.

\begin{table*}
\renewcommand{\arraystretch}{1.5}
\begin{minipage}[c]{0.395\textwidth}\vspace{0pt}
\centering
    \includegraphics[width=0.95\textwidth]{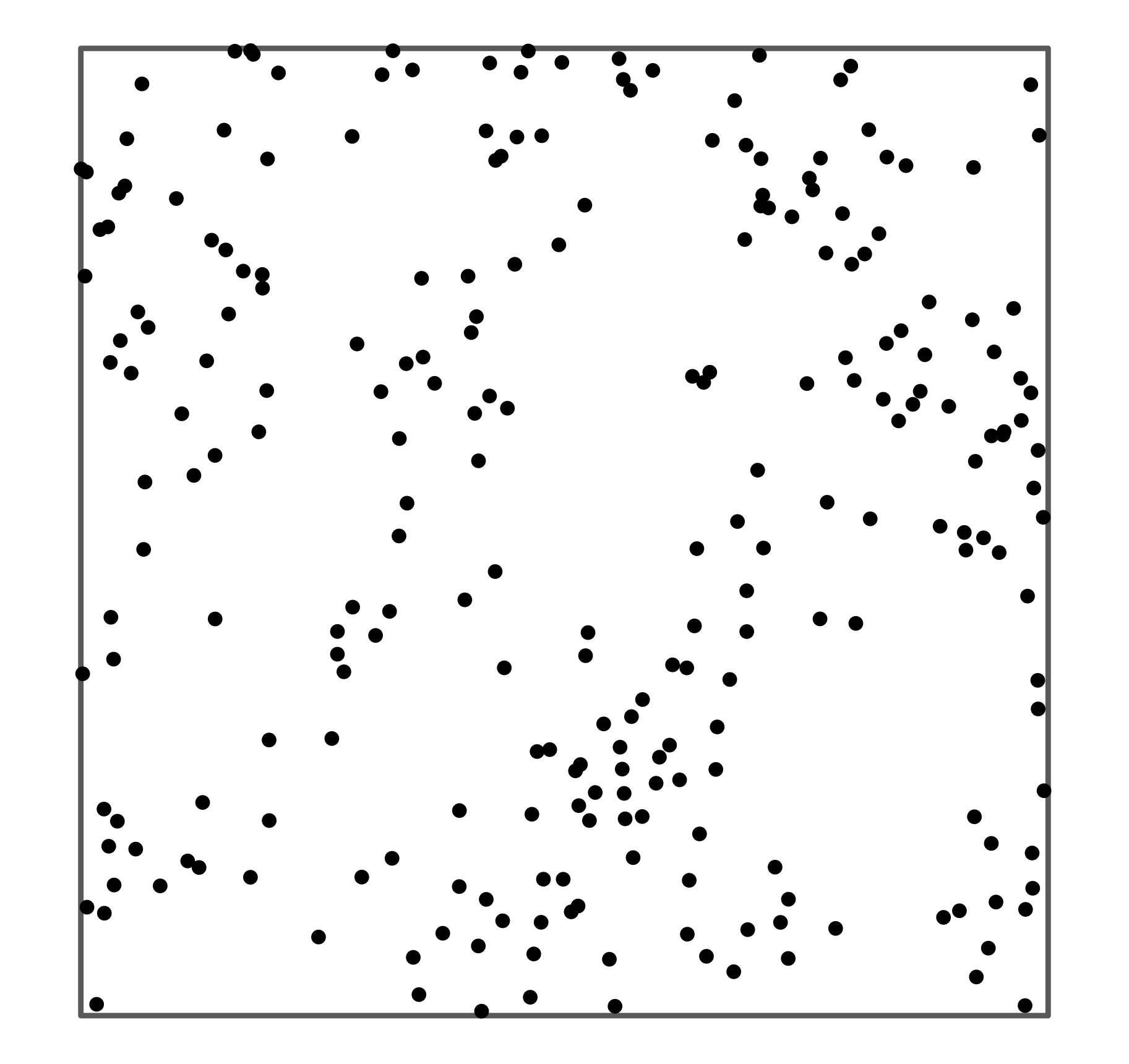}
    \captionof{figure}{Realization of a Matérn cluster point process with $\kappa=50$, $\mu=5$ and $R=0.1$ in the observation window $W=[0,1]^2$.}
    \label{fig:get-obs-pattern}
\end{minipage}\hfill
\begin{minipage}[c]{0.595\textwidth}
\centering\vspace{0pt}
\begin{tabularx}{\linewidth}{Xllll}\toprule
Summary Statistic & $D_{\operatorname{FUN}}$ & $D_{\operatorname{MAD}}$ & $D_{\operatorname{QDIR}}$ & $D_{\operatorname{SCORE}}$\\ \midrule \relax
$L$-function & 0.002 & 0.002 & 0.002 & 0.002\\
$J$-function & 0.162 & 0.250 & 0.184 & 0.348 \\
$1$-dim Betti curve $\beta_1$ & 0.002 & 0.002 & 0.018 & 0.002 \\
\bottomrule
\end{tabularx}
\captionof{table}{Monte Carlo $p$-values of the global envelopes shown in Figure~\ref{fig:get-example}. Both $D_{\operatorname{FUN}}$ and $D_{\operatorname{SCORE}}$ use the extreme rank length ordering (ERL). A $p$-value estimate of $0.002$ indicates that the observed pattern shown in Figure~\ref{fig:get-obs-pattern} was the single most extreme in the set with $499$ simulations of complete spatial randomness.}
\label{tab:pvals}
\end{minipage}
\end{table*}  

Table~\ref{tab:pvals} additionally lists the estimated $p$-values obtained for each of the visualized tests. In all tests with either the $L$-function or the $1$-dimensional Betti curve $\beta_1$, the null hypothesis of complete spatial randomness is rejected at significance level $\alpha=0.05$. When using the $1$-dimensional Betti curve, we observe large differences in the envelopes between the scaled and unscaled maximum absolute deviation tests. In particular the significant differences at large spatial scales are not seen with the unscaled test statistic $D_{\operatorname{MAD}}$.

Several other test statistics have a graphical representation, but not necessarily as an envelope of a summary statistic. \citet{baddeley2014} mention that one can represent the $D_{\operatorname{DCLF}}$ test statistic and the corresponding acceptance region as a function of the upper bound of the integration domain. With this representation one can visually investigate the \textit{gap} between the observed test statistic and the border of the acceptance region. It is possible to state for which upper bounds the null hypothesis would be rejected or has to be accepted. This allows to identify the spatial scales where the test decision changes and consequently where the null model and the observed point pattern are closer or further away from each other. The same construction is also possible for other scalar-valued test statistics that involve integrals over a transformation of the pointwise differences $\widehat{T}(\x,r)-T(\Y,r)$ such as the integrated continuous ranked probability score $D_{\operatorname{CRPS}}$, the scaled variants of the $D_{\operatorname{DCLF}}$ test statistic or the direct integral test statistic $D_{\operatorname{INT}}$. 

As mentioned in Section~\ref{sec:test-no-deviation}, the pointwise critical values of a test based on $D_{\operatorname{POINT}}$ can also be represented as envelopes. To obtain a valid test, the test statistic should be compared to the envelope only at the single, a priori chosen scale $r^*$.

\subsection{Combining functional summary statistics and test statistics}
\label{sec:combi}

The framework of global envelope tests even allows to construct combined envelopes for multiple functional summary statistics.
This is particularly helpful when the alternative is not specified. In such cases, one does not know in advance what the relevant characteristics are and consequently what summary statistic should be chosen. For instance, \citet{diggle2013} recommends using a combination of the $K$-, $F$- and $G$-functions in the exploratory analysis of a point pattern since the individual functions complement each other. 
\citet{krebs2022} mention that conclusions should not be based on single TDA-based summary statistics, as their power depends heavily on the given setting. \citet{mrkvicka2009} simultaneously uses multiple functional summary statistics and additionally different types of test statistics for the individual functional summary statistics in the context of goodness-of-fit testing for random closed sets.

One should keep in mind that using multiple functional summary statistics simultaneously is again a multiple testing problem. Adding many summary statistics that do not detect a certain deviation from the null hypothesis may blur the overall detection. Consequently, the power of a combined test is not necessarily higher than the power of a test with a single functional summary statistic.

In the following, we will review the two approaches for simultaneously using multiple functional summary statistics for goodness-of-fit testing that are available in the R-package \texttt{GET} \citep{GET}. 

The first approach introduced by \citet{mrkvicka2017} is called the one-step combining procedure in \texttt{GET} and belongs to the class of global envelope tests. The idea consists of concatenating the discretized functional summary statistics for each point pattern into a long vector. Then, the extreme rank ordering is applied to the set of these vectors which gives a single $p$-value estimate together with the corresponding global envelope. This procedure only requires that all summary statistics are computed at the same number of evaluation points. This is needed to ensure that all summary statistics have the same impact on the overall test decision. Instead of the extreme rank measure, one can choose any of the measures that have a global envelope representation. In particular, also the scaled maximum absolute deviation based measures can be used. 

As a second option, the so-called two-step combining procedure is available in \texttt{GET}. In its first step, either the scalar-valued test statistics or the scalar-valued depth measures obtained from multiple Monte-Carlo tests based on different summary statistics are computed. Here, one does not have to choose the same test statistic for all summary statistics. The important aspect is that every test statistic is summarized in a scalar value, and that extremeness under the null hypothesis is indicated by either small values for all test statistics or large values for all test statistics. The reason for this requirement is that the second step consists of applying the one-sided Monte Carlo test based on the extreme rank length ordering on the set of vectors containing the individual test summaries. Since the extreme rank length ordering is only based on the pointwise ranks, the (possibly different) ranges of the individual measures are irrelevant. 

For example, it is possible to use a scalar-valued deviation based test statistic such as $D_{\operatorname{DCLF}}$ for the first functional summary statistic and the vector-valued test statistic $D_{\operatorname{FUN}}$ with some depth measure for the second functional summary statistic. By construction, large values are extreme for the first statistic while small values are extreme for the second measure. To obtain the same type of extremeness it is possible to invert the scale for the depth measure whose range is the interval $[0,1]$. We can then perform the one-sided test where large values are extreme.

On the other hand, it is not straight-forward how to combine the test statistic $D_{\operatorname{FUN}}$ with the scalar-valued test statistics of type B like the integral test statistics $D_{\operatorname{INT}}$ since the type of extremeness differs.

\begin{figure}
    \centering
	\includegraphics[width=\textwidth]{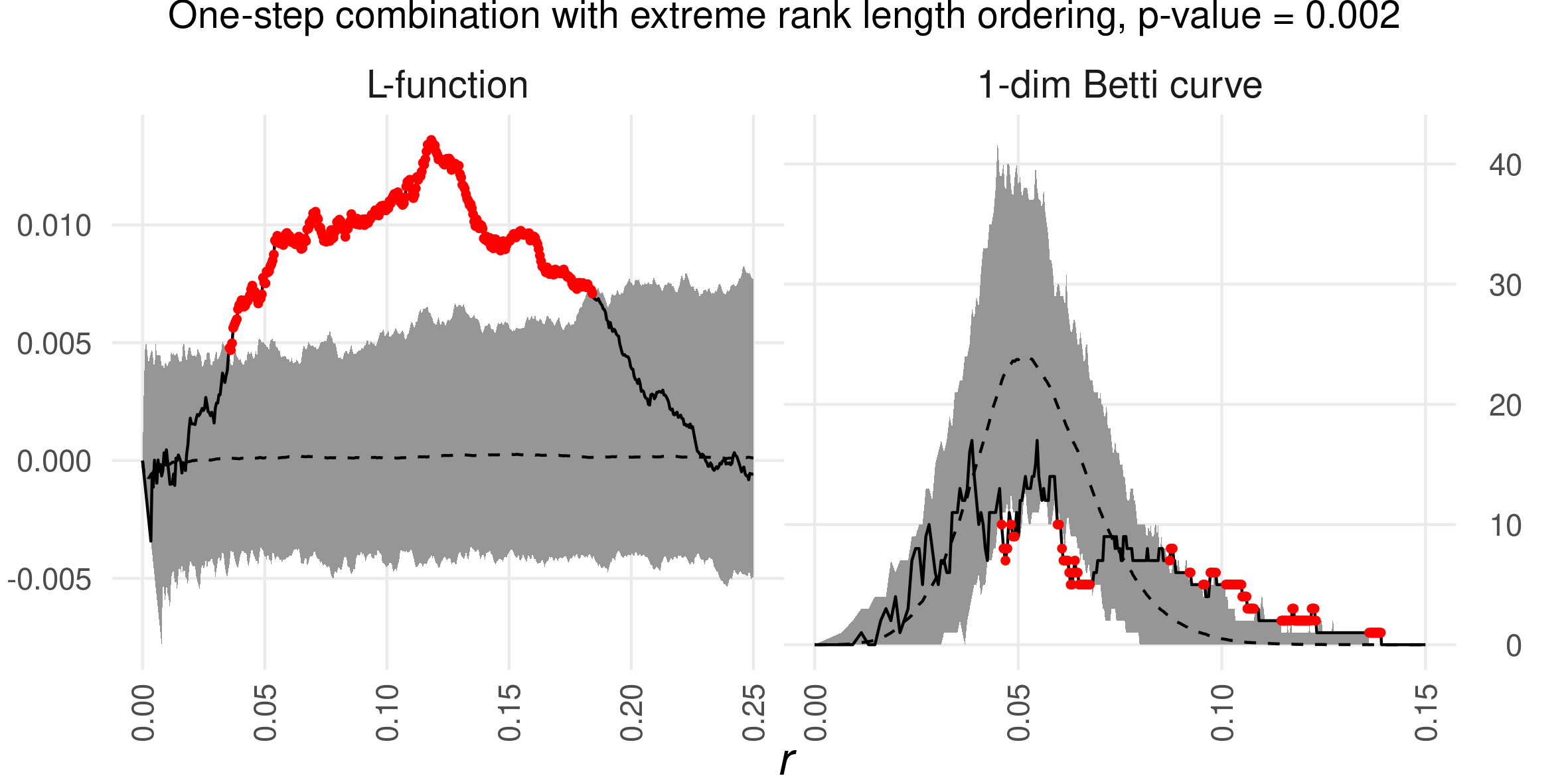}
	\caption{Combined global envelopes for testing the complete randomness hypothesis for the realization of a Matérn cluster point process shown in Figure~\ref{fig:get-obs-pattern}. Both functional summary statistics, the $L$-function and the one-dimensional Betti curve $\beta_1$, are evaluated at $n=513$ evaluation points. The solid black/red line corresponds to the empirical summary function of the observed pattern and the dashed line represents the pointwise sample mean over all computed empirical summary functions. The envelope bounds are computed using $m=499$ simulations.}
	\label{fig:get-example-combi}
\end{figure}

For both the one-step and the two-step combining procedure it is possible to construct global envelopes as outlined in Section~\ref{sec:envelope}. Figure~\ref{fig:get-example-combi} shows the graphical representation of the one-step combining procedure based on the extreme rank length measure for the example discussed in the previous section. This representation is an adaption of the plots produced by the function \texttt{global\_envelope\_test} in \texttt{GET} to better visualize the concatenation of the two individual vectors.

The one-step combining procedure with the extreme rank ordering is used in \citet{biscio2019} for the accumulated persistence functions $\operatorname{APF}_0$ and $\operatorname{APF}_1$ as summary statistics while the extreme rank length ordering is used in \citet{mrkvicka2017} with any combination of $L$-, $F$-, $G$- and $J$-functions. The latter study showed that the combined global envelope tests have a power that is at least comparable to the single most powerful summary statistic (which was different for each of the alternatives that were considered). 

To the best of our knowledge, the two-step combining procedure based on different measures for the individual summary statistics has not yet been investigated in simulation studies.

An alternative approach to combining summary statistics is used in the tests in \citet{biscio2020}. The authors consider a linear combination with fixed linear coefficients of two scalar-valued test statistics of TDA-based summary statistics. The first one is the integral measure $D_{\operatorname{INT}}$ of the number of deaths of $0$-dimensional topological features $\operatorname{ND}_0$ and the second one is the point evaluation $D_{\operatorname{POINT}}$ of the $1$-dimensional accumulated persistence function $\operatorname{APF}_1$. 
Their simulations indicate that the linear combination of the test statistics outperformed the individual test statistics in the asymptotic goodness-of-fit test when testing for complete spatial randomness. Additionally, the pointwise linear combination of both summary statistics is used in a Monte Carlo test based on $D_{\operatorname{FUN}}$ with the extreme rank measure for the same hypothesis. Again, the linear combination performed better than the individual summary statistics. \citet{biscio2020} point out, that these results indicate that it is worthwhile to consider further research regarding suitable combinations of summary statistics, both in terms of linear combinations and in the combined envelope approach mentioned before. This aspect is consequently relevant for our forthcoming comparative power study.

\subsection{Number of simulations}
\label{sec:nsim}

One important parameter of simulation-based goodness-of-fit tests is the number of simulations $m$ of the null model that is used to estimate the $p$-value in a Monte Carlo test setting or the quantiles and moments of the (limiting) distribution of the summary statistic estimator. We focus on reviewing approaches for selecting the number of simulations in Monte Carlo tests. This aspect is also discussed in \citet{mrkvicka2016} with a focus on global envelope tests based on the extreme rank ordering.

Assume in the following the fixed significance level $\alpha=0.05$. 
Based on the Monte Carlo $p$-values in \eqref{eq:mc-pval}, one needs at least $m=19$ simulations to be able to reject the null hypothesis at this level. This minimal number was used in the comparative studies in \citet{baddeley2014} and \citet{baddeley2017}.  \citet{baddeley2014} mention that a higher number of simulations usually increases the power of the goodness-of-fit test, but their study focuses on other aspects that affect the power.

\begin{figure}[t]
\centering
\includegraphics[width=0.8\textwidth]{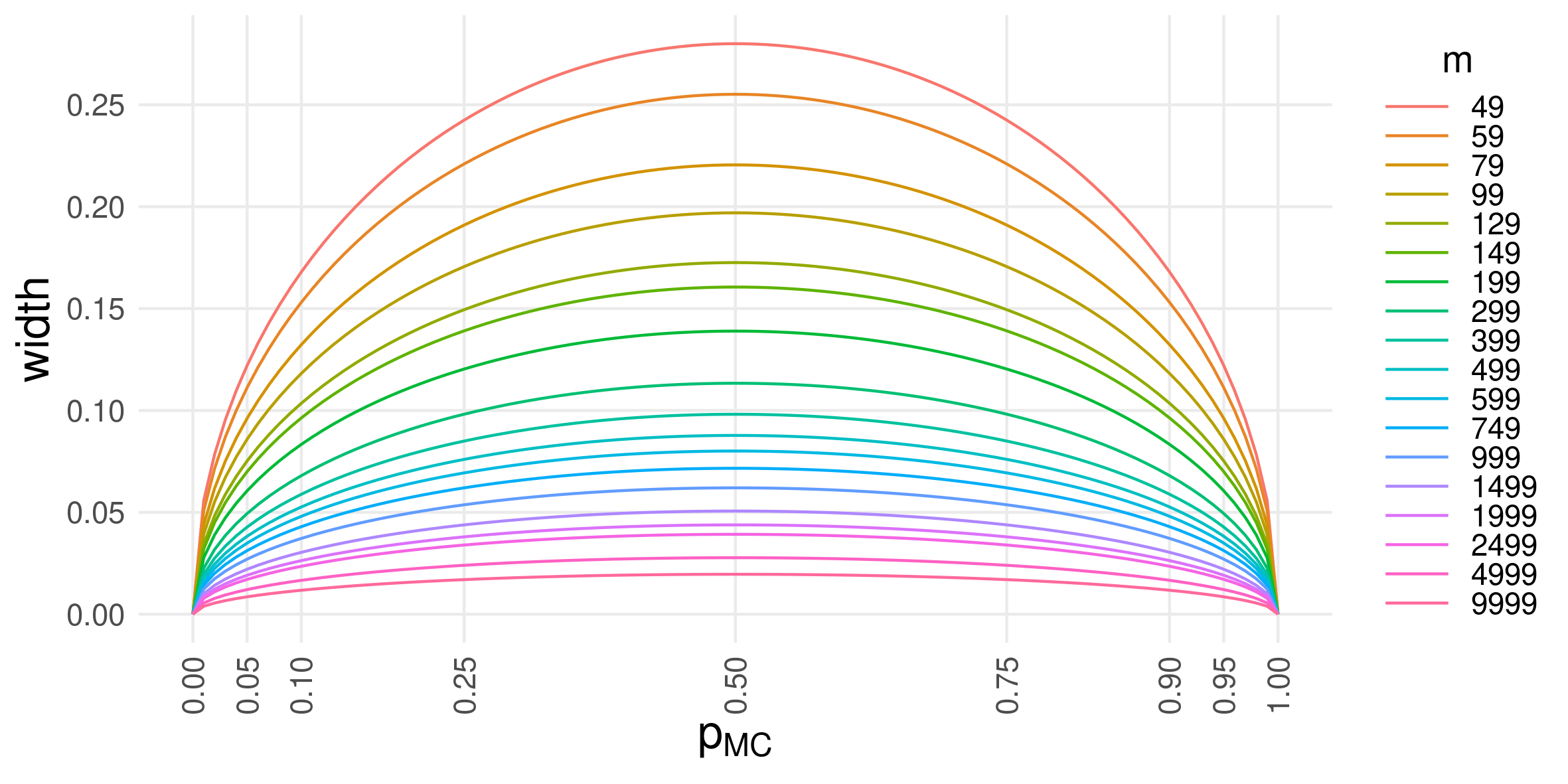}
\caption{Width of the approximate $95$~\% confidence interval for the true $p$-value of a point pattern fulfilling the null hypothesis.}
\label{fig:cipval}
\end{figure}

In the literature, the most common choice is $m=99$ \citep{besag1977, diggle1979, thonnes1999, baddeley2000, ho2006, ho2007, robins2016, baddeley2017}. In these works, mostly deviation based test statistics of type A were investigated. \citet{diggle1979} mentions that $m=99$ usually suffices as beyond that point there are no large increases in the power. \citet{davison1997} advise $m \geq 99$ for general Monte Carlo tests where the test statistic is continuous. Their conclusion is based on the derivation of an upper bound for the difference in power when using the Monte Carlo $p$-value or the true $p$-value. Furthermore, they state that $m = 999$ should in general be more than enough.
\citet{loosmore2006} discuss the problem of characterizing the uncertainty of Monte Carlo $p$-value estimates. They derive that under the null hypothesis $mp_{\text{MC}}$ approximately follows a binomial distribution $\operatorname{Bin}(mp,p)$. Using the central limit theorem, one obtains the approximate $95\%$ confidence interval for the true $p$-value $p$ as \begin{equation}
   I_{0.95} = p_{\text{MC}} \pm q_{0.975} \sqrt{\frac{p_{\text{MC}}(1-p_{\text{MC}})}{m}}.
\end{equation}
The number of simulations $m$ influences the width of the confidence interval which is \begin{equation}
\label{eq:width-ci}
    2\cdot  q_{0.975} \cdot \sqrt{\frac{p_{\text{MC}}(1-p_{\text{MC}})}{m}}.
\end{equation}
Figure~\ref{fig:cipval} shows the width of this interval depending on the values of $p_{\text{MC}}$ and $m$.

\citet{loosmore2006} also derive a suggestion for the number of simulations from the confidence intervals: Consider the significance level $\alpha=0.05$. Assume that a Monte Carlo $p$-value of $p_{\operatorname{MC}}=0.1$ is observed, implying acceptance of the null hypothesis. The probability of the true $p$-value $p$ being larger than $\alpha$ should be high to not change the test decision when using the $p_{\operatorname{MC}}$ instead of $p$. The high probability is concretized by using the approximate $95\%$ confidence interval for $p$ derived above. \citet{loosmore2006} suggest that the width of the interval should be at most $0.04$ if $p_{\operatorname{MC}}=0.1$ as the lower bound for this choice is equal to $0.08 > \alpha$.
Plugging these values into Equation~\eqref{eq:width-ci} we obtain $m \geq 865$. Based on this reasoning, \citet{loosmore2006} recommend using $m=999$ simulations. 

For global envelope tests with the test statistic $D_{\operatorname{FUN}}$, \citet{myllymaki2017} recommend $m=2499$ when using the extreme rank measure, which only considers the most extreme pointwise rank. Their recommendation is based on a simulation study of the number of ties in the extreme rank measure under the null hypothesis of complete spatial randomness with either the $L$-, $F$-, $G$- or the $J$-function. 

In the same setting, \citet{mrkvicka2017} discuss the number of simulations when combining several summary statistics using the extreme rank measure. They recommend $m=k\times2499$ simulations when using $k$ summary statistics simultaneously. Furthermore, both \citet{mrkvicka2017} and \citet{myllymaki2017} remark that the suggested number of $m=2499$ can be reduced if the extreme rank length measure is chosen instead of the extreme rank measure. The same conclusion applies for the continuous and the area rank measure, see \citet{myllymaki2020}. Specific power studies can be used to give better recommendations for the choice of $m$. This is done e.g. in \citet{myllymaki2020} for all four measures and the quantile scaled maximum absolute deviation statistic $D_{\operatorname{QDIR}}$ in the setting of Gaussian processes. The power increased only in a few cases when more than $m=599$ simulations were used. In fact, for the extreme rank length measure and the area measure, there was a decrease in power for increasing $m$ when the observed data was extreme over a large part of the domain.

According to \citet{myllymaki2017}, the common choice $m=99$ is reasonable for the test statistics $D_{\operatorname{MAD}}, D_{\operatorname{ST}}$ and $D_{\operatorname{QDIR}}$.

The minimal $m$ for the global envelope approaches of \citet{wiegand2016} can be computed explicitly as outlined in Section~\ref{sec:envelope}. In general, the necessary number for the simulation-based global envelopes is even higher compared to the global envelope test of \citet{myllymaki2017} with the extreme rank ordering.

Overall, there is always the trade-off between computational speed and a good approximation of the true $p$-value. Figure~\ref{fig:cipval} can help to choose the right number of simulations for a particular setting. If one wants to make sure that the lower bound of the approximate $95$~\% confidence interval when observing a (hypothetical) $p$-value estimate of $p_{\operatorname{MC}}=0.1$ is larger than $0.05$, then one should choose at least $m=139$. This recommendation can be used in all Monte Carlo tests where the probability of having ties in the test statistic is negligible. Note that the confidence intervals are not exact by construction, and thus for small numbers of $m$ additional approximation errors occur. 

\section{Power studies for goodness-of-fit tests}
\label{sec:powerstudy}

Power studies in the literature usually compare the power of the tests for the null hypothesis of complete spatial randomness. Some papers have also examined more complicated null models. The studies differ not only in the subset of possible combinations of summary statistic and test statistic that are compared but also in the general setting. This includes different intensities, window sizes and different types of alternatives, i.e. the true point process models from which the data are sampled. This makes it difficult to compare the conclusions of different studies, especially if they have nothing in common.
Certain combinations of test statistic and functional summary statistic have not yet been investigated.

\begin{sidewaystable}
    \centering
    \renewcommand{\arraystretch}{1.5}
   \begin{tabularx}{\textwidth}{cp{3cm}p{5cm}p{4cm}X}\toprule
    ID & Study & Comparisons & Alternatives & Key Findings\\ \midrule \relax
    A &\citet{diggle1979} & $\operatorname{MAD}$ with $K$, $L$, $F$, $G$; classical scalar indices & Poisson cluster, SSI processes & $L$ more powerful than $K$; $F$ and Quadrat-Counts more powerful against clustering; $G$ against regularity; advocate use of functional summary statistics\\
    B& \citet{ripley1979} & Classical scalar indices of nearest-neighbor distances; variance function; $\operatorname{MAD}$ of $L$ & Strauss, Matérn cluster processes & Tests based on variance function are powerful against clustering; squared nearest neighbor distances and $L$ against regularity\\
    C& \citet{zimmerman1993} & Cramér--von Mises test statistic $\bar{\omega}^2$ for bivariate uniform distribution; classical scalar indices of nearest-neighbor distances; variance function; $\operatorname{MAD}$ of $L$ & Poisson cluster, SSI, inhomogeneous Poisson process & $\bar{\omega}^2$ less powerful than the others against clustering, regularity; $\bar{\omega}^2$ more powerful against inhomogeneity\\
    D&\citet{thonnes1999} & $\operatorname{MAD}$, $\operatorname{DCLF}$ with $F$, $G$, $J$; influence of the upper bound & Thomas cluster, Matérn Type II hardcore, Area-interaction processes & Tests with $J$ comparable to $G$; $F$ only in case of weak clustering more powerful than $J$ and $G$; drop in power when increasing the upper bound for $J$ especially with $\operatorname{MAD}$\\
    E&\citet{baddeley2000} & Several estimators for $J$ with integrated studentized deviations, comparison with \citet{thonnes1999} & Matérn Type II hardcore, Matérn cluster process & Choice of test statistics higher influence on power compared to choice of estimator\\
    F& \citet{mugglestone2001} & Spectral tests; classical scalar indices; test of \citet{zimmerman1993} ($\bar{\omega}^2$), $\operatorname{MAD}$ with $L$  & Poisson cluster, SSI, inhomogeneous Poisson process & $L$ more powerful than spectral method against regularity; certain spectral tests are competitive against clustering; $\bar{\omega}^2$ most powerful against inhomogeneity\\
    G&\citet{grabarnik2002} & $Q^2$ of \citet{grabarnik2002}; $\operatorname{MAD}$ with $L$, $G$, third-order characteristic & Mixture of Matérn cluster and Strauss process & $L$ more powerful against clustering; third-order characteristic against moderate clustering; $Q^2$ and $G$ against strong regularity; $Q^2$ against the mixture of clustering and regularity\\
    H&\citet{rajala2010} & $\operatorname{MAD}$, $\operatorname{DCLF}$ with graph-based functional summary statistics and $K$, $G$, $F$, $J$, $pcf$; influence of the upper bound  & Matérn Type II hardcore, Thomas cluster, Baddeley-Silverman cell process, Line segments-Cox process, Random intrusion model & Clustering function competitive to $K$-function and more powerful for large upper bounds; connectivity functions have no discriminatory power \\
    I&\citet{baddeley2014} & $\operatorname{MAD}$, $\operatorname{DCLF}$ with $L$, $G^\bigstar$; influence of the upper bound & Strauss, Matérn cluster process & $\operatorname{DCLF}$ in most cases more powerful than $\operatorname{MAD}$ for $G^\bigstar$ and $L$; exception $\operatorname{MAD}$ with upper bound close to range of interaction of regular process \\
    \bottomrule
    \end{tabularx}
    \caption{Summary of previous power studies comparing tests for the null hypothesis of complete spatial randomness}
\label{tab:lit-overview-1}
\end{sidewaystable}

\begin{sidewaystable}
    \centering
    \renewcommand{\arraystretch}{1.5}
   \begin{tabularx}{\textwidth}{cp{3cm}p{5cm}p{4cm}X}\toprule
    ID & Study & Comparisons & Alternatives & Key Findings\\ \midrule \relax
    J&\citet{robins2016} & TDA-based rank functions and classical functional summary statistics with (weighted) $\operatorname{DCLF}$ & Strauss, Matérn cluster, Baddeley-Silverman cell process & Rank functions more powerful than $K$, $L$, $G$, $F$ against Baddeley-Silverman and Strauss; competitive against Matérn cluster; $1$-dim rank function performs worse than $0$-dim rank function due to boundary effects \\
    K&\citet{myllymaki2017} & Extreme rank length envelope tests; other depth functions; (scaled) deviation-based test statistics for $L$, $J$; varying upper bound & Strauss, Matérn cluster processes & Extreme rank length and $\operatorname{QDIR}$, $\operatorname{ST}$ most robust w.r.t. upper bound; unscaled deviation-based test statistics and other depth function less powerful\\
    L&\citet{mrkvicka2017} & All one-step combinations of $L$, $F$, $G$, $J$ with extreme rank length envelope tests; other multiple testing corrections & Strauss, Matérn cluster, Non-overlapping Matérn cluster, Superposition of two indep. Matérn cluster processes & Power of the combined tests not much smaller than for the most powerful summary statistic alone\\
    M& \citet{ebner2018} & Asymptotic tests using the Minkowski functionals; classical scalar indices; $\operatorname{DCLF}$ with $L$; varying number of points & inhomogeneous Poisson processes, Baddeley-Silverman, Matérn cluster process & Minkowski functionals more powerful than $L$ and Quadrat-Counts against Baddeley-Silverman; optimizing of tuning parameters is needed for the Minkowski functional approaches; test combining the different Minkowski functionals outperforms the individual tests\\
    N&\citet{biscio2019} & Extreme rank global envelope tests with accumulated persistence functions & Baddeley-Silverman cell, Matérn cluster process, Bessel-type DPP & $\operatorname{APF}$s with extreme rank ordering powerful against Baddeley-Silverman but less powerful compared to test of \citet{robins2016}; combination of $\operatorname{APF}_0$ and $\operatorname{APF}_1$ more powerful than individual summary statistics\\
    O&\citet{biscio2020} & Asymptotic tests using integral of $\operatorname{ND}_0$ and $L$, point evaluation of $\operatorname{APF}_1$; extreme rank global envelope tests with same summary statistics & Matérn cluster, Strauss, Baddeley-Silverman cell & TDA-based statistics individually less powerful than $L$ except for Baddeley-Silverman; linear combination of $0$- and $1$-dim features more powerful than individual tests \\
    P&\citet{heinrichmertsching2024} & Continuous ranked probability score for $K$ in two-sample tests & inhomogeneous Poisson, Strauss, inhomogeneous Thomas & Mean score able to differentiate the different models even Baddeley-Silverman and Poisson \\
    Q&\citet{biscio2022} & Asymptotic $\operatorname{MAD}$ tests with $K$; varying size of the observation window and upper bound & log-Gaussian Cox process with exponential covariance, Strauss process & Large number of observed points are necessary to detect deviations from CSR using the asymptotic tests\\
    R&\citet{botnan2022} & Asymptotic tests for scalar summaries of multiparameter TDA approaches, integral of $K$ & Matérn cluster, Strauss, Baddeley-Silverman cell process & $K$ more powerful for simple point processes but TDA offers insights for processes with complex interactions such as Baddeley-Silverman\\
    S&\citet{krebs2022} & Asymptotic tests using integral of $\operatorname{ND}_0$ and $L$, point evaluation of $\operatorname{APF}_1$ & Matérn cluster processes, Strauss processes & Performance of TDA-based tests varies depending on the specific setting; relying on a single TDA-based test is discouraged \\
    \bottomrule
    \end{tabularx}
    \caption{Summary of previous power studies comparing tests for the null hypothesis of complete spatial randomness}
\label{tab:lit-overview-2}
\end{sidewaystable}

\begin{figure}[ht]
	\subfloat[\citet{baddeley2014}]{\includegraphics[width=0.245\textwidth]{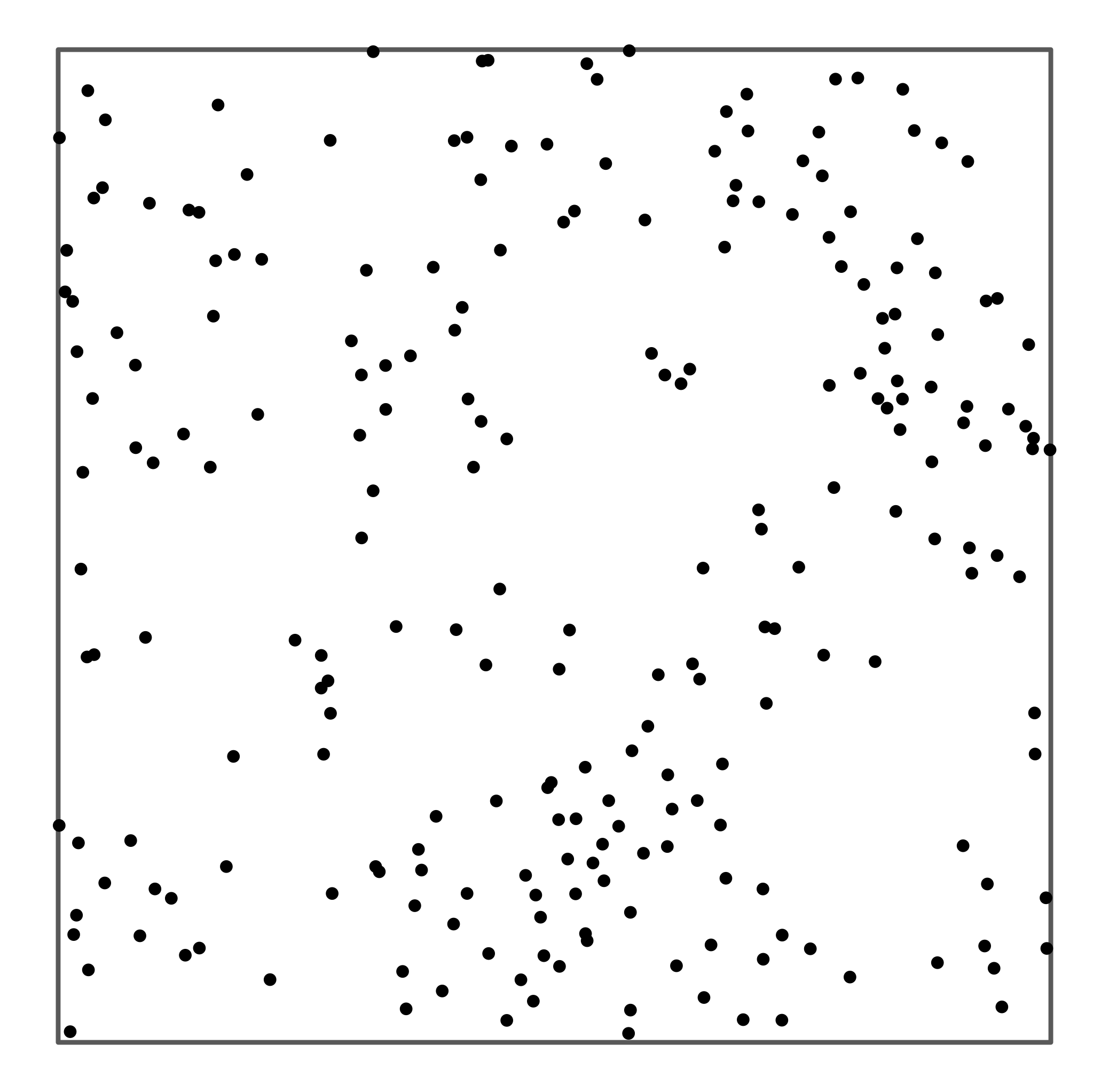}} \hfill
	\subfloat[\citet{robins2016}]{\includegraphics[width=0.245\textwidth]{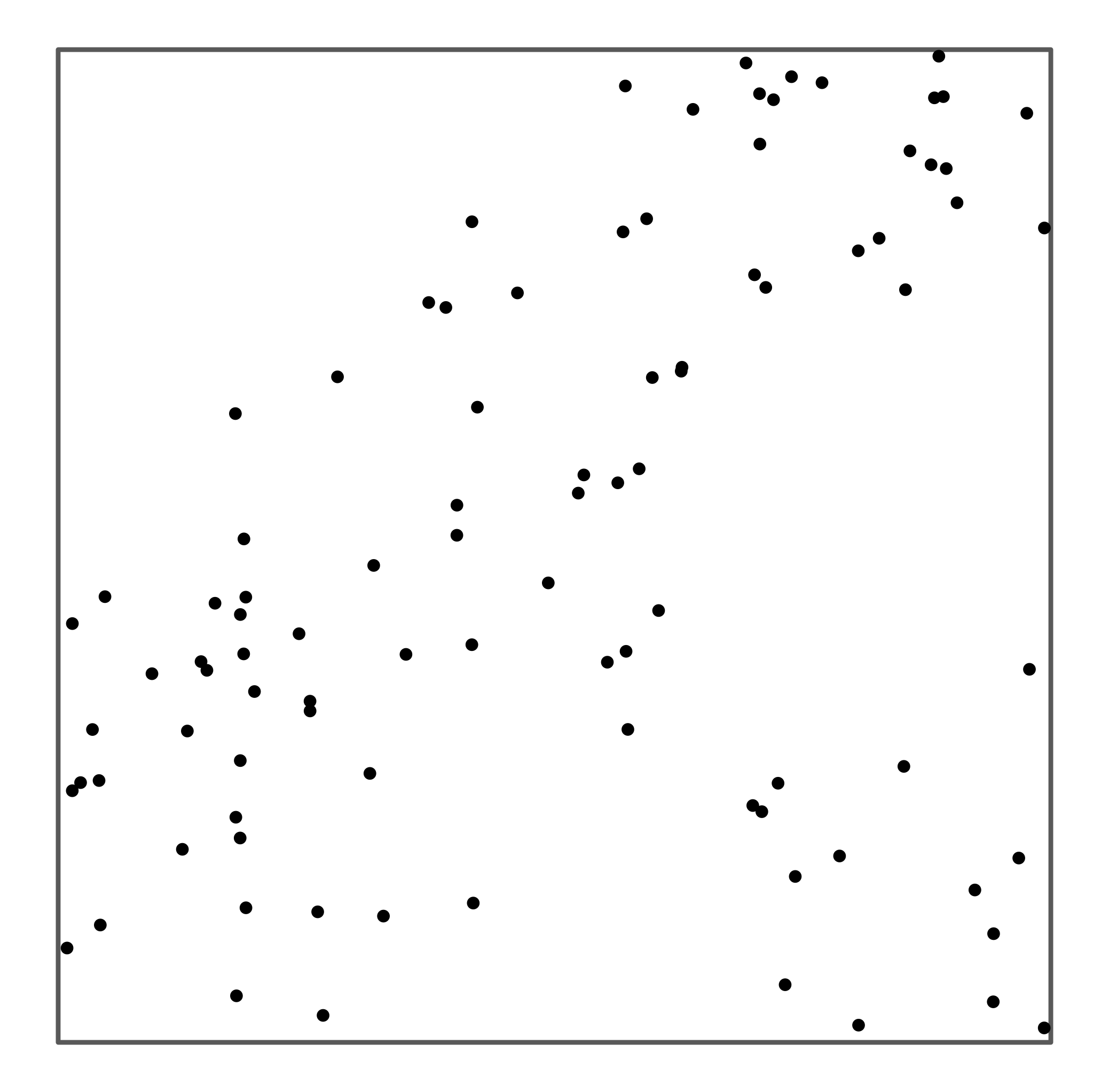}}\hfill 
	\subfloat[\citet{myllymaki2017}]{\includegraphics[width=0.245\textwidth]{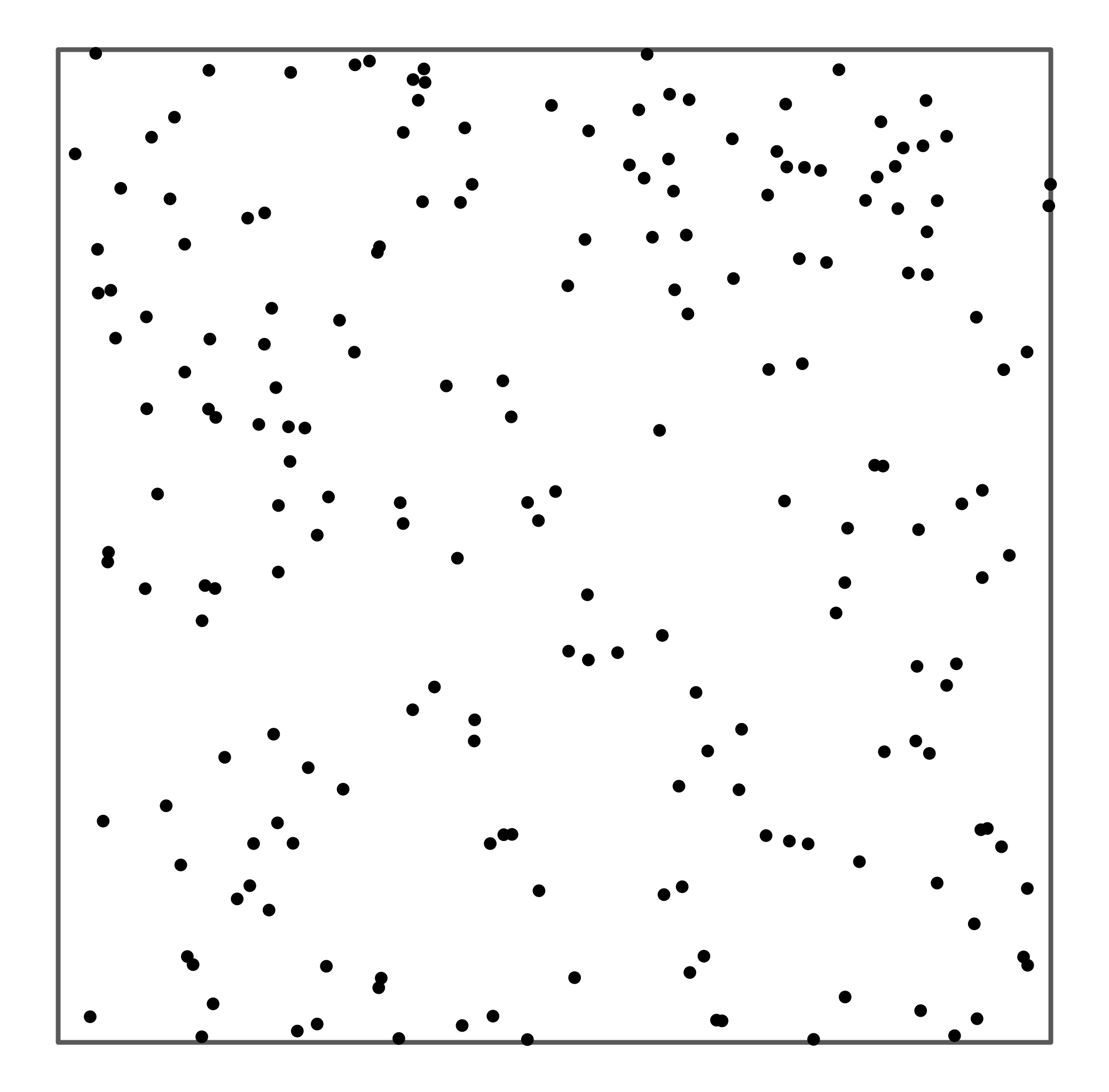}}\hfill
	\subfloat[\citet{biscio2019}]{\includegraphics[width=0.245\textwidth]{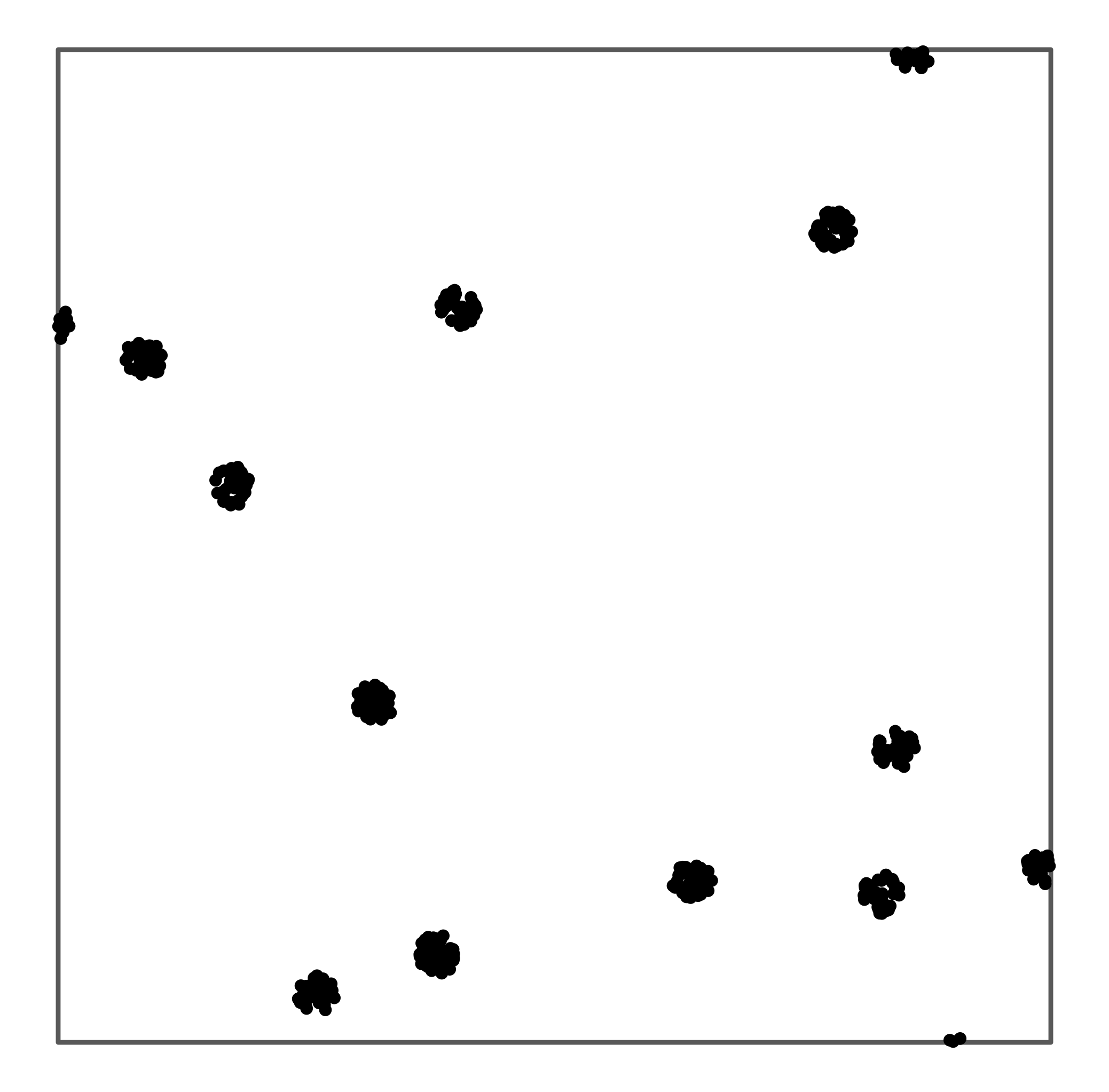}}
	\caption{Four realizations of Matérn cluster point processes MatClust($\kappa$, $R$, $\mu$) that are used as alternatives. From left to right: MatClust($0.005$, $14$, $5$), MatClust($10$, $0.2$, $10$), MatClust($1000$, $0.015$, $0.2$) and MatClust($10$, $0.02$, $40$). Pattern (a) is observed in $[0,100]^2$, (b) - (d) in the unit square $[0,1]^2$.}
	\label{fig:sim-study-models}
\end{figure}

Tables~\ref{tab:lit-overview-1} and ~\ref{tab:lit-overview-2} provide an overview of relevant papers that include power studies for the null hypothesis of complete spatial randomness. We list only works that compare at least two tests. Some of the tests that were used for the comparison are specialized tests for the CSR hypothesis. All of them are discussed in Section~\ref{sec:csr-hypo}. The functional summary statistics are introduced in Section~\ref{sec:summary-stat}. Although many studies use the same type of alternative model, the setting is often very different due to the parameters chosen. For example, a Matérn cluster process can be configured so that its realizations resemble a Poisson point process while another set of parameters yields realizations with very pronounced clusters. Four realizations of the models used are shown in Figure~\ref{fig:sim-study-models}. 

As individual studies sometimes compare many tests that fit into our proposed general framework, we also provide Table~\ref{tab:lit-combi} to help identify  studies that use the same tests (but usually in different test settings). The letters in Table~\ref{tab:lit-combi} correspond to the corresponding entries of Tables~\ref{tab:lit-overview-1} and ~\ref{tab:lit-overview-2}.

The main findings of the previous studies are the following:
\begin{itemize}
    \item Tests based on functional summary statistics should be preferred over any of the classical scalar-valued indices.
    \item When using functional summary statistics derived from associated structures (i.e. rank functions, accumulated persistence functions, Minkowski functionals, clustering function, \dots) one should not rely on a single characteristic but combine multiple aspects to obtain powerful tests.
    \item TDA-based functional summary statistics are especially good when dealing with processes with more complex interactions.
    \item Many possible combinations of test statistic and functional summary statistic have not yet been investigated.
\end{itemize}

\begin{table*}\centering
\begin{threeparttable}
\renewcommand{\arraystretch}{1.5}
\setlength{\tabcolsep}{3pt}
\rowcolors{3}{gray!20}{white}
\begin{tabularx}{\textwidth}{>{\,}llLLLLLLLLLLLLL} \toprule
\hiderowcolors 
\multicolumn{2}{l}{\multirow{3}{*}{\textbf{Test Statistic} $D$}} &  \multicolumn{13}{c}{\textbf{Summary Statistic} $T$} \\  \cmidrule{3-15}
& & \multicolumn{3}{c}{Second order} &  \multicolumn{4}{c}{Distance-based} & \multicolumn{6}{c}{TDA-based}\\
\cmidrule(r){3-5} \cmidrule(r){6-9} \cmidrule{10-15}
& & $K$ & $L$ & $pcf$ & $F$ & $G$ & $G^{\bigstar}$ & $J$ & $\operatorname{APF}_0$ & $\operatorname{APF}_1$ & $\operatorname{ND}_0$ & $\beta_0$ & $\beta_1$ & $\chi$\\ \midrule
\multicolumn{2}{l}{\rowgroup{\textbf{Type A}}}\\ \cmidrule{1-2}
\showrowcolors
\multicolumn{2}{l}{\cellcolor{white}$\operatorname{MAD}$} & A, H, Q\tnote{*} & A, B, C, F, G, I, K & H & A, D, H & A, D, H, G & I & D, H, K & & & & & &  \\
\multicolumn{2}{l}{\cellcolor{white}$\operatorname{ST}$} & & K & & & & & K  & & & & & &  \\
%$\operatorname{Q}$ & & & & & & & & & & & & &  \\
\multicolumn{2}{l}{\cellcolor{white}$\operatorname{QDIR}$} & & K & &  & & & K  & & & & & &  \\[0.75ex]
\multicolumn{2}{l}{\cellcolor{white}$\operatorname{DCLF}$} & H, J & D, I, J, K, M& H& D, H, J& D, H, J & I & D, H, K & & & & J\tnote{\textdaggerdbl} & J\tnote{\textdaggerdbl} &  \\
\multicolumn{2}{l}{\cellcolor{white}$\operatorname{ST,DCLF}$} & & K & & & & & E\tnote{\P}, K  & & & & & &  \\
%$\operatorname{Q,DCLF}$ & & & & & & & & & & & & &  \\
 \multicolumn{2}{l}{\cellcolor{white}$\operatorname{QDIR,DCLF}$} & & K & & & & & K  & & & & & & \\[0.75ex]
\multicolumn{2}{l}{\cellcolor{white}$\operatorname{CRPS}$} & P & & & & & & & & & & & &  \\[1.25ex] %\cmidrule{1-14}
\hiderowcolors 
\multicolumn{2}{l}{\rowgroup{\textbf{Type B}}}\\ \cmidrule{1-2}
\showrowcolors
\multicolumn{2}{l}{\cellcolor{white}$\operatorname{POINT}$} & & & & & & & & R\tnote{*\S} & O\tnote{*}, R\tnote{*\S} S\tnote{*} & & & & M\tnote{*} \\
\multicolumn{2}{l}{\cellcolor{white}$\operatorname{INT}$} & R, S\tnote{*} & O\tnote{*} & & & & & & & & O\tnote{*}, S\tnote{*} & & &  \\[1.25ex] %\cmidrule{1-14}
\hiderowcolors 
\multicolumn{2}{l}{\rowgroup{\textbf{Type C}}}\\ \cmidrule{1-2}
\showrowcolors
\cellcolor{white} & \cellcolor{white}$\operatorname{RANK}$ & & O& & & & & & N\tnote{\textdagger} & N\tnote{\textdagger}, O & O& & &  \\
\cellcolor{white}  & \cellcolor{white}$\operatorname{ERL}$ & & K, L\tnote{\textdagger} & & L\tnote{\textdagger} & L\tnote{\textdagger} & & K, L\tnote{\textdagger} & & & & & &  \\
\cellcolor{white}  & \cellcolor{white}$\operatorname{AREA}$ & & & & & & & & & & & & &  \\
\multirow{-4}{0.8cm}{\cellcolor{white}$\operatorname{FUN}$  with} & \cellcolor{white}$\operatorname{CONT}$ & & & & & & & & & & & & &  \\[0.75ex]
\multicolumn{2}{l}{\cellcolor{white} $\operatorname{SCORE}$} & & & & & & & & & & & & &  \\
\bottomrule
\end{tabularx}
\begin{tablenotes}
\item[*] asymptotic tests based on limit theorems
\item[\textdagger] including combined envelope tests
\item[\textdaggerdbl] test statistic includes weight function
\item[\S] test statistic is a linear combination of multiple test statistics
\item[\P] pure studentized deviations instead of squared studentized deviations 
\end{tablenotes}
\caption{Overview of the combinations of test statistic and summary statistic in the power studies stated in Table~\ref{tab:lit-overview-1} and Table~\ref{tab:lit-overview-2}. The letters indicate the corresponding entry of the two summary tables. If not marked otherwise Monte Carlo tests were conducted.}
\label{tab:lit-combi}
\end{threeparttable}
\end{table*}

\section{Discussion and future research}
\label{sec:conclusion}

This paper presents a review on both classical approaches and recent developments in goodness-of-fit testing for spatial point processes. Our main contribution is the introduction of a unifying notation which allows to establish a general framework for goodness-of-fit tests based on functional summary statistics. These summaries are based on either probabilistic, geometric or topological descriptors of point processes. Several options for test statistics and corresponding orderings are introduced. Thus, our framework contains a large number of different goodness-of-fit tests. Additionally, we also discussed several practical aspects of the tests such as the estimation of unknown quantities and the necessary number of simulations of the null model.

We reviewed power studies on goodness-of-fit testing for spatial point processes found in the literature. While these studies cover only a comparatively small number of tests that can be constructed using our framework, some key findings are already possible, as outlined in Section~\ref{sec:powerstudy}. In our opinion, the following topics have not yet been sufficiently investigated.

The first one is the combination of different types of functional summary statistics, e.g. a classical functional summary statistics with a TDA-based one, see Section~\ref{sec:combi}. In particular, it is not known, which of the TDA-based functional summary statistics complements which classical one. Furthermore, other combination techniques such as linear combinations are of interest. The main question is whether there are combinations that provide more powerful tests against several alternatives than the individual tests. In the available simulation studies, already a combination of two TDA-summaries led to tests whose powers depend less on the specific alternative.

For many of the functional summary statistics, only selected test statistics have been considered. It is therefore not clear whether one can obtain better performing tests by using one of the other test statistics. Table~\ref{tab:lit-overview-2} lists the combinations that have been investigated in the literature, which clearly shows several gaps.

Which functional summary statistics one might consider depends on the null hypothesis in question. Consequently, it is important and necessary to know which type of deviations each test can detect.

The last aspect is still the number of simulations required in each setting. Some null models are more expensive to simulate than others, and thus it is beneficial to know lower bounds that give valid and powerful tests. This is in particular of interest if we aim to use the graphical representation. A high number of several thousands is recommended for the extreme rank ordering. For the other vector orderings, smaller numbers are possible in theory but guidelines for the point process setting are not available.

Our forthcoming power study will investigate the above-mentioned topics.

\bibliography{bibliography} 
\end{document}